\documentclass[longauth]{aa/aa}

\usepackage{lineno} 
\usepackage{graphicx}
\usepackage{txfonts}
\usepackage{lipsum}
\usepackage{subcaption}         
\usepackage{lscape}             
\usepackage{longtable}
\usepackage{placeins}           

\usepackage{CJKutf8}     
\usepackage{tabularx}
\usepackage{threeparttable}

\def\gridline#1{\vskip6pt\hbox to\hsize{#1}\vskip6pt}

\def\fig#1#2#3{\hfill\vbox{\parskip=0pt\hsize=#2
\includegraphics[width=#2]{#1}\vskip2pt\vtop{\centering
\footnotesize
\hsize=#2
#3\vskip1pt
}}\hfill}

\def\leftfig#1#2#3{\vbox{\parskip=0pt\relax\hsize=#2
\includegraphics[width=#2]{#1}\vskip2pt\vtop{\hsize=#2
\centering
#3\vskip1sp\vskip1sp}}\hfill}

\newcommand*{\orcid}[1]{
    \href{https://orcid.org/#1}{\,\raisebox{0.2em}{
    \hspace{-1em}
    \includegraphics[height=0.6em,width=0.6em]{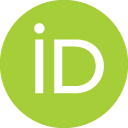}
}}}

\newcommand{\numGalaxies}{40}
\newcommand{\numGMCs}{8984}
\usepackage{xspace}
\newcommand{\Icokpc}{$I_{\mathrm{CO, 1kpc}}$\xspace}

\newcommand{\rgal}{$r_{\mathrm{gal}}$\xspace}
                                
\usepackage[pdfpagelabels=false]{hyperref}      
\hypersetup{colorlinks=true,linkcolor=blue,citecolor=blue,filecolor=blue,urlcolor=blue,}

\begin{document}

   \title{The Structure of Molecular Gas in PHANGS-ALMA Galaxies: Cloud Spacing, Two-Point Correlation and Stacked Intensity Profiles}
    \titlerunning{Clustering of molecular gas in PHANGS--ALMA}



\newcommand{\OSU}{\label{OSU} Department of Astronomy, The Ohio State University, 140 West 18th Avenue, Columbus, Ohio 43210, USA}

\newcommand{\Alberta}{\label{Alberta} Department of Physics, University of Alberta, Edmonton, AB T6G 2E1, Canada}

\newcommand{\ANU}{\label{ANU} Research School of Astronomy and Astrophysics, Australian National University, Canberra, ACT 2611, Australia}

\newcommand{\IPAC}{\label{IPAC} Caltech-IPAC, 1200 E. California Blvd. Pasadena, CA 91125, USA}

\newcommand{\Carnegie}{\label{Carnegie} Observatories of the Carnegie Institution for Science, 813 Santa Barbara Street, Pasadena, CA 91101, USA}

\newcommand{\CCAPP}{\label{CCAPP} Center for Cosmology and Astroparticle Physics, 191 West Woodruff Avenue, Columbus, OH 43210, USA}

\newcommand{\CfA}{\label{CfA}Harvard-Smithsonian Center for Astrophysics, 60 Garden Street, Cambridge, MA 02138, USA}

\newcommand{\CITEVA}{\label{CITEVA} Centro de Astronomía (CITEVA), Universidad de Antofagasta, Avenida Angamos 601, Antofagasta, Chile}

\newcommand{\CNRS}{\label{CNRS} CNRS, IRAP, 9 Av. du Colonel Roche, BP 44346, F-31028 Toulouse cedex 4, France}

\newcommand{\ESO}{\label{ESO} European Southern Observatory, Karl-Schwarzschild Stra{\ss}e 2, D-85748 Garching bei M\"{u}nchen, Germany}

\newcommand{\ESOChile}{\label{ESOChile} European Southern Observatory, Avenida Alonso de Cordoba 3107, Casilla 19, Santiago 19001, Chile}

\newcommand{\HD}{\label{HD} Astronomisches Rechen-Institut, Zentrum f\"{u}r Astronomie der Universit\"{a}t Heidelberg, M\"{o}nchhofstra\ss e 12-14, D-69120 Heidelberg, Germany}

\newcommand{\ICRAR}{\label{ICRAR} International Centre for Radio Astronomy Research, University of Western Australia, 35 Stirling Highway, Crawley, WA 6009, Australia}

\newcommand{\IRAM}{\label{IRAM} Institut de Radioastronomie Millim\'{e}trique (IRAM), 300 Rue de la Piscine, F-38406 Saint Martin d'H\`{e}res, France}

\newcommand{\ITA}{\label{ITA} Universit\"{a}t Heidelberg, Zentrum f\"{u}r Astronomie, Institut f\"{u}r Theoretische Astrophysik, Albert-Ueberle-Str 2, D-69120 Heidelberg, Germany}

\newcommand{\IWR}{\label{IWR} Universit\"{a}t Heidelberg, Interdisziplin\"{a}res Zentrum f\"{u}r Wissenschaftliches Rechnen, Im Neuenheimer Feld 205, D-69120 Heidelberg, Germany}

\newcommand{\JHU}{\label{JHU} Department of Physics and Astronomy, The Johns Hopkins University, Baltimore, MD 21218, USA}

\newcommand{\Leiden}{\label{Leiden} Leiden Observatory, Leiden University, P.O. Box 9513, 2300 RA Leiden, The Netherlands}

\newcommand{\Maryland}{\label{Maryland} Department of Astronomy, University of Maryland, College Park, MD 20742, USA}

\newcommand{\MPE}{\label{MPE} Max-Planck-Institut f\"{u}r extraterrestrische Physik, Giessenbachstra{\ss}e 1, D-85748 Garching, Germany}

\newcommand{\MPIA}{\label{MPIA} Max-Planck-Institut f\"{u}r Astronomie, K\"{o}nigstuhl 17, D-69117, Heidelberg, Germany}

\newcommand{\Nagoya}{\label{Nagoya} Department of Physics, Nagoya University, Furo-cho, Chikusa-ku, Nagoya, Aichi 464-8602, Japan}

\newcommand{\NRAO}{\label{NRAO} National Radio Astronomy Observatory, 520 Edgemont Road, Charlottesville, VA 22903-2475, USA}

\newcommand{\OAN}{\label{OAN} Observatorio Astron\'{o}mico Nacional (IGN), C/Alfonso XII, 3, E-28014 Madrid, Spain}

\newcommand{\ObsParis}{\label{ObsParis} Sorbonne Universit\'{e}, Observatoire de Paris, Universit\'{e} PSL, CNRS, LERMA, F-75014, Paris, France}

\newcommand{\Princeton}{\label{Princeton} Department of Astrophysical Sciences, Princeton University, Princeton, NJ 08544 USA}

\newcommand{\UToledo}{\label{UToledo} University of Toledo, 2801 W. Bancroft St., Mail Stop 111, Toledo, OH, 43606}

\newcommand{\Toulouse}{\label{Toulouse} Universit\'{e} de Toulouse, UPS-OMP, IRAP, F-31028 Toulouse cedex 4, France}

\newcommand{\UBonn}{\label{UBonn} Argelander-Institut f\"ur Astronomie, Universit\"at Bonn, Auf dem H\"ugel 71, 53121 Bonn, Germany}

\newcommand{\UChile}{\label{UChile} Departamento de Astronom\'{i}a, Universidad de Chile, Camino del Observatorio 1515, Las Condes, Santiago, Chile}

\newcommand{\UConn}{\label{UConn} Department of Physics, University of Connecticut, Storrs, CT, 06269, USA}

\newcommand{\UCSD}{\label{UCSD}Center for Astrophysics and Space Sciences, Department of Physics,  University of California, San Diego, 9500 Gilman Drive, La Jolla, CA 92093, USA}

\newcommand{\UGent}{\label{UGent} Sterrenkundig Observatorium, Universiteit Gent, Krijgslaan 281 S9, B-9000 Gent, Belgium}

\newcommand{\ULyon}{\label{ULyon} Univ Lyon, Univ Lyon 1, ENS de Lyon, CNRS, Centre de Recherche Astrophysique de Lyon UMR5574,\\ F-69230 Saint-Genis-Laval, France}

\newcommand{\UMass}{\label{UMass} University of Massachusetts—Amherst, 710 N. Pleasant Street, Amherst, MA 01003, USA}

\newcommand{\UWyoming}{\label{UWyoming} Department of Physics and Astronomy, University of Wyoming, Laramie, WY 82071, USA}

\newcommand{\LAM}{\label{LAM} Aix Marseille Univ, CNRS, CNES, LAM (Laboratoire d’Astrophysique de Marseille), Marseille, France}

\newcommand{\UHawaii}{\label{UHawaii} Institute for Astronomy, University of Hawaii, 2680 Woodlawn Drive, Honolulu, HI 96822, USA}

\newcommand{\UCM}{\label{UCM} Departamento de F\'{\i}sica de la Tierra y Astrof\'{\i}sica, Universidad Complutense de Madrid, E-28040, Spain}

\newcommand{\IPARC}{\label{IPARC} Instituto de F\'{\i}sica de Part\'{\i}culas y del Cosmos IPARCOS, Facultad de Ciencias F\'{\i}sicas, Universidad Complutense de Madrid, E-28040, Spain}

\newcommand{\STScI}{\label{STScI} Space Telescope Science Institute, 3700 San Martin Drive, Baltimore, MD 21218, USA}

\newcommand{\McMaster}{\label{McMaster} Department of Physics and Astronomy, McMaster University, 1280 Main Street West, Hamilton, ON L8S 4M1, Canada}

\newcommand{\INAF}{\label{INAF} INAF -- Osservatorio Astrofisico di Arcetri, Largo E. Fermi 5, I-50157, Firenze, Italy}

\newcommand{\Sydney}{\label{Sydney} Sydney Institute for Astronomy, School of Physics A28, The University of Sydney, NSW 2006, Australia}

\newcommand{\CITA}{\label{CITA} Canadian Institute for Theoretical Astrophysics (CITA), University of Toronto, 60 St George St, Toronto, ON M5S 3H8, Canada}

\newcommand{\ASIAA}{\label{ASIAA} Institute of Astronomy and Astrophysics, Academia Sinica, No. 1, Sec. 4, Roosevelt Road, Taipei 10617, Taiwan}

\newcommand{\TKU}{\label{TKU} Department of Physics, Tamkang University, No.151, Yingzhuan Rd., Tamsui Dist., New Taipei City 251301, Taiwan}

\newcommand{\PSMA}{\label{PSMA} Penn State Mont Alto, 1 Campus Drive, Mont Alto, PA  17237, USA}

\newcommand{\ILL}{\label{ILL} ILL}

\newcommand{\stromlo}{\label{stromlo} Research School of Astronomy and Astrophysics, Australian National University, Mt Stromlo Observatory, Weston Creek, ACT 2611, Australia}

\newcommand{\UCatolica}{\label{UCatolica} Instituto de Astronom\'ia, Universidad Cat\'olica del Norte, Av. Angamos 0610, Antofagasta, Chile}

\newcommand{\UT}{\label{UT} McDonald Observatory, The University of Texas at Austin, 1 University Station, Austin, TX 78712-0259, USA}

\newcommand{\Vanderbilt}{\label{Vanderbilt} Department of Physics and Astronomy, Vanderbilt University, VU Station 1807, Nashville, TN 37235, USA}

\newcommand{\UNF}{\label{UNF} Department of Physics, University of North Florida, 1 UNF Dr. Jacksonville FL 32224}

\newcommand{\NAOC}{\label{NAOC} Chinese Academy of Sciences South America Center for Astronomy, National Astronomical Observatories, CAS, Beijing 100101, China}

\newcommand{\CASA}{\label{CASA} Center for Astrophysics and Space Astronomy, University of Colorado, 389 UCB, Boulder, CO 80309-0389, USA}

\newcommand{\UNAM}{\label{UNAM} Universidad Nacional Aut\'onoma de M\'exico, Instituto de Astronom\'ia, AP 106, Ensenada 22800, BC, M\'exico}

\newcommand{\UDP}{\label{UDP} Instituto de Estudios Astrof\'isicos, Facultad de Ingenier\'ia y Ciencias, Universidad Diego Portales, Av. Ej\'ercito Libertador 441, Santiago, Chile}

\newcommand{\Steward}{\label{Steward} Steward Observatory, University of Arizona, 933 N. Cherry Ave., Tucson, AZ 85721-0065, USA}  

\newcommand{\APO}{\label{APO} Apache Point Observatory and New Mexico State University, P.O.\ Box 59,
Sunspot, NM 88349-0059, USA}

\newcommand{\UNAMCU}{\label{UNAMCU} Universidad Nacional Aut\'onoma de M\'exico, Instituto de Astronom\'ia, AP 70-264, CDMX 04510, M\'exico}

\newcommand{\UWash}{\label{UWash}Department of Astronomy, University of Washington, Seattle, WA, 98195}

\newcommand{\CC}{\label{CC}Department of Physics, Colorado College, Colorado Springs, CO 80903}

\newcommand{\Utah}{\label{Utah}Department of Physics and Astronomy, University of Utah, 115 S. 1400 E., Salt Lake City, UT 84112, USA}

\newcommand{\UConcepcion}{\label{UConcepcion}Departamento de Astronom\'ia, Universidad de Concepci\'on, Casilla 160-C, Concepci\'on, Chile}

\newcommand{\FCLA}{\label{FCLA}Franco-Chilean Laboratory for Astronomy, IRL 3386, CNRS and Universidad de Chile, Santiago, Chile}

\newcommand{\Oklahoma}{\label{Oklahoma}Homer L. Dodge Department of Physics and Astronomy, University of Oklahoma, Norman, OK 73019, USA}

\newcommand{\UIUC}{\label{UIUC}Department of Astronomy, University of Illinois, Urbana, IL 61801, USA}

\newcommand{\Harvard}{\label{Harvard}Harvard-Smithsonian Center for Astrophysics, Cambridge, MA 02138, USA}

\newcommand{\caltech}{\label{caltech}Department of Astronomy, California Institute of Technology, Pasadena, CA 91125, USA}

\newcommand{\UOA}{\label{UOA}Department of Physics, University of Arkansas, 226 Physics Building, 825 West Dickson Street, Fayetteville, AR 72701, USA}

\newcommand{\units}{\label{units}Department of Physics, Astronomy Section, University of Trieste, Via G.B. Tiepolo, 11, I-34143 Trieste, Italy}

\newcommand{\Rad}{\label{Rad}{Elizabeth S. and Richard M. Cashin Fellow at the Radcliffe Institute for Advanced Studies at Harvard University, 10 Garden Street, Cambridge, MA 02138, USA}}

\newcommand{\UCT}{\label{UCT}{Department of Astronomy, University of Cape Town, Rondebosch 7701, Cape Town, South Africa}}

\newcommand{\LJMU}{\label{LJMU}{Astrophysics Research Institute, Liverpool John Moores University, IC2, Liverpool Science Park, 146 Brownlow Hill, Liverpool L3 5RF, UK}}

\newcommand{\COOL}{\label{COOL}{Cosmic Origins Of Life (COOL) Research DAO, \href{https://coolresearch.io}{https://coolresearch.io}}}

\newcommand{\Ox}{\label{Ox}{Sub-department of Astrophysics, Department of Physics, University of Oxford, Keble Road, Oxford OX1 3RH, UK}}

\newcommand{\UShizuokaGlobal}{\label{UShizuokaGlobal}{Faculty of Global Interdisciplinary Science and Innovation, Shizuoka University, 836 Ohya, Suruga-ku, Shizuoka 422-8529, Japan}}

\newcommand{\NAOJ}{\label{NAOJ}{National Astronomical Observatory of Japan, 2-21-1 Osawa, Mitaka, Tokyo, 181-8588, Japan}}

\newcommand{\StU}{\label{StU}{SUPA, School of Physics and Astronomy, University of St Andrews, North Haugh, St Andrews, KY16 9SS}}


\author{
       Hao He \begin{CJK*}{UTF8}{gbsn}(何浩)\end{CJK*}\inst{\ref{UBonn}} \thanks{\email{hhe@astro.uni-bonn.de}} \orcid{0000-0001-9020-1858}  
       \and  Adam K. Leroy\inst{\ref{OSU}, \ref{CCAPP}} \orcid{0000-0002-2545-1700} 
       \and Erik  Rosolowsky \inst{\ref{Alberta}} \orcid{0000-0002-5204-2259}  
       \and Annie Hughes \inst{\ref{Toulouse}} \orcid{0000-0002-9181-1161} 
       \and Jiayi~Sun \begin{CJK*}{UTF8}{gbsn}(孙嘉懿)\end{CJK*}\inst{\ref{Princeton}}\fnmsep\thanks{NASA Hubble Fellow} \orcid{0000-0003-0378-4667}
       \and  Joshua Machado \inst{\ref{OSU}, \ref{CCAPP}} 
       \and Frank Bigiel \inst{\ref{UBonn}} \orcid{0000-0003-0166-9745}
       \and Ashley Barnes \inst{\ref{ESO}} \orcid{0000-0003-0410-4504}
       \and Zein Bazzi \inst{\ref{UBonn}} \orcid{0009-0001-1221-0975}
       \and Yixian Cao \inst{\ref{MPE}} \orcid{0000-0001-5301-1326}
       \and M\'elanie Chevance\inst{\ref{ITA}, \ref{COOL}} \orcid{0000-0002-5635-5180}
       \and Dario Colombo\inst{\ref{UBonn}} \orcid{0000-0001-6498-2945}
       \and Simon C. O. Glover \inst{\ref{ITA}} \orcid{0000-0001-6708-1317}
       \and Jonathan~D.Henshaw \inst{\ref{LJMU}, \ref{MPIA}} \orcid{0000-0001-9656-7682}
       \and Eric~W.~Koch\inst{\ref{NRAO},\ref{CfA}}\orcid{0000-0001-9605-780X}
       \and Sharon E. Meidt \inst{\ref{UGent}} \orcid{0000-0002-6118-4048}
       \and Hsi-An Pan \inst{\ref{TKU}} \orcid{0000-0002-1370-6964}
       \and Toshiki Saito \inst{\ref{UShizuokaGlobal}, \ref{NAOJ}} \orcid{0000-0002-2501-9328}
       \and Sumit K. Sarbadhicary \inst{\ref{JHU}}\orcid{0000-0002-6313-4597}
       \and Eva Schinnerer \inst{\ref{MPIA}}\orcid{0000-0002-3933-7677}
       \and Rowan J. Smith \inst{\ref{StU}} \orcid{0000-0002-0820-1814}
       \and Antonio Usero \inst{\ref{OAN}} \orcid{0000-0003-1242-505X}
       \and David H. Weinberg\inst{\ref{OSU}, \ref{CCAPP}} \orcid{0000-0001-7775-7261} 
       \and Thomas G. Williams\inst{\ref{Ox}}\orcid{0000-0002-0012-2142}
}

\institute{\tiny
\UBonn     \and 
\OSU \and 
\CCAPP  \and
\Alberta \and
\Toulouse \and
\Princeton \and
\ESO \and 
\MPE \and
\ITA \and
\COOL \and
\LJMU \and 
\MPIA \and
\NRAO \and
\CfA \and
\UGent \and
\TKU \and
\UShizuokaGlobal \and
\NAOJ \and
\JHU \and
\StU \and
\OAN \and
\Ox
}

\date{Received September 30, 20XX}

 
\abstract
{The spatial distribution of giant molecular clouds (GMCs) at sub-kpc scales encodes information about cloud formation and evolution. However, we still lack a general quantitative characterization of molecular gas structure at this scale.}
{We aim for a quantitative description of molecular gas structure at 150 -- 1000 pc for a typical star-forming main sequence galaxy. We analyse how GMCs cluster together and how CO emission is spatially correlated with bright GMCs using a sample of \numGMCs\ GMCs from \numGalaxies\ galaxies observed by PHANGS-ALMA. } 
{We homogenize our data to a common spatial resolution of 150~pc and mass sensitivity of 2.5 M$_{\odot}$ pc$^{-2}$ to remove observational bias. We then calculate nearest neighbour distances, neighbour number density, and two-point correlation functions for the catalogued GMCs in each galaxy.  When analysing the two-point correlation function, we generate several control samples that reflect different null hypotheses on large spatial scales. We stack integrated intensity CO emission profiles around the position of catalogued GMCs to probe the gas distribution on scales between the observational resolution and the typical GMC-GMC spacing. }
{Our measurements of cloud spacing and number of neighbours show that GMC clustering follows the large-scale gas distribution. Once we account for this contribution, the peak excess clustering relative to the null hypothesis in the two-point correlation function drops from 
$1+\omega \sim $ 2.3 to 1.3, with the power-law slope flattened from -0.25 to 0. Stacks of CO intensity around local maxima show a strong clustering signal on scales smaller than the typical GMC-GMC separation. We show that this is largely the same signal captured by the ``GMC size'' measured by \texttt{CPROPS}, with an additional $\sim$20\% of the flux in an extended component beyond 500~pc. We find that our stacked profiles can be fit with a double Gaussian function plus a constant offset. The broad Gaussian component accounts for 70\% of the over-density power above the constant background, and is stronger around massive and gravitationally bound GMCs. 
}
{
Our measurements yield a general statistical description of the structure of CO emission from $\approx 150$~pc to galactic scales that can serve as a benchmark for 
simulations of molecular cloud formation and destruction in galaxy disks. Our results indicate that galactic structure exerts a strong influence on the GMC distribution in galaxy disks, and the formation of massive, gravitationally bound GMCs is related to strong local gas clustering.  
}


\maketitle

\section{Introduction}
\label{sec:intro}

Molecular gas is the interstellar medium (ISM) component that forms stars \citep[e.g.,][]{shu_87,bigiel_star_2008, kennicutt_star_2012}. The spatial distribution of this gas reflects the complex interplay between galactic dynamics, stellar feedback, and gravity. On small scales, turbulent theories of star formation predict that the fragmentation of molecular gas leads to a scale-free, hierarchically structured molecular phase \citep[references in][]{guszejnov_universal_2018} . On large scales, galactic structure and dynamics set the characteristic scales of instability and structure formation \citep[e.g.][]{elmegreen_1983,meidt_molecular_2022, meidt_bottom_2024} 
Large-scale dynamical features like bars and spiral arms can also create converging flows where molecular clouds can form \citep{dobbs_formation_2014}. Together with supernova feedback, these factors shape the gas distribution.  

Observations of the CO emission in the Milky Way 
and nearby galaxies 
show that the molecular phase is mainly composed of individual clumpy structures called giant molecular clouds (GMCs). 
There have been extensive studies of the size, mass, surface density, and dynamical state of these clouds \citep[e.g., see reviews in][]{heyer_molecular_2015, chevance_life_2023, schinnerer_molecular_2024}, as well as many studies of the global gas content of galaxies \citep[see reviews in][]{kennicutt_star_2012,saintonge_cold_2022}. The structure of the molecular gas on intermediate scales has been less explored \citep[but see ][ for a recent analysis of the clustering of molecular clouds in the Large Magellanic Cloud]{grishunin_lmc_2024}. Despite the potential utility of clustering metrics as a diagnostic of molecular cloud formation and destruction processes, we thus still lack a comprehensive statistical characterization of the spatial distribution of GMCs and CO emission in galaxy disks on scales of 100 -- 1000 pc, i.e. larger than individual GMCs but smaller than the bulk molecular gas distribution. 

In this study, we use the PHANGS--ALMA survey \citep{leroy_phangs-alma_2021} to address this gap.  We measure the clustering of CO(2-1) emission using three complementary techniques common in other parts of the astronomical literature. 
In addition to maps of the CO(2-1) integrated intensity, we analyse \texttt{CPROPS} catalogues of GMCs \citep[][A.~Hughes et al., in prep.]{rosolowsky_giant_2021}. Our three methods are:
\begin{enumerate}

\item \textbf{GMC Spacing and Number Density}: The typical distance between an object and its ($n$th) nearest neighbour is a simple measurement of the spatial clustering \citep{nearest_nbrs}.  A common tool in fields like galaxy population studies \citep[e.g.][]{nearest_nbrs_galaxy}, nearest neighbour analysis has some precedent in the analysis of molecular cloud substructure \citep[e.g.][]{isf_spacing}. 
In Section \ref{sec:spacing}, we examine how the GMC spacing varies as a function of the data characteristics and the local galactic environment.

\smallskip

\item \textbf{The Two-Point Correlation Function (2PCF)}: The 2PCF reflects the distribution of all GMC-GMC distances, and is thus sensitive to a broader range of structures than only the nearest neighbour distance. The 2PCF has been developed for and widely applied to large scale galaxy studies \citep{peebles_large-scale_1980}, but it has also been used in nearby galaxies to quantify the spatial distribution of clouds, and the relative distribution of clouds and young clusters \citep[e.g.,][]{zhang_antennae_2001, grasha_connecting_2018, grasha_spatial_2019, peltonen_clusters_2023}.  These works usually focus on a single galaxy \citep[but see][ for previous analysis using PHANGS--ALMA]{turner_phangs_2022}. These studies suggest that GMCs are not uniformly distributed across galaxies, but instead exhibit hierarchical clustering, with more massive GMCs associated with large scale gas concentrations, such as centres and spiral arms \citep[consistent with studies using other techniques including][]{sun_molecular_2022, stark_giant_2006, colombo_pdbi_2014, hirota_whole-disk_2024}. The 2PCF is most useful for assessing correlation relative to a ``control'' population, which allows us to explore the relationship of the GMC population to the large-scale CO emission distribution across our galaxy disks. Section \ref{sec:tpcf} presents our 2PCF analysis.


\smallskip

\item \textbf{Intensity Profiles around GMCs}: We also examine the average radial distribution of CO emission around the catalogued centres of GMCs. This approach assesses structure on scales below the cloud-cloud spacing and accesses some of the same information as the cloud sizes recorded in the GMC catalogues. This kind of spatial stacking analysis has been used extensively across many fields \citep[e.g.][]{qso_stacking} as well as in molecular cloud analysis \citep{radial_profiles_mw}. It can be viewed as a relative of the 2PCF, i.e., cross-correlation analysis between a set of points (the GMC centres) and the emission map. Section \ref{sec:stacking} presents our stacking analysis.
\end{enumerate}

Though we refer to the objects we study as ``Giant Molecular Clouds'', formally \texttt{CPROPS} yields a catalogue of emission concentrations at the resolution of the input data. For our main catalogue at 150 pc resolution, the mass sensitivity is $\sim 10^{6}$ M$_{\odot}$. Opinions vary as to whether these objects are massive individual clouds or ``Giant Molecular Associations'' of smaller clouds. The data-driven view is that these are peaks of emission at $150$~pc resolution identified from homogenized data in a reproducible manner. 

\section{Data} 
\label{sec:data}

PHANGS-ALMA observed CO(2-1) emission from 90 nearby galaxies ($d \lesssim$ 22~Mpc) at a resolution of $1{-}1.5''$ ($\approx 50{-}150$ pc). We refer to \citet{leroy_phangs-alma_2021} for a presentation of the sample selection and properties of the targets. We draw galaxy centres, distances, orientations, and exponential scale lengths from that paper, which in turn adopts distances from \citet{anand_distances_2021} and orientations from \citet{lang_phangs_2020}. 

\subsection{GMC catalogues}
\label{subsec:GMC_cat}

We analyse the PHANGS GMC catalogues derived by \citet{rosolowsky_giant_2021} and A. Hughes et al. (in prep.) and analysed in \citet{sun_molecular_2022}. GMCs are extracted using the \texttt{CPROPS} algorithm \citep{rosolowsky_bias-free_2006, rosolowsky_giant_2021}. \texttt{CPROPS} identifies significant local maxima in position-position-velocity space, with ``significant'' here defined as showing S/N $>$ 4 over two consecutive velocity channels and at least a 2$\times$RMS contrast against the local background. Then, pixels with significant emission are assigned membership in clouds using a watershed-based approach \citep{rosolowsky_giant_2021}. After segmentation, the algorithm determines the properties of each cloud using moment methods. These measurements are corrected for the finite angular and spectral resolution of the telescope via deconvolution and corrected for the finite sensitivity of the observations via extrapolation methods.

Despite attempts at corrections, the measured cloud properties depend on the resolution and the sensitivity of the data \citep[see][]{pineda_perils_2009,hughes_comparative_2013,leroy_portrait_2016,rosolowsky_giant_2021}. To control for this, the GMC catalogue includes GMCs extracted from data cubes homogenized to have fixed physical resolutions (60, 90, 120 and 150~pc, as allowed by the data) and sensitivities (two sensitivity thresholds). For details of the homogenization procedure see \citet{rosolowsky_giant_2021}. 

In our main analyses, we use data with 150~pc resolution and high sensitivity (46~mK per 2.5~km~s$^{-1}$ channel). We choose 150~pc resolution to include as many galaxies as possible. The choice to use the ``high sensitivity'' subset reduces our sample from $90$ to $40$ galaxies, but the sensitivity is two times better than the value achieved across the full sample at this physical resolution. This minimizes the impact of low completeness, i.e., the catalogued clouds contain a large fraction of the total flux in the input CO data cubes. Specifically, these catalogues have median flux recovery of $\sim$ 82\%, though the completeness level varies some with environment. For example, the flux recovery fraction in the spiral arms is $\sim80\%$, while that of the interarm regions is $\sim60$\% (see Hughes et al. in prep.). 


\begin{figure*}[ht!]
\centering
\includegraphics[width = 0.8\textwidth]{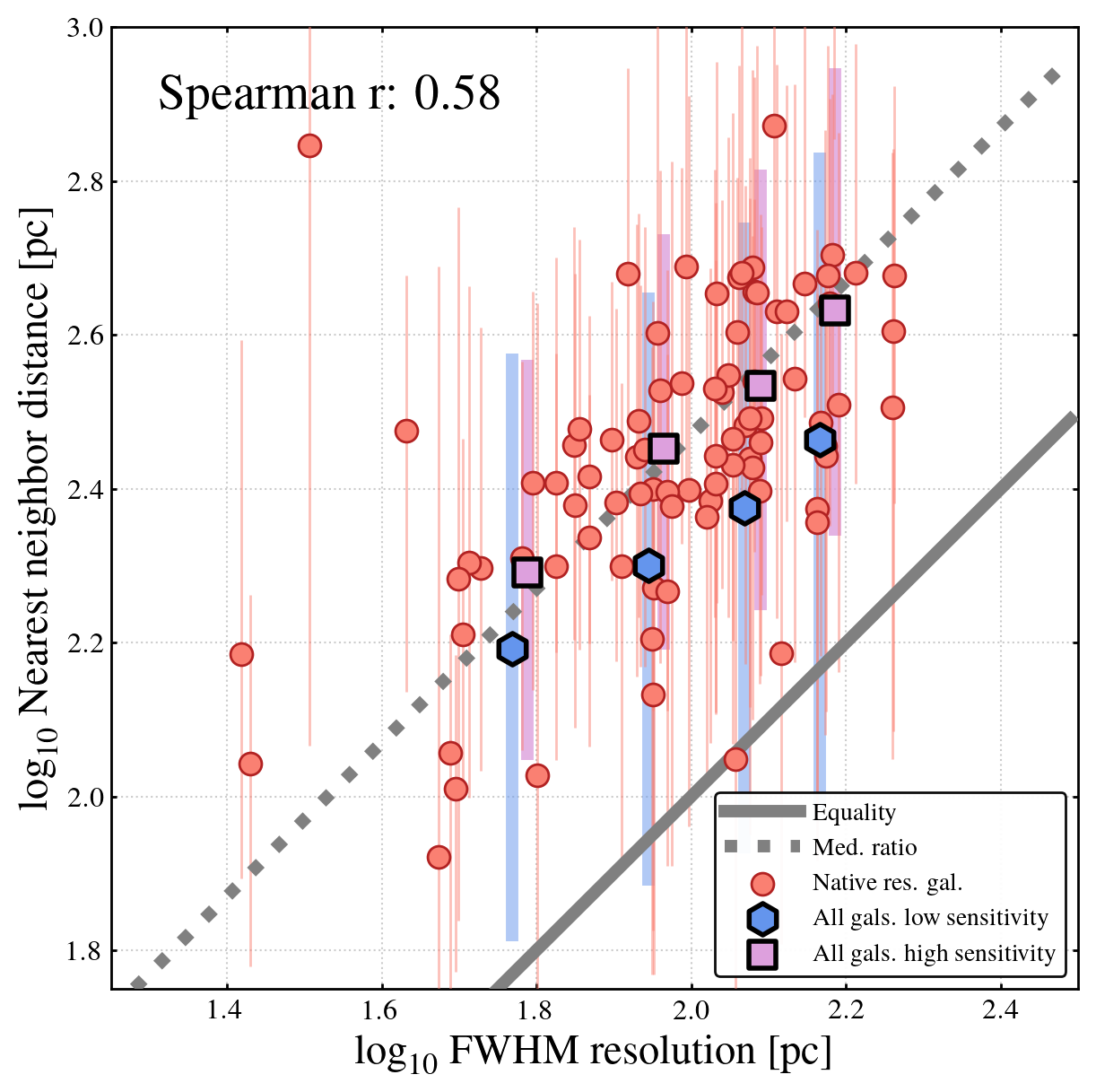}
\caption{\textit{Distance to a GMC's nearest neighbour as a function of the resolution of the data used to produce the GMC catalogue.} Salmon circles show the median nearest neighbour distance ($d_{\mathrm{sep, 1st}}$) for GMCs in individual galaxies at their native resolution and sensitivity. The Spearman coefficient for individual galaxies at their native resolution and sensitivity is labelled at top left. 
Error bars indicate the 16-84th percentile range of $d_{\mathrm{sep, 1st}}$ for that galaxy. Purple squares and blue hexagons show the median $d_{\mathrm{sep, 1st}}$ 
for GMCs in all galaxies after homogenizing the data to the same resolution and fixed high or low sensitivity. The solid line indicates the one-to-one relation and the dashed line indicates the typical ratio ($\sim$3) between the median $d_{\mathrm{sep, 1st}}$ and the resolution for the native resolution catalogues. The median nearest neighbour distance depends on the sensitivity and resolution of the data. In the rest of the paper, we work with a homogenized data set with 150~pc resolution and 46~mK noise.}
\label{fig:spacing_and_res}
\end{figure*}

\subsection{CO integrated intensity maps and radial profiles}
\label{subsec:data:maps}

When analyzing the intensity distribution about each peak (\S~\ref{sec:stacking}), we use the PHANGS-ALMA CO(2-1) integrated intensity maps at 150~pc resolution constructed using the PHANGS-ALMA pipeline \citep{leroy_phangs-alma_2021-1} with the \texttt{broad} masks. These have high completeness and include almost all CO emission. When using low resolution integrated CO maps as a control for the 2PCF, we use maps convolved with an elliptical Gaussian so that they have FWHM resolution of $1$~kpc in the plane of the galaxy (i.e., we account for inclination in the convolution).
We also use radial profiles of mean CO intensity from \citet{sun_molecular_2022} to construct controls when calculating the 2PCF. These record the average kpc-scale CO(2-1) intensity in 500~pc wide radial bins.

\section{Spacing and number density of GMCs}
\label{sec:spacing}

\begin{figure*}[thb]
\centering
\gridline{
\fig{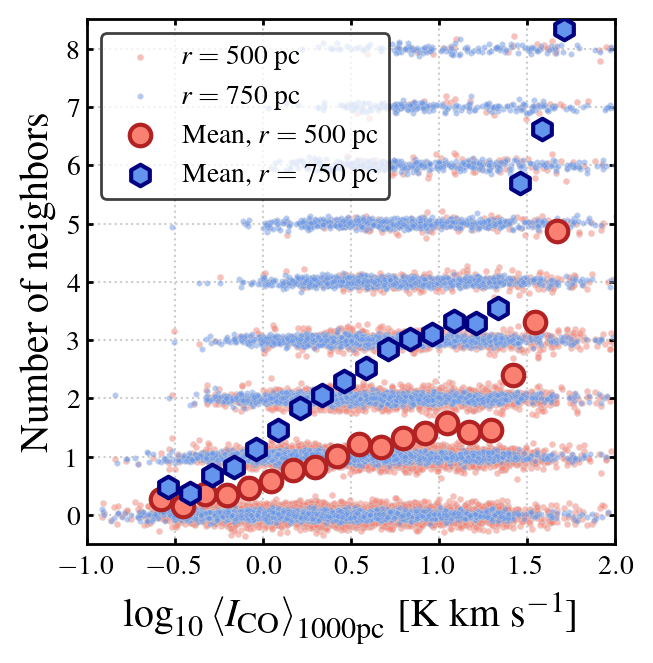}{0.435\textwidth}{}
\hspace{-4\baselineskip}
\fig{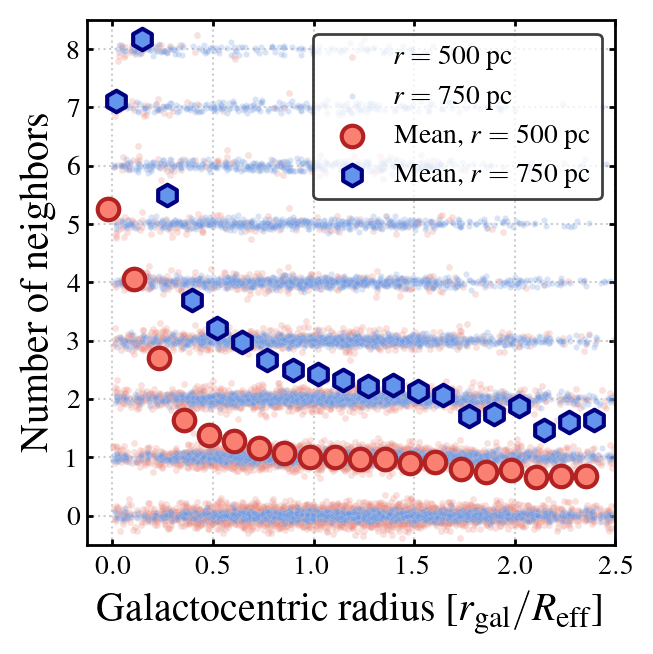}{0.435\textwidth}{}
}
\vspace{-3\baselineskip}
\gridline{
\fig{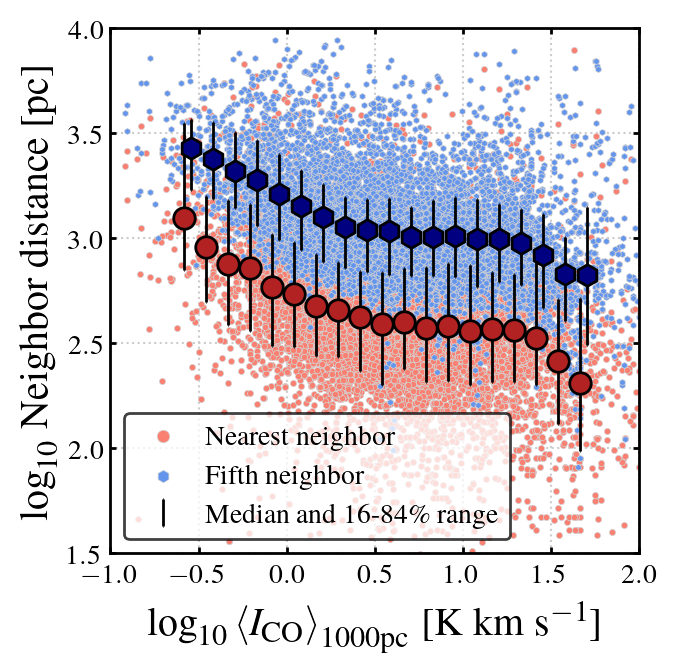}{0.45\textwidth}{}
\hspace{-4\baselineskip}
\fig{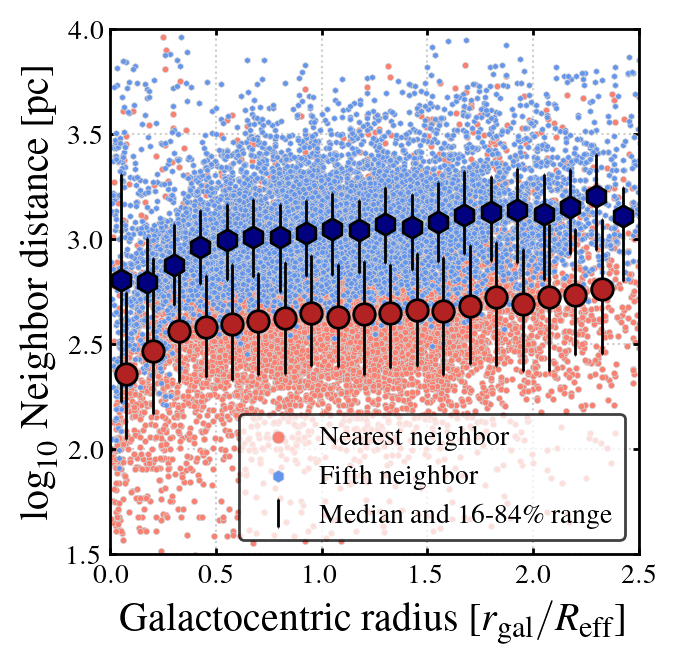}{0.45\textwidth}{}
}
\vspace{-2\baselineskip}
\caption{\textit{Spacing and number of neighbouring GMCs as a function of galactic environment.} For each cloud (represented by a small individual point), the plots in the top row show the number of neighbouring GMCs within 500~pc ($N_{\mathrm{{neighb, 500pc}}}$, pink) or 750~pc ($N_{\mathrm{{neighb, 750pc}}}$, blue). The bottom row shows the distance to the nearest GMC ($d_{\mathrm{sep, 1st}}$, pink) or the fifth-nearest GMC ($d_{\mathrm{sep, 5th}}$, blue). The left column shows these metrics as a function of the mean CO(2-1) intensity (\Icokpc) measured at low 1~kpc resolution. The right column shows both metrics as a function of galactocentric radius (\rgal), normalized to the stellar mass effective radius ($R_{\mathrm{eff}}$). The larger, bold symbols show the mean number of neighbours (top panels) or median spacing (bottom panels) in bins of \Icokpc or \rgal. All panels show that the spacing and space density of GMCs are related to the large-scale environment. The space density of GMCs anti-correlates with \rgal and correlates with \Icokpc, while the nearest neighbour distance exhibits opposite behaviour.}
\label{fig:spacing_and_nn}
\end{figure*}

\begin{figure}
\gridline{
\fig{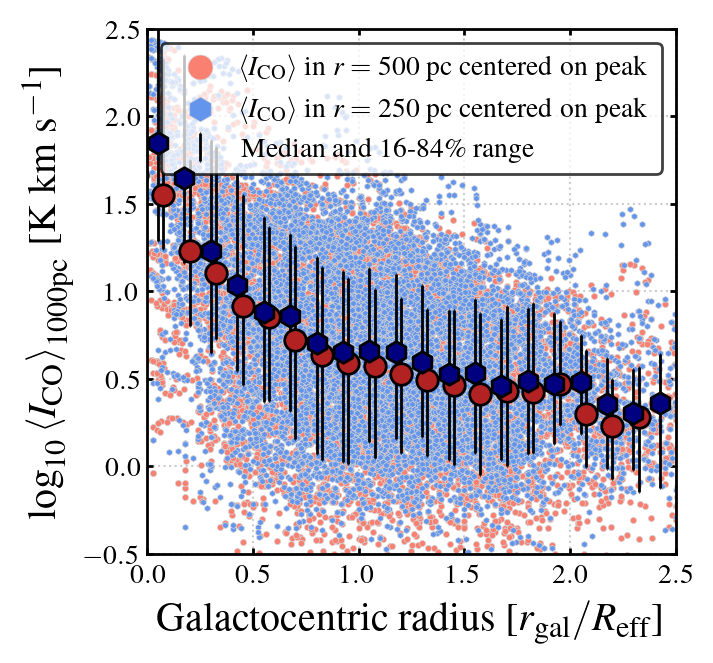}{0.5\textwidth}{}
}
\vspace{-2\baselineskip}
\caption{The kpc-scale intensity of CO(2-1) emission as a function of galactocentric radius normalized to the stellar effective radius of the galaxy. Each point shows a measurement centred on an individual GMC. Large symbols with error bars show the median and 16{-}84\% range of the data binned by galactocentric radius. The radial decrease of the kpc-scale CO intensity 
manifests in measurements of the GMC spacing, space density, and the two point correlation function.}
\label{fig:Ico_rgal}
\end{figure}

The simplest metrics of GMC clustering are the spatial separation between neighbouring GMCs (the `cloud-cloud spacing') and the number of neighbouring GMCs within a certain aperture. To calculate these, we first deproject the sky position of each GMC into a position in the galactic disk, taking into account the distance, inclination, and position angle of the galaxy. Then we calculate the distances between each GMC and its 1${\mathrm{st}}$ and 5${\mathrm{th}}$ nearest neighbours, $d_{\mathrm{sep, 1st}}$ and $d_{\mathrm{sep, 5th}}$. We also draw apertures with fixed radii of 500~pc and 750~pc and count the number of neighbouring GMCs within each aperture, $N_{\mathrm{neighb, 500pc}}$ and $N_{\mathrm{neighb, 750pc}}$. We repeat this exercise for each galaxy and each available resolution and noise level (\S \ref{subsec:GMC_cat}).

\subsection{Impact of resolution and sensitivity}
\label{subsec:spacingres}

The resolution and sensitivity of the data significantly affect estimtes of the cloud-cloud spacing and number of neighbours. Fig.~\ref{fig:spacing_and_res} shows the nearest neighbour distance for GMC catalogues as a function of spatial resolution. For the native resolution catalogues, the median nearest neighbour distance for each galaxy (pink circles) correlates well with the resolution of the data (Spearman correlation coefficient 0.58). When we repeat the same calculations using datasets homogenized to a fixed resolution (\S~\ref{subsec:GMC_cat}), we again see that the median nearest neighbour distance correlates with the observational resolution (blue diamonds and purple squares, which aggregate all galaxies at fixed resolution).

The strong correlation in Fig.~\ref{fig:spacing_and_res}  indicates that the observational resolution sets the lower limit of the spacing between two identified GMC structures, and that \texttt{CPROPS} and similar algorithms \citep[e.g., \texttt{CLUMPFIND};][]{williams_determining_1994} tend to find beam-scale objects. Several previous works have emphasized that the sizes and hence masses of clouds identified by these algorithms depend on the resolution  \citep[e.g.,][]{pineda_perils_2009,hughes_comparative_2013,leroy_portrait_2016,rosolowsky_giant_2021}. Here we show that the same bias affects the cloud-cloud spacing measurements\footnote{\citet{kim_environmental_2022} find that the spacing of local maxima they identify from PHANGS--ALMA CO and PHANGS--H$\alpha$ maps correlates with galaxy distance, which is a manifestation of the same resolution bias.}. For PHANGS--ALMA and \texttt{CPROPS}, the median nearest neighbour distance is $\approx 3$ times the FWHM beam size. We expect clustering to be suppressed below this scale because the algorithm will struggle to ``pack'' clouds together more densely than this, unless clouds are well-separated along the velocity axis (as sometimes occurs in galaxy centres). We indeed observe this effect in \S~\ref{sec:tpcf}.

We also observe a dependence of spacing on sensitivity, finding that our low sensitivity data (high noise) exhibit a smaller characteristic spacing than high sensitivity data (low noise). This arises because our low sensitivity data fail to capture GMCs in CO-faint regions like outer galaxy disks. GMCs tend to be both faint and more sparsely distributed in these regions (\S~\ref{subsec:homogspacing}). With better sensitivity, GMCs are detected in these regions and thus the median cloud-cloud spacing increases.


These results motivate our use of a homogenized dataset with fixed resolution and sensitivity in the rest of our analyses.

\subsection{Environmental dependence of GMC clustering}
\label{subsec:homogspacing}

\begin{table}[t!]
\centering
\caption{Spearman correlation$^{\dagger}$ between clustering metrics and \Icokpc and \rgal}
\label{tab:homospacing}
\begin{threeparttable}
\begin{tabularx}{0.3\textwidth}{ccc}
\hline \hline
 & $d_{\mathrm{sep, 1st}}$ & $N_{\mathrm{neighb, 500pc}}$ \\ 
\hline
\Icokpc & -0.33 & 0.35 \\
\rgal & 0.23 & -0.25  \\
\hline
\end{tabularx}
\end{threeparttable}
\begin{tablenotes}
\item 
The p-values for all correlations are smaller than 10$^{-100}$. 
\end{tablenotes}
\end{table}

In Fig.~\ref{fig:spacing_and_nn}, we use the homogenized dataset to correlate the cloud-cloud spacing and number of neighbours with tracers of the large-scale galactic environment: the kpc-scale CO velocity-integrated intensity (\Icokpc, top panel) and the galactocentric radius ($r_{\mathrm{gal}}$, bottom panel). \Icokpc\ captures the large scale structure of CO emission in the galaxy disk, while galactocentric radius correlates with stellar surface density, star formation rate, H$_2$/H I ratio, and numerous other galactic properties. \Icokpc and \rgal are themselves related: 
in Fig.~\ref{fig:Ico_rgal}, we show \Icokpc around each GMC as a function of \rgal. \Icokpc shows a strong declining trend towards larger \rgal \citep[consistent with many literature studies, e.g., see][]{young_molecular_1991,leroy_star_2008}. 

Both the nearest neighbour distance ($d_{\mathrm{sep, 1st}}$, $d_{\mathrm{sep, 5th}}$) and the number of neighbours within a fixed aperture ($N_{\mathrm{neighb, 500pc}}$, $N_{\mathrm{neighb, 750pc}}$) vary with galactic environment. The number of neighbours in a fixed aperture increases with increasing \Icokpc\ and decreasing \rgal, while the nearest neighbour distances decrease with increasing \Icokpc\ and decreasing \rgal. 
Both metrics indicate that GMCs are more clustered towards the central regions of galaxies and in regions rich in molecular gas. Table~\ref{tab:homospacing} reports the the Spearman correlation coefficient between both clustering metrics and \Icokpc and \rgal, calculated using all GMCs in the homogenized catalogue. We see a stronger correlation between the clustering metrics with \Icokpc than with \rgal, which suggests that the radial dependence of GMC clustering originates from the \Icokpc dependence. We record the median, 16th and 84th values of \Icokpc, nearest neighbour distance $d_{\mathrm{sep, 1st}}$ and number of neighbours $N_{\mathrm{neighb, 500pc}}$ at different \rgal in Table~\ref{tab:spacing_nn}.



The two clustering metrics complement one another. In crowded regions, the nearest neighbour distance approaches the minimum value set by the resolution and hence becomes insensitive. This can be seen in the lower left panel of Fig.~\ref{fig:spacing_and_nn}, where the trend in spacing as a function of \Icokpc\ flattens for $\log I_{\mathrm{CO, 1kpc}} > 0.5$. This sentence is a bit ambiguous. The cloud-cloud spacing then drops at $\log I_{\mathrm{CO, 1kpc}} > 1.3$, which corresponds to the drop in the central regions of galaxies (\rgal $< 0.5 R_{\mathrm{eff}}$, right panel). In these regions, clouds can be well-separated in velocity space and more easily distinguished by cloud-finding algorithms. On the other hand, the number of neighbours becomes an insensitive metric in sparse regions where the lower limit of zero neighbours is frequently reached. This can be seen in the upper right panel of Fig.~\ref{fig:spacing_and_nn}, where the number of neighbours drops to very low values at large \rgal. \footnote{84\% (50\%) of PHANGS-ALMA galaxies have full azimuthal field coverage out to 
1.3 (1.7) R$_{\mathrm{eff}}$.
} 



We conclude that GMC spacing reflects the large-scale structure of the galactic disk. For both metrics, the central regions of galaxies stand out as the most crowded. More clustering in regions with higher large-scale \Icokpc\ is reasonable
but not required. Regions with higher \Icokpc\ could host a larger number of GMCs with similar brightness, the individual GMCs located in this regions could be brighter and more massive, or both. \citet{sun_molecular_2022} showed that gas-rich regions harbour brighter and more massive GMCs on average. Our measurement shows that in such regions, the GMCs are also more strongly clustered. 

\begin{figure*}
\centering
\gridline{
\fig{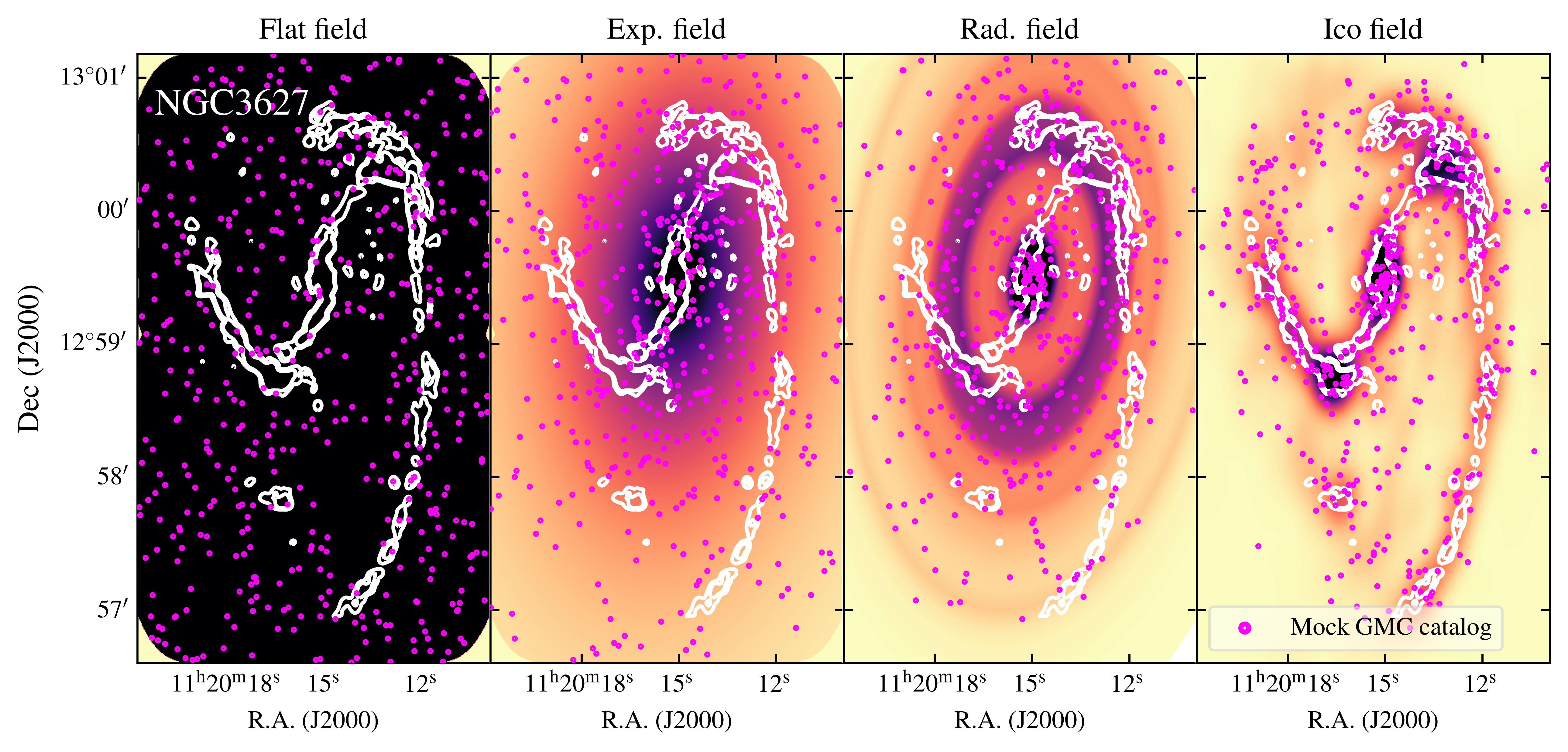}{\textwidth}{}
}
\vspace{-2\baselineskip}
\caption{Control distributions used to calculate the two-point correlation function. The images show normalized probability distributions. The colour reflects the chance of placing a GMC when constructing our random control catalogues (\S \ref{sec:tpcf}). From left to right the models are: ``Flat'' --- equal probability across the map; ``Exp'' --- an exponential azimuthally symmetric disk with scale length matched to the stellar mass scale length; ``Rad.'' -- probability proportional to the radial profile of CO intensity; and ``\Icokpc'' -- probability proportional to the intensity of CO emission at $1$~kpc resolution. In each panel, magenta circles show examples of mock GMC locations drawn from the probability maps and contours show the CO(2-1) integrated intensity at 150~pc resolution. }
\label{fig:density_field}
\end{figure*}

\section{Two-point correlation function}
\label{sec:tpcf}

\begin{figure*}
\centering
\includegraphics[width = \textwidth]{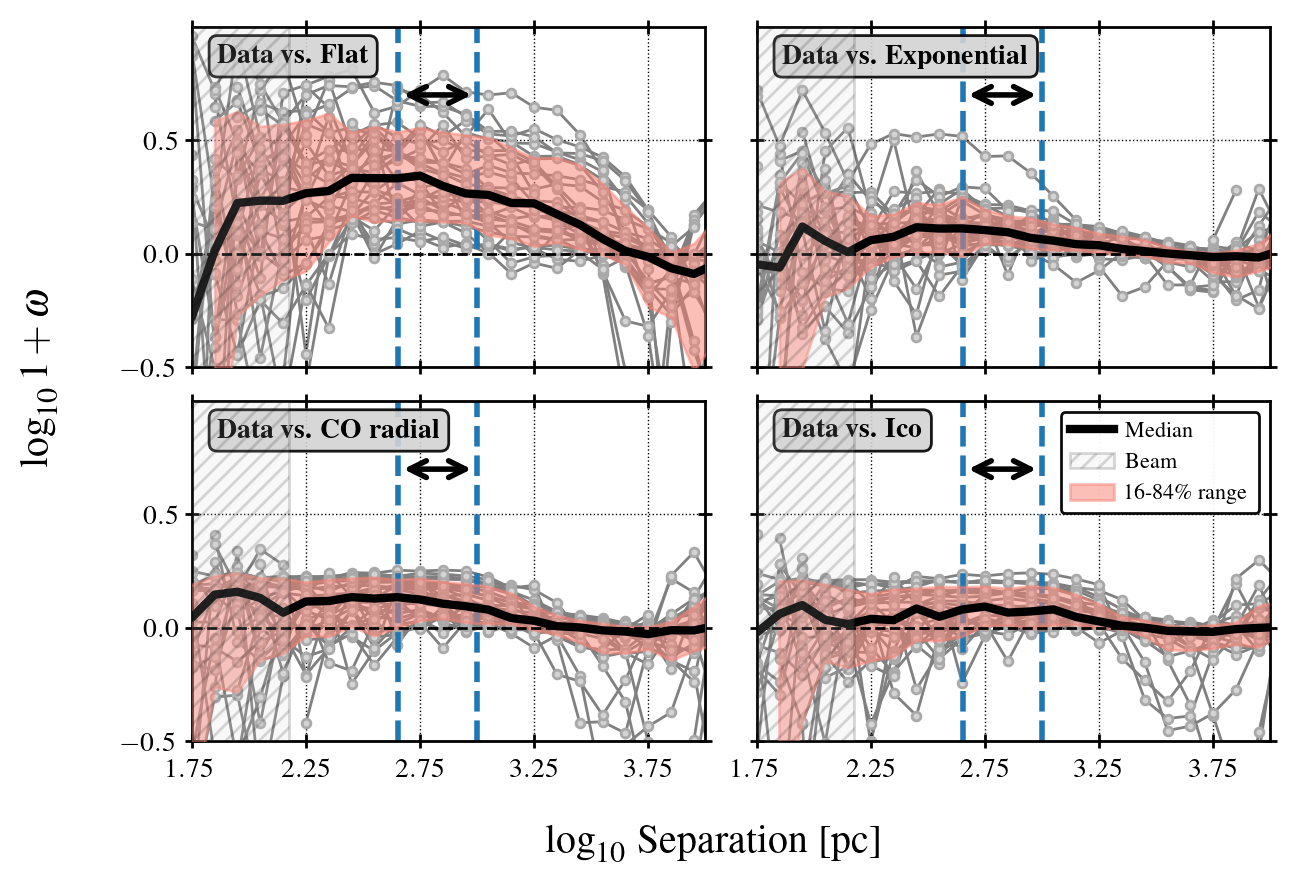}
\caption{Two point correlation function (2PCF) expressed as $\log_{10} 1+\omega$ (\S \ref{sec:tpcf}) for GMCs in PHANGS--ALMA. Each gray trace shows the 2PCF for an individual galaxy, with points indicating the bin centres used in the calculation. We also show the median (black line), $16{-}84\%$ range (pink). The gray hatched region shows the resolution of the data, and we expect clustering to be suppressed by the cloud-finding algorithm near this value (\S~\ref{sec:spacing}).The area between the two vertical dashed lines indicates the scale range that is not affected by observational and our sampling bias (\S~\ref{subsec:2pcf_results}). 
Each panel shows the 2PCF generated using a different control (\S~\ref{subsec:mock_cat}, Fig.~\ref{fig:density_field}). The first panel, illustrating clustering relative to a uniform control that does not include any disk structure, shows the strongest clustering. The other controls, which do account for the large scale structure of the disk, show much weaker but still significant clustering. We compare these in more detail in Fig.~\ref{fig:tpcf_zoomin}). The contrast between the first and the other three panels highlights that the dominant signal is the radial structure of the galactic disk, which must be accounted for to reveal finer scale structure. }
\label{fig:twopt_multipanel}
\end{figure*}

\begin{figure*}
\centering
\gridline{
\fig{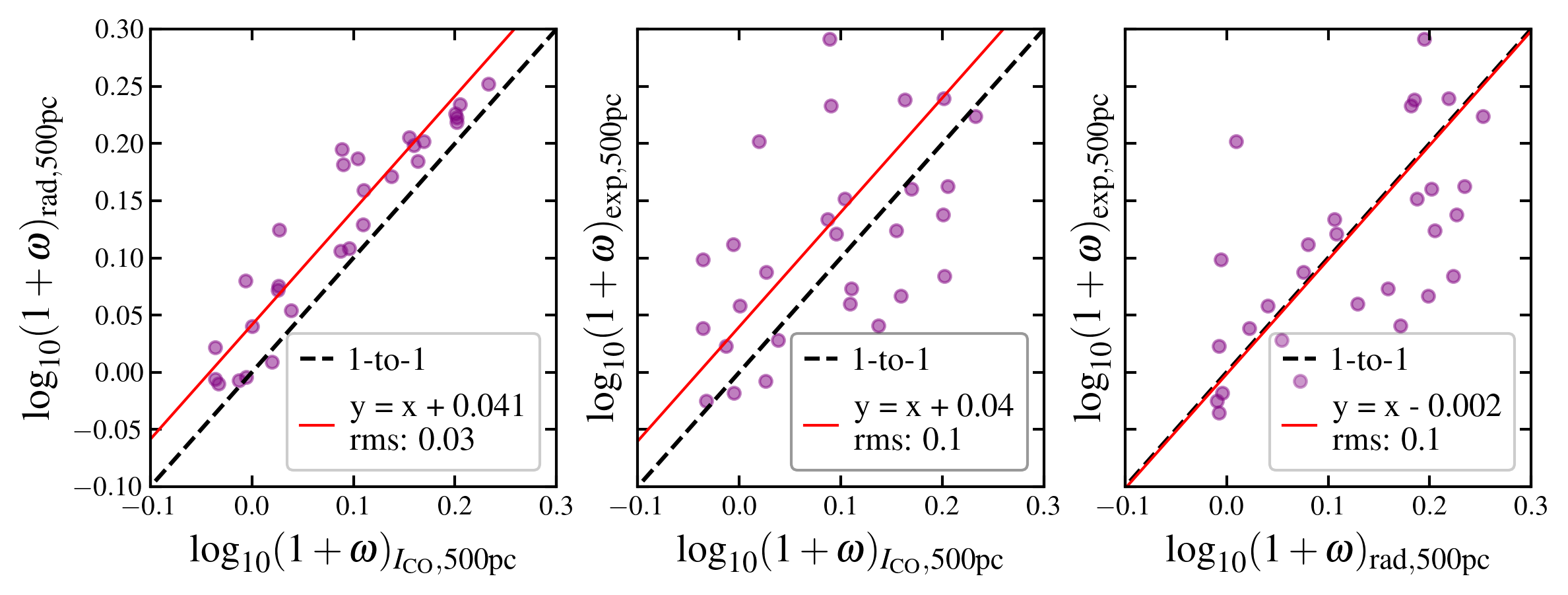}{\textwidth}{}
}
\vspace{-2\baselineskip}
\caption{
Amplitude $1+\omega$ at 500~pc (dashed vertical line) for the 2PCF calculated relative to the three controls of ``Exp'', ``Rad'' and ``$I_{\rm CO}$''. Each point shows the result for one galaxy (we exclude galaxies with $N_{\mathrm{GMC}} < 50$, for which we consider our 2PCF unreliable). The dashed lines indicate the one-to-one relation and the red solid lines indicate the linear fit to the correlation. Results for all three controls correlate well and the amplitude appears to correlate with morphology of the galaxy (Fig.~\ref{fig:mom0_sort_by_tpcf}, \S~\ref{subsec:galtogal2pcf}). We observe higher clustering relative to the radial profiles (``Rad'') than relative to the low resolution CO map (``$I_{\mathrm{CO}}$''), indicating that about \textbf{30\%} of the clustering relative to the radial profile comes from the overall azimuthal structure. Outliers in the right two panels indicate where galaxies are not well-described by exponential disks.}
\label{fig:tpcf_zoomin}
\end{figure*}

\begin{figure*}[ht!]
\centering
\gridline{
\fig{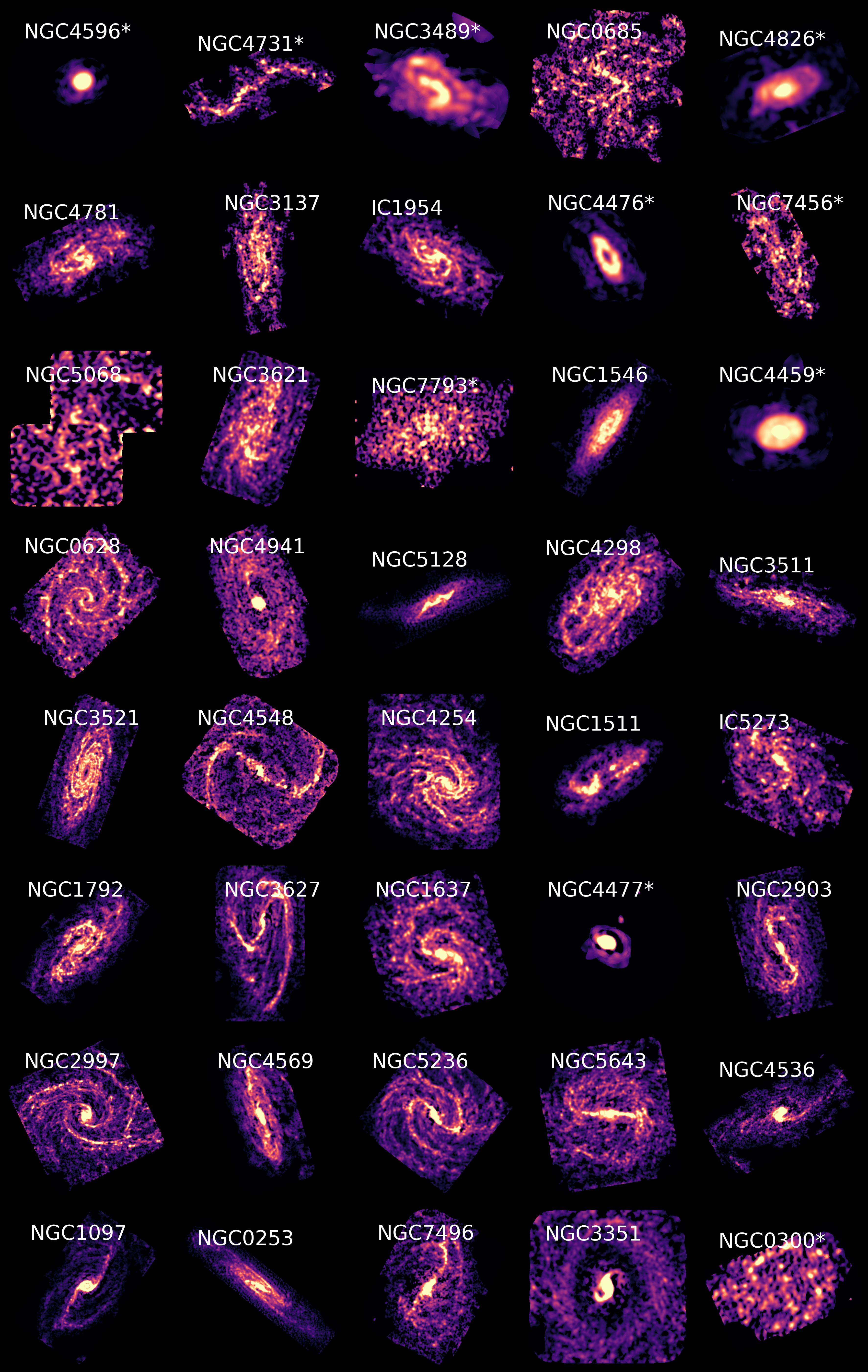}{0.7\textwidth}{}
}
\vspace{-1.5\baselineskip}
\caption{Integrated CO (2-1) intensity maps of our targets sorted by amplitude of the 2PCF at 500 pc calculated using the ``$I_{\mathrm{CO}}$ '' reference. The amplitude increases from upper left (lowest $1+\omega$) to lower right (highest $1+\omega$). Galaxies with a $*$ next to their name have fewer than 50 GMCs and we consider them unreliable. Galaxies with low 2PCF amplitudes often show flocculent structure, while galaxies with higher amplitudes frequently show prominent centres, bars and sharp spiral arm features. These features 
play a key role in setting the clustering of GMCs once the largest scale structure is removed.}
\label{fig:mom0_sort_by_tpcf}
\end{figure*}

The two-point correlation function (hereafter `2PCF') captures information on all cloud-cloud spacings, not only nearest neighbours. The 2PCF is used extensively in studies of large-scale structure to describe the degree of clustering as a function of spatial scale \citep{peebles_large-scale_1980}. Several studies have applied the 2PCF to study the spatial clustering of molecular clouds and stellar clusters \citep[e.g.,][]{grasha_connecting_2018,grasha_spatial_2019,turner_phangs_2022,peltonen_clusters_2023}. Here we construct a general measurement of the 2PCF describing the autocorrelation of GMCs.



We use the two-dimensional projected two-point correlation function $\omega(\theta)$ to describe the clustering of GMCs within the galactic plane. $\omega(\theta)$ is defined as the excess probability of finding two GMCs with an angular separation $\theta$ compared to what is expected for an uncorrelated random Poisson distribution. 

We calculate $\omega (\theta)$ using the popular Landy-Szalay (L-S) estimator \citep{landy_bias_1993}, which has low variance and little systematic bias \citep{kerscher_comparison_2000}. The L-S estimator requires a data catalogue ("D") as well as a "random objects" catalogue (R), which is used as a control distribution. Using these catalogues, one calculates the distances between all pairs of real objects, all pairs of real with random objects, and all pairs of random objects. The number of object-pair separations in each spatial bin are counted and then normalized to control for differing sizes of the real and random control catalogues. These normalized terms can be expressed as 
\begin{equation}
\begin{split}
DD(\theta) &= N_{DD}(\theta) / N_D^2 \\
RR(\theta) &= N_{RR}(\theta) / N_R^2 \\
DR(\theta) &= N_{DR}(\theta)/ (N_D N_R)~,
\end{split}
\end{equation}
where $N_{DD}(\theta)$, $N_{RR}(\theta)$ and $N_{DR}(\theta)$ are the number of data--data, data--random and data-random pairs within a small bin around a separation $\theta$, and $N_D$ and $N_R$ are the number of objects in the data and random catalogues. Then $\omega(\theta)$ can be expressed as 
\begin{equation}
\omega(\theta) = \frac{DD(\theta)-2DR(\theta)+RR(\theta)}{RR(\theta)}~.
\end{equation}
In our analysis, we defined logarithmic bins of $\theta$ 0.1 dex wide and estimate the uncertainty on the 2PCF using error propagation from Poisson statistics (see Appendix \ref{app:unc}). 

\subsection{Random source catalogue generation}
\label{subsec:mock_cat}

We generate the random source catalogues used as controls using Monte Carlo simulations. This requires us to adopt a two-dimensional probability distribution that describes where a random source is likely to appear. Most 2PCF studies of large scale structure assume a uniform likelihood distribution, because the galaxy distribution appears to be isotropic at large cosmological scales. For studies of structure within galaxies, however, a uniform distribution no longer represents an ideal null hypothesis. We consider the following models:

\begin{itemize}
    \item ``\textbf{Flat}'': In this model, clouds appear anywhere within the ALMA field of view with uniform probability. 
    This is the traditional set-up and we expect the 2PCF measured using this control to be dominated by clustering signal caused by the overall structure of the galaxy.
    
    \item ``\textbf{Exp}'': In this model, the probability of finding a cloud at a given location follows an exponential disk with scale length equal to that inferred from the radial profile of stellar mass surface density. 
    Compared to the \textbf{``Flat''} distribution, the 2PCF using this control will capture only clustering relative to the overall structure of an exponential disk.
    
    \item ``\textbf{Rad}'': In this model, the probability of finding a cloud at a location follows a radial profile of kpc-scale CO intensity (\S \ref{subsec:data:maps}). This set-up resembles the \textbf{``Exp''} distribution, but better captures clustering in excess of the actual large-scale molecular gas distribution when this distribution deviates from an exponential profile. 
    
    \item ``$\mathbf{\mathit{I}_{CO}}$'': This model sets the probability of finding a cloud proportional to the kpc-scale CO intensity. Compared to the \textbf{``Exp''} and \textbf{``Rad''} distributions, this set-up captures non-radial variations in kpc-scale structure.
\end{itemize}

We generate normalized probability distribution maps and then generate mock catalogues. For each galaxy, we set the number of random sources $N_{R} = 100 N_{D}$. The large $N_{R}/N_{D}$ ratio ensures the estimated 2PCF is close to the true 2PCF function without systematic bias \citep{kerscher_comparison_2000}. Fig.~\ref{fig:density_field} shows example probability density fields and random source catalogues.

\subsection{Measured two-point correlation function}
\label{subsec:2pcf_results}

Fig.~\ref{fig:twopt_multipanel} shows the 2PCFs of all galaxies in our sample calculated for each control distribution. We show the 2PCF of individual galaxies as well as the median and 16-84\% range of $1+\omega$ in each bin across the whole sample. We exclude galaxies with fewer than 50 GMCs ($N_{\mathrm{GMC}} < 50$), since those 2PCF measurements have large uncertainties due to small number statistics. We record the 2PCF for individual galaxies in Table~\ref{tab:2pcf_indvd} and the overall median, 16th and 84th values in Table~\ref{tab:tpcf_meds}. These empirical measurements are one of our main results.

The 2PCF measured using a flat control (first panel) shows strong clustering out to scales of several kpc. The median $1+\omega$ rises from large to small scales until $\theta \approx 500$~pc, below which the $1+\omega$ remains high but relatively flat.
Individual galaxies show large scatter in their amplitude, but almost all show significant clustering out to $\gtrsim 1$~kpc scales. The $1 + \omega$ that we measure roughly resembles that found by \citet{turner_phangs_2022}, who analysed GMC clustering in 11 PHANGS--ALMA targets and found peak amplitudes of $\sim$ 2--8. 

The subsequent panels of Fig.~\ref{fig:twopt_multipanel} show that the GMC clustering signal is significantly weaker when we use controls that account for the overall structure of the galaxy. When we switch the control from ``Flat'' to either ``Exp'', Rad'' or ``$I_{\mathrm{CO}}$'', the median peak amplitude of clustering drops 
from $\sim 2.3$ to $\sim1.3$ (i.e., the clustering excess $\omega$ drops from 130\% to 30\%). The key difference between ``Flat'' and these other three controls is that  ``Exp'', ``Rad,'' and ``$I_{\mathrm{CO}}$'' each account for the radial decline in molecular gas content exhibited by almost all galaxies (e.g., see Fig.~\ref{fig:Ico_rgal}). The significant drop between the first panel and the subsequent three panels indicates that most of the GMC clustering  on $\gtrsim 500$~pc scales stems from the large-scale radial distribution of molecular gas, which is close to the radial distribution of a stellar exponential disk.
 
The controls ``Exp,'' ``Rad,'' and ``$I_{\mathrm{CO}}$'' represent the large-scale structure of CO emission with increasing fidelity. ``Rad'' captures variations in the CO radial profile missed by the exponential approximation, while  ``$I_{\mathrm{CO}}$'' also controls for azimuthal variations. In Fig.~\ref{fig:tpcf_zoomin}, we compare the amplitude of the 2PCF using these three controls at $\theta = 500$~pc for individual galaxies.  We see that all three amplitudes are correlated.  Our fit indicates that the 2PCF amplitudes using ''Exp'' and ''Rad'' controls have no systematic
difference, which is consistent with radial profile measurements
in literature studies \citep[e.g.,][]{brown_vertico_2021} that show a good match between CO and stellar scale lengths. The larger scatter in the correlations involving ''Exp'' (Fig \ref{fig:tpcf_zoomin} middle and right panel) is consistent with observations that the radial profile of CO emission often has a more complex structrue than an exponential.

On the other hand, we still see a systematic lower clustering degree for ``Data vs $I_{\mathrm{CO}}$'' compared to ``Data vs Rad'', which demonstrates the universal impact of large-scale azimuthal variation. The clustering excess $\omega$ drops from 30\% using ``Rad'' control to 20\% using ``$I_{\rm CO}$ control, indicating 30\% of the clustering relative radial profile comes from the overall azimuthal structure.  The visualization of the controls in Fig. \ref{fig:density_field} illustrates this point well: the real data are clearly more clustered than what might be inferred from a radial profile due to the concentration of molecular gas into arms and bars. Overall, azimuthal structure in the CO distribution makes a subdominant, second-order contribution to the 2PCF signal compared to the radial distribution.

In Fig. \ref{fig:twopt_multipanel} the amplitude of clustering remains roughly flat below $\theta \sim 500$~pc for all controls. A priori, this is unexpected: previous 2PCF studies of star formation tracers, such as the distribution of stellar clusters \citep[e.g.,][]{menon_dependence_2021} have reported a power-law behaviour from $\theta \sim 10{-}1000$ pc, with rising amplitude towards small scales. This is interpreted as an indication of scale-free GMC fragmentation, which is theoretically expected (\S \ref{sec:intro}). Based on our analysis of the nearest neighbour cloud spacings (\S~\ref{sec:spacing}), we observe that \texttt{CPROPS} struggles to distinguish GMCs on scales approaching the beam size. We expect similar algorithms to also suffer the same bias. We interpret the lack of rise in the 2PCF below \textbf{$\theta = 500$}~pc, i.e. roughly the same scale as the typical cloud-cloud spacing, to reflect the impact of this bias. 

Information on smaller scales (but still above the beam size) is still contained in our maps and the GMC catalogues. Here we access this information via the stacked profiles (i.e., the intensity-peak cross-correlations) that we construct in \S~\ref{sec:stacking}. In the GMC catalogues, this information is recorded as cloud sizes. In \S~\ref{sec:stacking}, we show that the emission is clustered around the local CO peaks as expected. Another natural next step, but beyond the scope of this paper, would be to conduct the 2PCF analysis directly on the map of CO intensity itself.

A final subtlety in Fig. \ref{fig:twopt_multipanel} is that $1+\omega$ is often positive at scales $> 1$~kpc even for the ``Rad'' and ``$I_{\rm CO}$'' controls, which should control for structure on these scales. Based on inspecting the data, this is driven primarily by an imperfect mapping between CO luminosity and cloud locations. In the bright, inner regions of galaxies, \texttt{CPROPS} finds clouds that contain more luminosity. Meanwhile in low density regions like spurs and outer spiral arms the identified clouds tend to be lower luminosity. In contrast, the ``$I_{\rm CO}$'' control will put relatively more clouds than the catalogues in the high luminosity regions and fewer in the low luminosity regions. This imperfect mapping leads to subtle difference in large-scale distributions of data and control catalogue, particularly in galaxies with large-scale structure traced by $I_\mathrm{CO}$.  This difference will introduce systematic clustering signal at large scales relative to the ``$I_{\rm CO}$'' control.

\subsection{Galaxy-to-galaxy variation}
\label{subsec:galtogal2pcf}

We summarize the clustering metrics for individual galaxies in Table~\ref{tab:clustering_summary}. In Fig.~\ref{fig:twopt_multipanel} and~\ref{fig:tpcf_zoomin}, we see significant galaxy-to-galaxy scatter in the amplitude of the 2PCF. We attribute the scatter relative to the flat control sample as reflecting variations in the large-scale distribution of CO. While the scatter among galaxies decreases, it nonetheless remains present when we control for radial structure. 


The handful of significant outliers when comparing the ``Exp'' control to other methods are galaxies where the large-scale molecular gas distribution does not follow the stellar radial distribution well. For example, the galaxy with the highest ``Data vs Exp'' amplitude is NGC~5128 (Centaurus~A), which shows a dense molecular disk embedded in an extended early-type galaxy \citep[e.g.,][]{espada_molecular_2018}. 



Aside from these outliers, the $1+\omega$ amplitudes at $500$~pc using the different controls correlate well with one another (Fig.~\ref{fig:tpcf_zoomin}). In Fig.~\ref{fig:mom0_sort_by_tpcf} we plot the $150$~pc resolution integrated CO(2-1) intensity maps of our targets sorted by the amplitude $1+\omega$ at $\theta = 500$~pc compared to the ``\Icokpc'' control. This Figure shows a clear morphological sorting. Galaxies with lower 2PCF amplitudes appear more flocculent, while the galaxies at the higher end tend to have more ordered structures, including prominent centres and bars, and sharp spiral arm features. These structures often have sizes or widths less than 1~kpc \citep[e.g.,][]{querejeta_spiral_2024}, which can contribute to the 2PCF signal relative to ``$I_{\rm CO}$'' control at sub-kpc scales. There are a few notable exceptions (e.g. NGC~0300) but these galaxies generally have fewer than 50~GMCs and we consider our 2PCF measurements for them to be unreliable.




\subsection{Power-law index of 2PCF}
\label{subsec:tpcf_slope}

\begin{figure}
    \centering
    \gridline{
    \fig{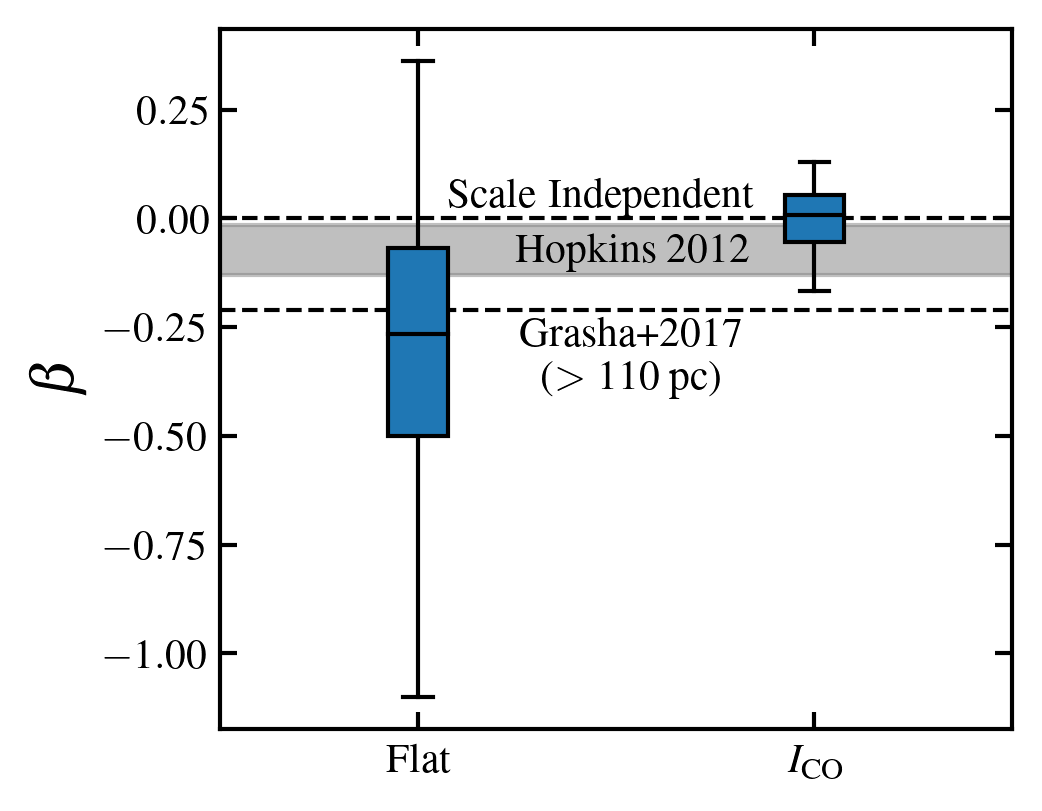}{0.45\textwidth}{}
    }
    \vspace{-2\baselineskip}
    \caption{The median of the calculated power-law slopes of individual galaxies at a scale range of 500 --- 1000 pc. The box range covers the 25th-75th percentile range and the whisker extend to 1.5 times the interquartile range. The dash lines indicate a slope of zero, and a representative slope of the power-law tail beyond 100 pc for young stellar clusters \citep{grasha_hierarchical_2017}. The gray shaded region indicates slope range predicted across 500 --- 1000 pc from \citet{hopkins2012}.  
    The slight power-law (power-law index of -0.25) trend with ``Flat'' 2PCF vanishes after we account for the large-scale CO distributions with ``$I_{\rm CO}$'' control, which suggests that GMC organization at scales of 500 -- 1000 pc is dominated by large-scale gravitational potential and dynamics.  }
    \label{fig:tpcf_slope}
\end{figure}

To quantify the scale dependence of the clustering, we fit a power-law function at our trusted scale range of 500 -- 1000 pc for 2PCFs of individual galaxies. Fig. \ref{fig:tpcf_slope} shows the power-law indices fit to our 2PCFs using the ``Flat'' and ``$I_{\rm CO}$'' controls. The ``Flat'' 2PCF has a median power-law index of $-0.25$  with a large galaxy-to-galaxy variation (25th-75th percentile range of $-0.1$ -- $-0.5$). The ``$I_{\rm CO}$'' 2PCF shows a median power-law index of $\sim$ 0, i.e., little or no scale dependence of the clustering, and the galaxy-to-galaxy variation is much reduced. This supports our argument that the majority of the signal in the ``Flat'' 2PCF is due to large-scale ($\gtrsim$1~kpc) galactic structures. 

So far, there is no systematic study of 2PCF on large-sample GMCs. Instead, most extragalactic studies focus on young stellar clusters, which form from GMCs and are expected to inherit their spatial correlation structure. Previous 2PCF studies of the distribution of young stellar clusters \citep[e.g.,][]{grasha_hierarchical_2017,menon_dependence_2021} have observed a rising amplitude towards small scales across spatial scales of 10 -- 1000~pc. Specifically, \citet{grasha_hierarchical_2017} suggest a double power-law behaviour for the observed 2PCF in six nearby galaxies. For scales below 100~pc, these previous measurements of the stellar cluster 2PCF shows a steep decline as a function of spatial scale, with a power law index of $\sim -0.7$. This value generally agrees with the theoretical prediction of slope $-1$ resulting from scale-free GMC fragmentation \citep{hopkins2012, gusev_hierarchy_2014}.
On scales of 100 -- 1000~pc, \citet{grasha_hierarchical_2017,menon_dependence_2021} have reported a shallower dependence of the stellar cluster 2PCF on spatial scale, with the 2PCF transitioning to have a power-law index of $\sim$ $-0.2$ -- $-0.3$ on larger scales. This appears consistent with our obtained value of $-0.25$ for the ``Flat'' 2PCF, which is the appropriate comparison because none of the above studies account for galactic structure in their control.

In addition, \citet{menon_dependence_2021} show that the 2PCF of older clusters exhibits a power-law index similar to the flatter power-law tail exhibited by young clusters on larger scales. Those authors also show a link between the near-IR scale length of the galaxy and the 2PCF scale that they measure for older clusters. Both these results support the idea that the 2PCF on $\sim 100{-}1000$~pc scales results from large-scale structures in the galaxy, in agreement with our comparison between the ``Flat'' and ``$I_{\rm CO}$'' 2PCFs. The slope of ``$I_{\rm CO}$'' 2PCF also generally agrees with the theoretical predictions from \citep{hopkins2012} across scales of 500 -- 1000 pc assuming a scale-free GMC collapse.



Accounting for our limiting resolution and adopted control distribution, our results are consistent with recent work on stellar clusters. However, those studies suggest that the clustering signal most sensitive to the physical processes governing cloud and cluster formation emerges on scales smaller than 100 pc that is not accessible with the current PHANGS--ALMA cloud-based 2PCF. Some theoretical works, such as \citet{hopkins2012}, predict a scaling relation of $\omega \propto R^{-1}$ extended to scales of 100 -- 1000 pc (note that our derived slope is from 1+$\omega$). However, the small $\omega$ values over our reliable scale range, together with the limited dynamical range in scale, make it difficult to robustly constrain this scaling relation from our measurements. A more suitable scale range for testing GMC formation models is 10–100 pc, which would require data with a resolution of $\sim$1 pc.

In the next section, we use a complementary technique to probe the clustering of CO emission on scales closer to our resolution limit. Ultimately, however, higher-resolution CO data will be required to place stronger constraints on cloud and stellar cluster formation using the 2PCF.

\begin{figure*}
\centering
\gridline{
\fig{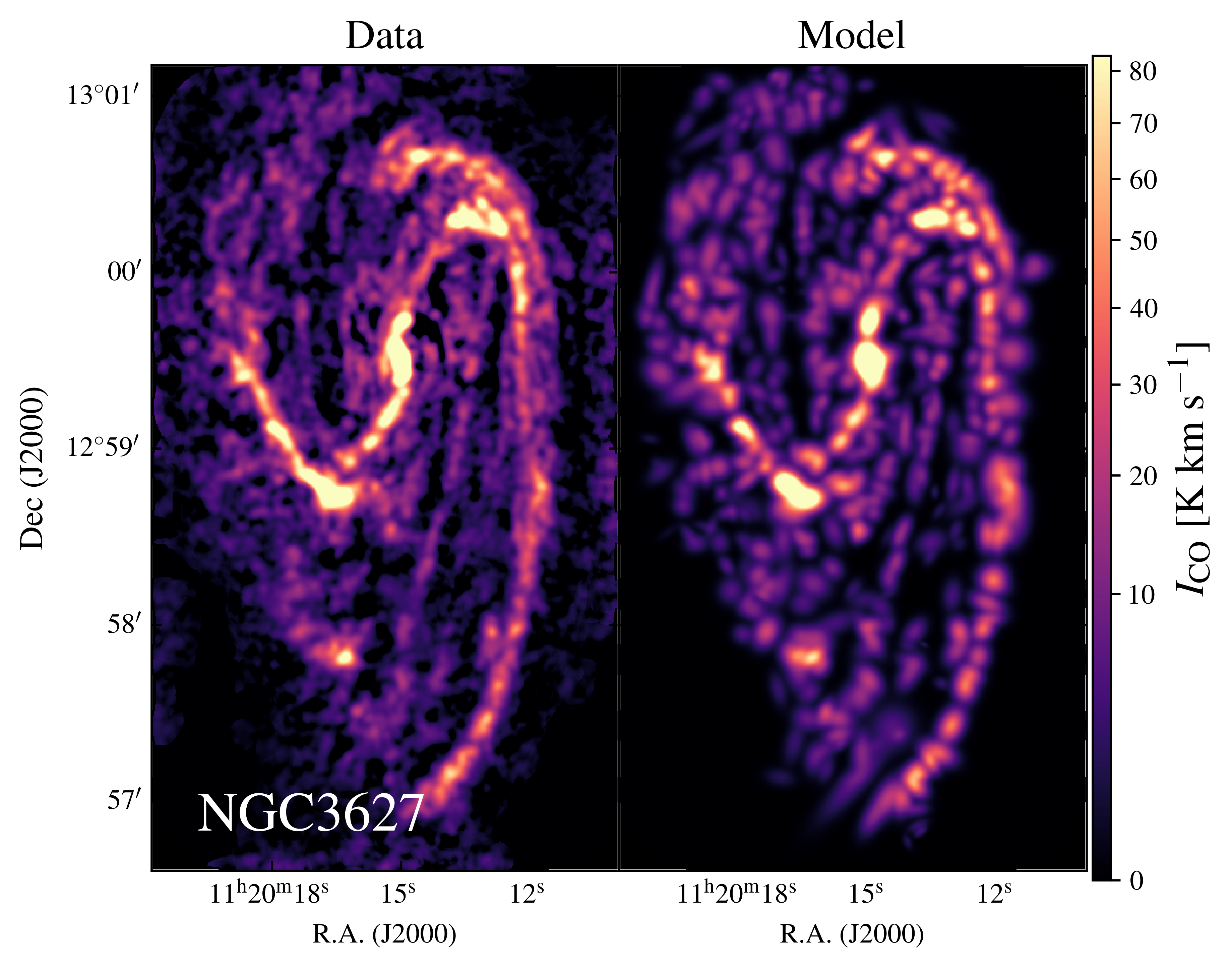}{0.6\textwidth}{}
\fig{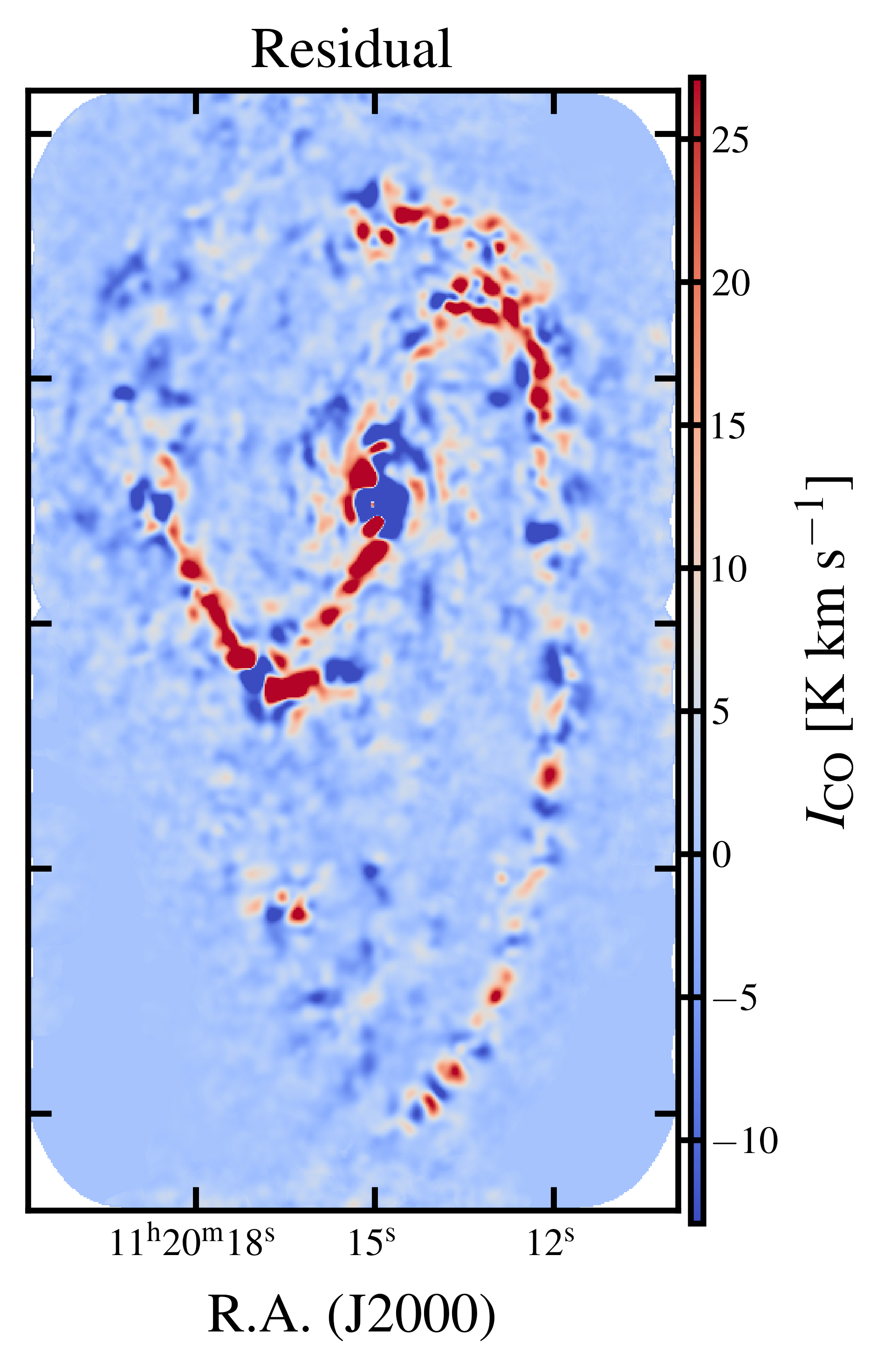}{0.31\textwidth}{}
}
\vspace{-2\baselineskip}
\caption{(\textit{Left}) Real and model integrated intensity maps for NGC~3627. The models convert the measured locations, intensity, and sizes of GMCs in the catalogue into a map (\S~\ref{subsec:model_image}). 
 (\textit{Right}) The residual map after subtracting the model image from the data. } 
\label{fig:model_images}
\end{figure*}


\begin{figure*}
\gridline{
\fig{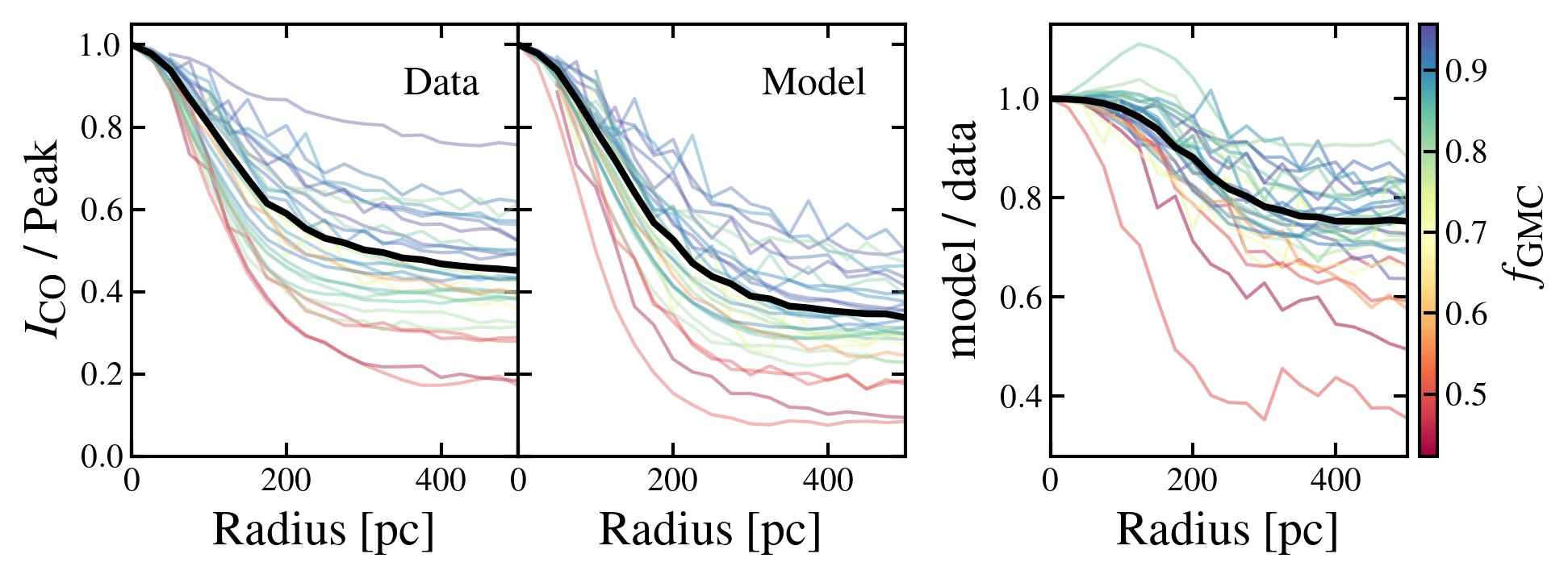}{\textwidth}{}
}
\vspace{-2\baselineskip}
\caption{(\textit{Left}) The stacked median of the normalized GMC radial profiles from all GMCs (black solid line) in the data (\textit{left panel}) and the model (\textit{right panel}), as well as the median of all GMCs in each galaxy (thin lines) color-coded by the the fraction of the total CO flux in the GMC catalogue for that galaxy (Hughes et al. in prep.). The flatter profiles correspond to galaxies with higher GMC flux fraction, which generally have brighter CO emission and stronger GMC clustering. (\textit{Right}) Median of the model/data ratio of all GMCs (\textit{black solid line}) and of individual galaxies color coded by the GMC flux fraction. The GMC completeness level affects our ability to recover the measured radial profile from the model.  }
\label{fig:stacking_radial_galaxies}
\end{figure*}



\begin{figure*}
\gridline{
\fig{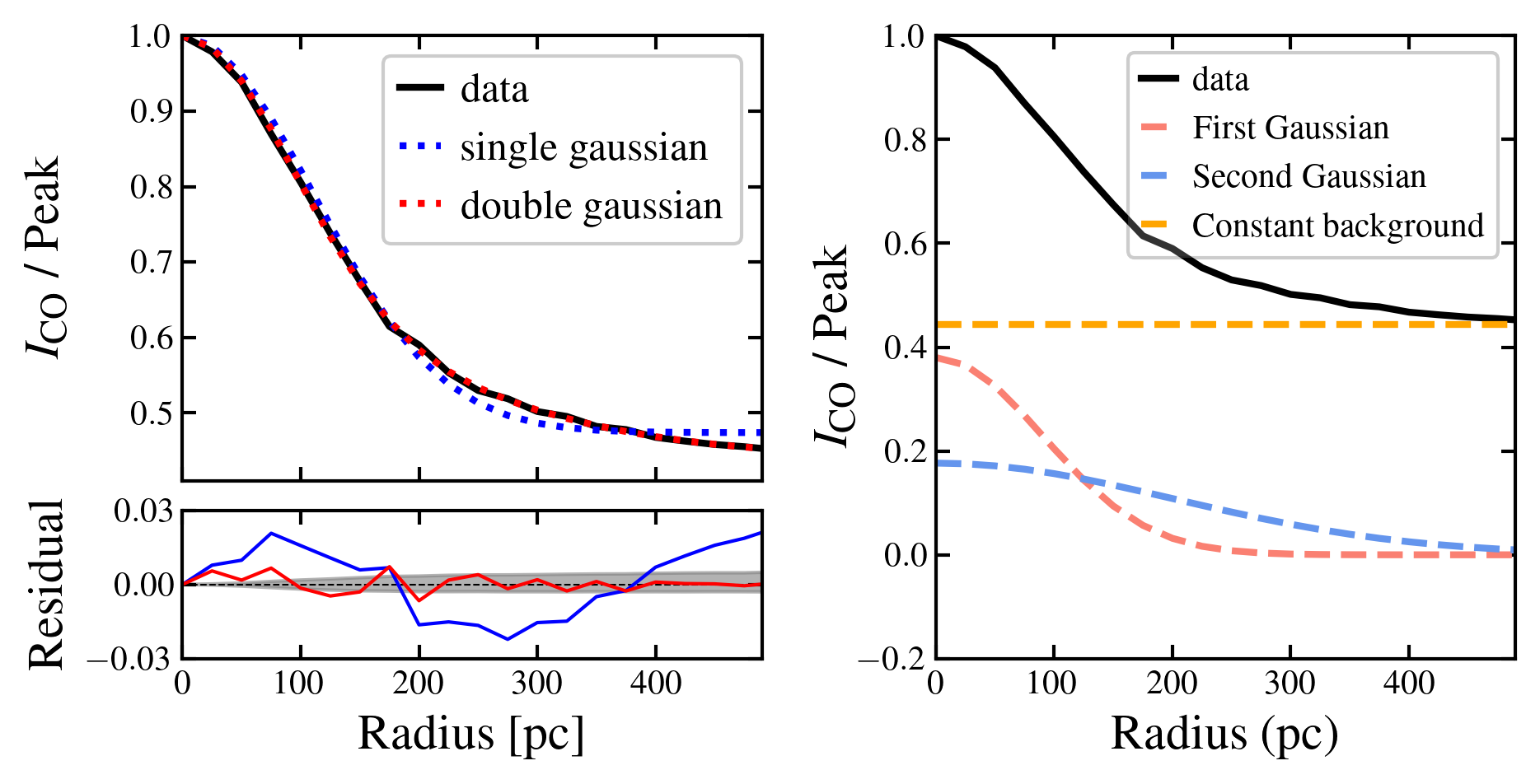}{0.9\textwidth}{}
}
\vspace{-2\baselineskip}
\caption{ The analytical fit to the stacked radial profile from the observed CO maps. (\textit{Left}) The single (\textit{blue dotted line}) and double (\textit{red dotted line}) Gaussian fit to the data. The bottom panel shows the residual ($\mathrm{data-fit}$) from the two fits. The gray shaded region indicates the uncertainty of the stacked median derived from error propagation. (\textit{Right}) The decomposition of our double Gaussian fitting, which includes two Gaussian (\textit{red and blue}) and a constant background (\textit{orange}).
 }
\label{fig:stacking_fitting}
\end{figure*}

\begin{figure*}
\gridline{
\leftfig{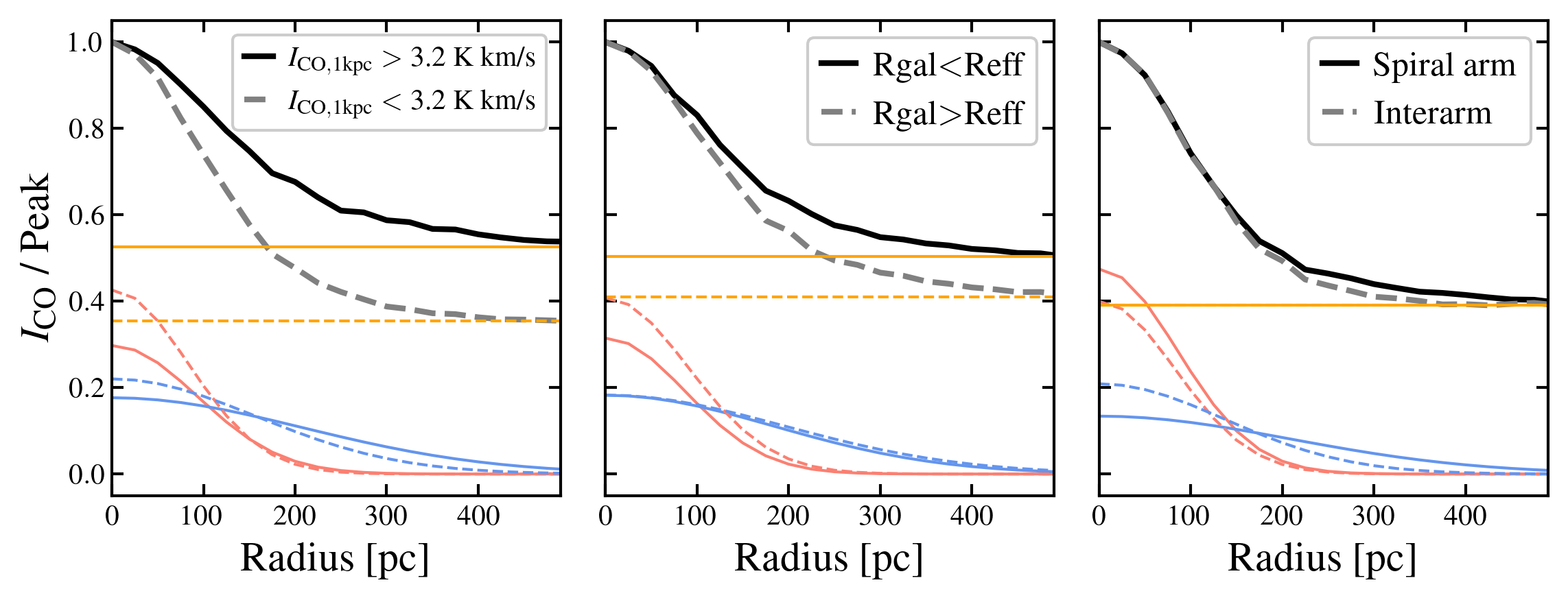}{\textwidth}{}
}
\vspace{-2.5\baselineskip}
\gridline{
\leftfig{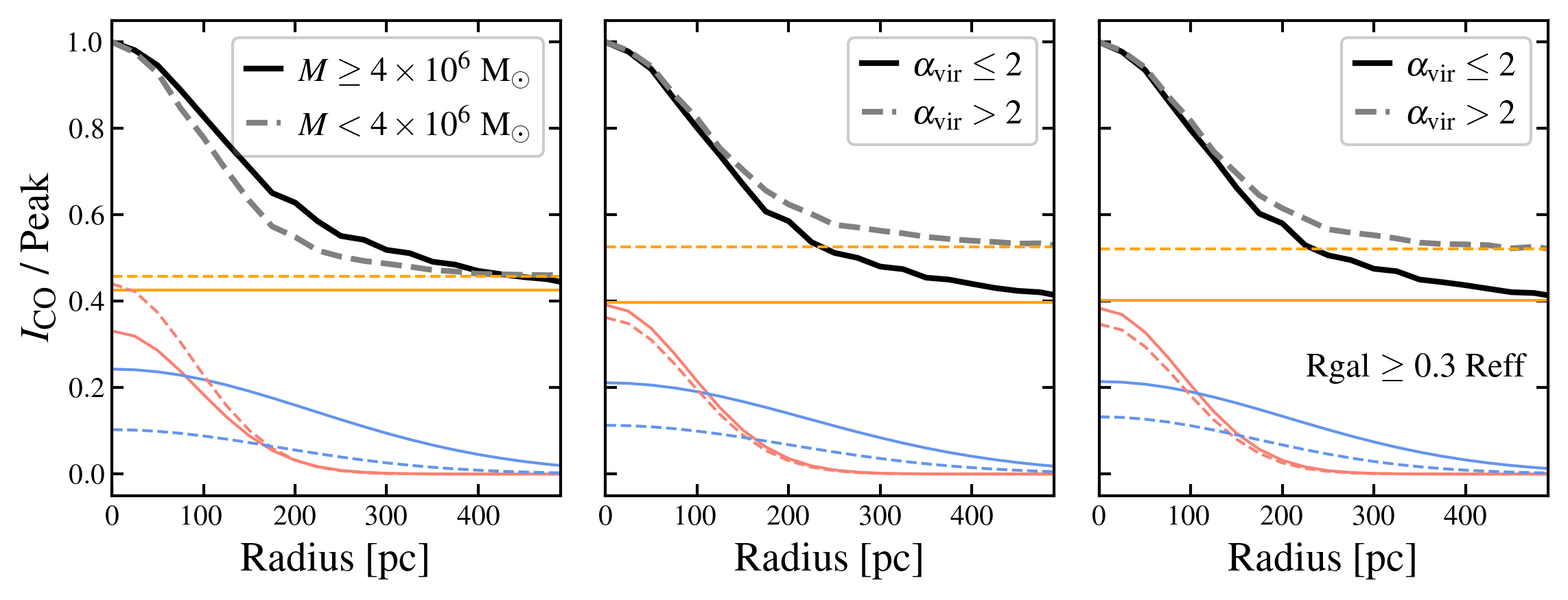}{\textwidth}{}
}
\vspace{-2.5\baselineskip}
\caption{(\textit{Top}) The stacked radial profiles of subgroups of GMCs within three different environmental categories, \Icokpc (\textit{left}), $r_{\mathrm{gal}}$ (\textit{middle}) and within or outside spiral arms (\textit{right}), with their corresponding decomposition (red for the narrow Gaussian, blue for the broad Gaussian, and yellow for the constant offset). (\textit{Bottom}) The stacked radial profiles of subgroups of GMCs of different properties, GMC mass (\textit{left}) and virial parameter (\textit{middle} and \textit{right}) with their corresponding decompositions. The middle panel includes all GMCs while the right panel only includes GMCs not in the centre (galactic radius greater than 0.3 times effective radius). }
\label{fig:stacking_div}
\end{figure*}

\section{Intensity profiles around GMCs}
\label{sec:stacking}

In the previous sections, we investigated the clustering of GMCs by approximating them as point sources. 
Because of the minimum spacing between extracted GMCs, this approach excludes information about how the CO emission 
may be clustered on spatial scales close to the resolution and below the typical cloud-cloud spacing (\S~\ref{sec:spacing}). To investigate this smaller scale clustering of CO emission, we construct a radial profile of CO emission around each GMC peak. We stack the profiles to obtain a characteristic radial profile of the CO intensity surrounding the GMCs in our sample, and investigate whether the typical profile varies between different galaxies and galactic environments. 

To measure the radial profile, we define annuli around each CO peak with bin width of 25~pc (this places three bins across the HWHM of the beam) out to a radius of 1~kpc. We then measure the average $I_{\mathrm{CO}}$ from the integrated intensity map within each annulus. We normalize these profiles by dividing each by its central value. Finally, we calculate the median of all normalized GMC radial profiles with each galaxy. By using the median value to construct these stacks, we avoid a few bright GMCs biasing the results. We construct these stacks for each galaxy, as well as combining all GMCs across all galaxies in our sample. 


Our stacking analysis shares a similar but not identical mathematical principle as the 2PCF. The 2PCF measures the probability of finding a pair of objects separated by $\theta$. Our stacking measures the probability of finding some CO emission at a distance of $\theta$ from a GMC peak. These stacks can thus be thought of as the peak-intensity two-point cross-correlation.



\subsection{Comparison to model images}
\label{subsec:model_image}

To interpret our results, we construct model images similar to the controls used for the 2PCF (\S~\ref{subsec:mock_cat}). To generate these, we place a model GMC at the location of each catalogued GMC. These model GMCs have a peak integrated intensity $I_0 \propto \sqrt{2\pi}\ T_{\mathrm{peak}}\ \sigma_v$, where $T_{\mathrm{peak}}$ is the peak brightness temperature and $\sigma_v$ is the velocity dispersion recorded for the GMC in the catalogue. They are elliptical Gaussians in shape, with the major axis, minor axis, and position angles measured by \texttt{CPROPS} (including the beam in order to match the observations). We add up all of the model GMCs in each galaxy to construct a model CO distribution. Finally, we found it necessary to moderately renormalize the model images to match the intensity values at the peak position in observations. To do this, we measured the intensity in both the model image and the data at each GMC peak. We then rescaled the model image by the median ratio of the data to the model for these peaks. We measure the radial profiles for all GMCs in each model image following the same procedure used for the observational data. We likewise calculate the stacked median radial profile for each galaxy and all GMCs in the sample using the model images. 

Fig.~\ref{fig:model_images} shows an example model image and a residual map constructed by subtracting the model from the data. Fig.~\ref{fig:stacking_radial_galaxies} shows the profiles constructed from the data and model images. These match well, especially at scales $\lesssim 200$~pc. Beyond $\sim 200$~pc and out to $\gtrsim 500$~pc, the data show a moderately higher intensity than our model, with the model $\approx 0.8$ times the data on average out to $\approx 500$~pc. In Fig.~\ref{fig:model_images} the bar and spiral arms stand out in the residual map, suggesting that the extended emission not captured in our model indicated in Fig.~\ref{fig:stacking_radial_galaxies} may primarily come from these structures.

Fig.~\ref{fig:stacking_radial_galaxies} suggests that this excess in the data relative to the model at $200{-}500$~pc is a general feature, but that its magnitude varies by galaxy. This variation appears linked to the fraction of the overall CO flux that is accounted for in the GMC catalogue, i.e., the flux completeness of the GMC catalogue ($f_{\mathrm{GMC}}=F_{\mathrm{GMC}} / F_{\mathrm{tot}}$).  The left panels of Fig.~\ref{fig:stacking_radial_galaxies} show that normalized profiles constructed using both the data and the reconstructed model images are more compact in galaxies with low $f_{GMC}$, i.e., clouds in galaxies with low $f_{GMC}$ 
are less likely to have nearby surrounding CO emission. Fig.~\ref{fig:stacking_radial_galaxies} shows that this decline is even more pronounced for the model than the data.
As a result, as $f_{\mathrm{GMC}}$ increases and more of the measured flux enters the catalogue, the data and model exhibit a better correspondence. This extended emission component results from faint clouds or diffuse gas that are not captured by \texttt{CPROPS}. Since the sensitivity threshold is fixed for our data set, this component represents a larger fraction of the total CO emission in lower mass, fainter galaxies.

\subsection{Shape of the stacked radial profile}
\label{subsec:fit_profile}

\begin{table*}
\centering
\caption{Parameters of double-Gaussian fit for the stacked radial profile}
\label{tab:gaus_fit}
\begin{threeparttable}
\begin{tabularx}{0.8\textwidth}{ccccccc}
\hline \hline 
 & $A_1$ & $\sigma_1$ & $A_2$ & $\sigma_2$ & C & $A_2\sigma_2^2/(A_1\sigma_1^2+A_2\sigma_2^2)$\\
\hline
all data & 0.38 $\pm$ 0.02 & 90 $\pm$ 3 & 0.18 $\pm$ 0.02 & 200 $\pm$ 20 & 0.46 $\pm$ 0.03 & 0.70 $\pm$ 0.04\\
high $I_{\mathrm{CO, 1kpc}}$ & 0.30 $\pm$ 0.04 & 93 $\pm$ 5 & 0.18 $\pm$ 0.03 & 210 $\pm$ 30 & 0.52 $\pm$ 0.05 & 0.75 $\pm$ 0.06\\
low $I_{\mathrm{CO, 1kpc}}$ & 0.44 $\pm$ 0.02 & 83 $\pm$ 2 & 0.21 $\pm$ 0.02 & 160 $\pm$ 10 & 0.35 $\pm$ 0.03 & 0.65 $\pm$ 0.06 \\ 
$R < R_{\mathrm{eff}}$ & 0.31 $\pm$ 0.04 & 87 $\pm$ 4 & 0.18 $\pm$ 0.03 & 180 $\pm$ 20 & 0.51 $\pm$ 0.05 & 0.72 $\pm$ 0.05\\
$R > R_{\mathrm{eff}}$ & 0.41 $\pm$ 0.03 & 90 $\pm$ 3 & 0.18 $\pm$ 0.03 & 200 $\pm$ 20 & 0.41 $\pm$ 0.05 & 0.68 $\pm$ 0.05\\
arm & 0.47 $\pm$ 0.02 & 85 $\pm$ 2 & 0.13 $\pm$ 0.02 & 210 $\pm$ 20 & 0.39 $\pm$ 0.03 & 0.63 $\pm$ 0.04\\
interarm & 0.4 $\pm$ 0.09 & 83 $\pm$ 5 & 0.21 $\pm$ 0.09 & 140 $\pm$ 20 & 0.39 $\pm$ 0.1 & 0.59 $\pm$ 0.1\\
$M \geq 4 \times 10^6$ M$_{\odot}$ & 0.33 $\pm$ 0.03 & 92 $\pm$ 4 & 0.24 $\pm$ 0.02 & 220 $\pm$ 20 & 0.43 $\pm$ 0.04 & 0.8 $\pm$ 0.03 \\
$M < 4 \times 10^6$ M$_{\odot}$  & 0.44 $\pm$ 0.03 & 87 $\pm$ 3 & 0.1 $\pm$ 0.03 & 180 $\pm$ 30 & 0.46 $\pm$ 0.04 & 0.5 $\pm$ 0.01 \\
$\alpha_{\rm vir} \leq$ 2 & 0.39 $\pm$ 0.03 & 91 $\pm$ 3 & 0.21 $\pm$ 0.02 & 220 $\pm$ 30 & 0.4 $\pm$ 0.04 & 0.76 $\pm$ 0.04 \\
$\alpha_{\rm vir} >$2 & 0.36 $\pm$ 0.02 & 90 $\pm$ 2 & 0.11 $\pm$ 0.02 & 200 $\pm$ 30 & 0.53 $\pm$ 0.03 & 0.6 $\pm$ 0.06 \\
\hline
\end{tabularx}
\begin{tablenotes}
    \item 
\end{tablenotes}
\end{threeparttable}
\end{table*}

To quantify the shape of the stacked profiles, we tested various analytic functions and find single or double Gaussians can describe the shape of the profile well:  
\begin{eqnarray}
\label{eq:doublegauss}
\begin{cases} 
A e^{-\frac{r^2}{2\sigma^2}} + C, & \mathrm{with}\ A+C=1 \\ 
A_1 e^{-\frac{r^2}{2\sigma_1^2}} + A_2 e^{-\frac{r^2}{2\sigma_2^2}} + C, &\mathrm{with}\ A_1+A_2+C=1,
\end{cases}
\end{eqnarray}
where $r$ is the radius from the peak and the remaining variables are free parameters. Fig.~\ref{fig:stacking_fitting} shows that double Gaussians better describe the shape of the stacked profile than the single Gaussian. The double Gaussian plus constant model gives a residual with a median amplitude of $\sim 0.003$ (i.e., $0.3\%$ of the peak $I_{\rm CO}$), compared to $0.02$ for the single Gaussian plus constant. Furthermore, the single Gaussian plus constant fit can deviate from the real profile by up to 10\% at 500 -- 1000 pc (not shown). The second Gaussian captures a slow decline out to large radius.

Table~\ref{tab:gaus_fit} reports our best fit parameters. For the full stack, we find a compact Gaussian with width $\sigma_{\rm 1} = 90$~pc, which corresponds to $\approx 63$~pc after deconvolving the beam, and an extended Gaussian with $\sigma_{\rm 2} = 200$~pc, corresponding to $\approx 190$~pc after deconvolution. We note that the deconvolved radius of the narrow Gaussian component is close to the beam size ($\sigma = {\rm FWHM} / 2.355=63$ pc), and hence the size measurement has a large uncertainty. 
The amplitude of the constant term, $C$, varies between $0.39$ and $0.52$, i.e., comparable to or greater than the peak amplitude of the two combined Gaussians. Both the large fraction of power in the second Gaussian component and the high constant background level highlight that there is significant power in the extended part of the profile. As Fig.~\ref{fig:stacking_radial_galaxies} shows, this extended structure exists in the model images constructed only from the \texttt{CPROPS} catalogue data, but the model images tend to have lower amplitudes at large scales than we observe in the real images. The $\approx 90$~pc sizes of the first (inner) Gaussian component of our fits resemble the $100$~pc median $1\sigma$ size of the catalogued GMCs. This reflects the good agreement between the catalogue-based model and the data at small $\lesssim 200$~pc scales in Fig.~\ref{fig:stacking_radial_galaxies}.


The high value of the constant background, $C \approx 0.5$ of the peak intensity, suggests that GMCs are closely packed around the identified \texttt{CPROPS} peaks. 
If each kpc region had GMCs of similar intensity, we would expect $C=0$ if only one GMC sits within the region, and $C=1.0$ if GMCs fully cover the region. Therefore, our measured background level suggests a typical filling factor (i.e., space density) of $\sim 50\%$ at $150$~pc resolution out to $\sim 500$~pc for regions centred on detected GMCs. This constant background also provides an excellent fit to the stacked GMC radial profile at 500–1000 pc. This interpretation is further supported by our two-point correlation function (2PCF) analysis, which shows almost no scale dependence across the same range once the contribution from large-scale structures is accounted for (\S \ref{subsec:tpcf_slope}). Physically, one should imagine this constant term as a field of molecular clouds unrelated to the peak. At sufficiently high resolution, the emission contributing to the constant offset in the stacked intensity profiles would resolve into discrete patches of CO emission. Since these patches show no preferred spatial relationship to the catalogued peaks, and have a sufficiently high surface density, there is typically some emission within roughly half of separation length in between any given pair of peaks, causing them to appear as a constant background level in the stacks. 

In addition to the background field captured by the constant, the profiles indicate an extended overdensity of CO emission around the peaks. In the double Gaussian fits, the second, extended Gaussian contributes $\sim$70\% of the total flux in the profile once the constant offset is subtracted. This broad component has a typical $1\sigma$ radius of $\sim$200 pc. This is much larger than the expected radius of individual GMCs based on highly resolved studies. This suggests that most of this power comes from the most close-by GMCs or extended CO-emitting gas structures like filaments or clusters of smaller clouds. Such clustering of gas  
may be qualitatively consistent with the sort of hierarchical structures that emerge from a scale-free GMC formation process \citep[e.g.,][]{guszejnov_universal_2018}, which predicts increasing clustering towards smaller scales. To test this, it would be interesting to construct stacked profiles around projected column density peaks in simulated data where such collapse is known to be occurring.

We summarize statistics measured from stacked profiles of the GMC populations in individual galaxies, along with the single/double Gaussian fitting results in Table~\ref{tab:stacking_summary}. The stacked profiles for the individual galaxies are shown in Fig.~\ref{fig:stacking_indvd}.

\subsection{Stacked profile and environment}
\label{sec:stack_env}

In the \textbf{top} panels of Fig.~\ref{fig:stacking_div}, we examine how the stacked profiles of the CO emission around GMCs vary according to galactic environment.
We divide our sample into halves based on three environmental factors: the kpc-scale CO intensity at the cloud's position \Icokpc, the galactocentric radius, and whether the GMC is located in a spiral arm or an interarm region \citep[from][]{querejeta_stellar_2021}. For the spiral arm versus interarm comparison, we exclude GMCs that fall in other dynamical environments (e.g., centre, bar, or disk of non-spiral galaxies). The best-fitting parameters for each environment are recorded in Table~\ref{tab:gaus_fit}. 

Fig.~\ref{fig:stacking_div} shows that GMCs in gas-rich regions and the inner parts of galaxies have broader stacked profiles than those in gas-poor regions. In high \Icokpc\ regions, $\sigma_1$, $\sigma_2$, and $C$ all appear larger. 
This reflects that GMCs in gas-rich regions and inner galaxies are more clustered (\S \ref{subsec:homogspacing}), with neighbouring GMCs contributing much of the flux that appears in the $C$ term and contributing to the extended Gaussian ($A_2$, $\sigma_2$). In principle, this could also reflect systematically larger GMC sizes in these regions, but we checked the GMC catalogues and found no systematic size difference among different environments, so the different primarily seems to reflect crowding. Spiral arms show a much weaker contrast with interarm regions, but we recall the different completeness in the two regions mentioned in \S \ref{sec:data}. Furthermore, our stacking measurement only measures the structure of emission around GMC peaks and does not distinguish, e.g., the mass function, dynamical state or other quantities.

\subsection{Stacked profile as a function of GMC properties}

In the lower panels of Fig. \ref{fig:stacking_div}, we divide GMCs based on their mass and virial parameter ($\alpha_{\rm vir}$). We find that high- and low-mass GMCs show a similar constant background level relative to the peak value. This suggests that GMCs of different masses reside in regions with similar GMC space density (i.e., GMC filling factor, \S \ref{subsec:fit_profile}). However, we find a much stronger broad Gaussian component for more massive GMCs, which suggests a link between formation of massive GMCs and strong local gas clustering.

We also compare stacked profiles of GMCs with different $\alpha_{\rm vir}$ (lower right panel of Fig. \ref{fig:stacking_div}). GMCs that appear less gravitationally bound ($\alpha_{\rm vir} >$ 2) show a higher constant background. One possible scenario is that a large fraction of high $\alpha_{\rm vir}$ GMCs are found in the centres of galaxies  \citep{sun_molecular_2020, rosolowsky_giant_2021}, where the GMC space density tends to be high (Fig. \ref{fig:spacing_and_nn}). We test this scenario by excluding GMCs in the centres (with galactic radius smaller than 0.3 $R_{\rm eff}$, bottom right panel of Fig \ref{fig:stacking_div}). We still see a similar contrast in the constant background between GMCs of different $\alpha_{\rm vir}$, which seems to rule out this scenario. Alternatively, it may reflect the observational bias that GMCs tend to be more likely spatially aligned in more crowded regions, where we are more likely to overestimate the velocity dispersion and thus $\alpha_{\rm vir}$. 
On the other hand, we see a stronger broad Gaussian component for more gravitationally bound GMCs. This is in tension with the environmental dependence that we find in \S \ref{sec:stack_env}, where stronger Gaussian component appears in regions with higher GMC space densities (or with higher constant offset). This could be explained by a close connection between gravitationally bound structures and neighbouring extended halos of smaller, lower density structures.   

Assuming that the gas surrounding GMCs is in hydrostatic equilibrium, a stronger broad Gaussian component implies a stronger ambient pressure acting on the target GMCs. This aligns with our expectation that a stronger external pressure can help with the formation of more massive and gravitationally bound GMCs. This pressure could come from global collapse of molecular gas \citep[e.g.,][]{vazquez-semadeni_global_2019}, dynamical impact of certain galactic structures \citep[e.g.,][]{dobbs2008,dobbs2013, meidt_bottom_2024} or intense stellar feedback activities \citep[e.g.,][]{inutsuka2015, kobayashi2017}. A more quantitative comparison with matched simulations can help us disentangle the relative contribution from those different physical mechanisms.


\subsection{Impact of resolution on the stacked profile}

\begin{figure*}
\gridline{
\fig{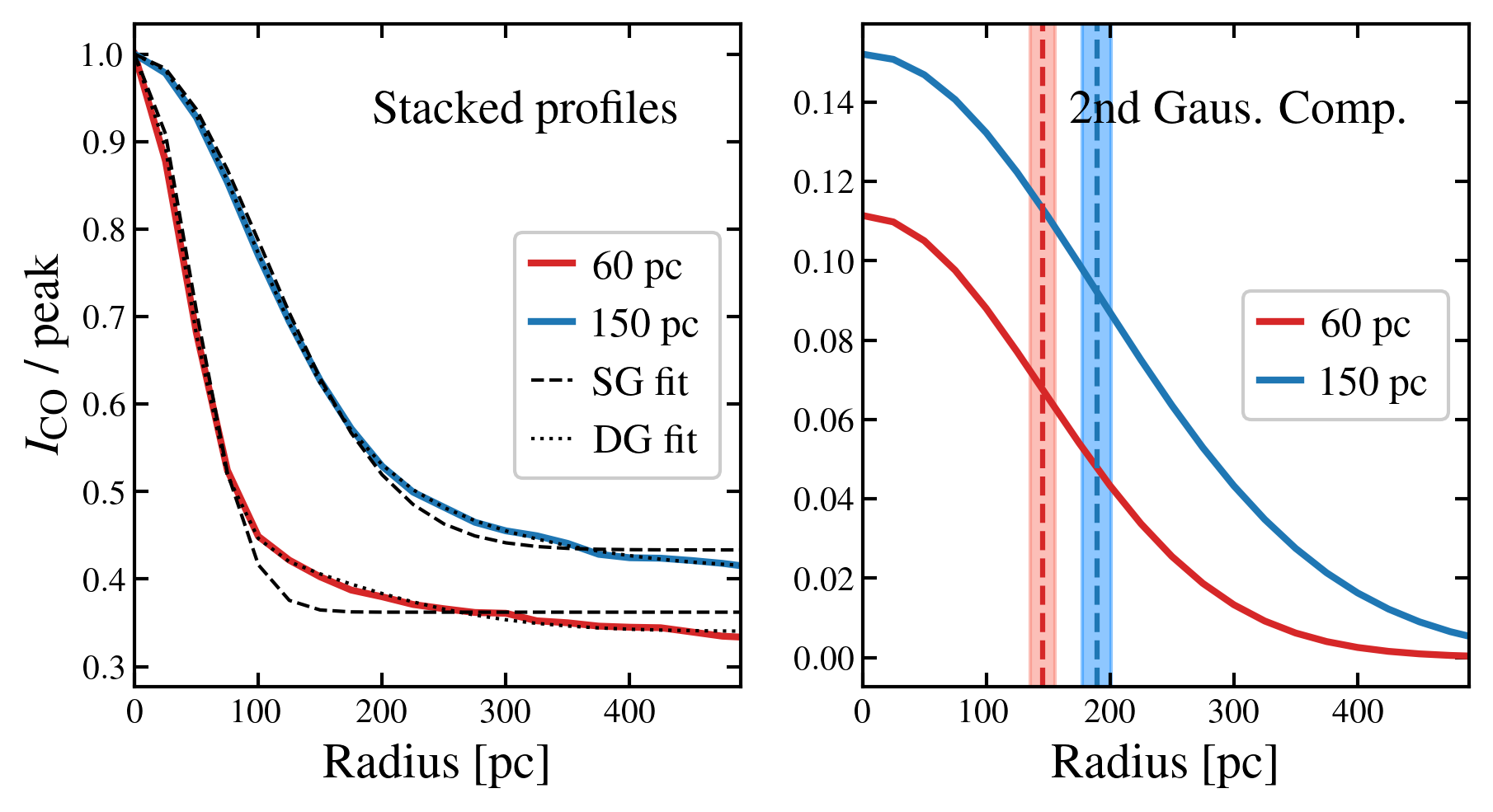}{0.9\textwidth}{}
}
\vspace{-2\baselineskip}
\caption{Comparison of stacked intensity profiles at different resolutions. (\textit{Left}) The measured stacked median of CO intensity profiles at 60 pc (red) and 150 pc (blue) resolution. The dashed and dotted lines indicate single Gaussian (SG) and double Gaussian (DG) fits (Eq. \ref{eq:doublegauss}). (\textit{Right}) The decomposed second Gaussian component from the double-Gaussians fits at the two resolutions. The vertical lines and shaded regions indicate the 1$\sigma$ values and the parameter error ranges of the second Gaussian components. We see that the size of the second Gaussian component is relatively independent of the spatial resolution. }
\label{fig:stacking_res}
\end{figure*}


Fig. \ref{fig:stacking_res} compares stacked profiles of CO emission around peaks at 60~pc and 150~pc resolution. Note that because the sensitivity and peaks identified also varies as a function of resolution, the two profiles do not reflect the same samples of peaks but we do restrict the comparison to the seven galaxies that have both catalogues with high sensitivity \citep[see][]{rosolowsky_giant_2021}. 

As expected, the stacked profile constructed at 60~pc resolution shows a much narrower core compared to the profile constructed at 150~pc resolution. To first order, this simply reflects the impact of beam dilution. Formally, our fits suggest that the narrow Gaussian component at 60~pc resolution is smaller than its counterpart at 150~pc, 1$\sigma = 38$~pc for the 60~pc data set compared to 1$\sigma = 88$~pc for the 150~pc data. In practice, both profiles indicate compact unresolved emission below the beam scale at the core of the profile. 


By contrast, the broad Gaussian component is much less affected by resolution (Fig. \ref{fig:stacking_res}, right panel). The broad Gaussian component at 60~pc has 1$\sigma = 145$~pc, while that at 150~pc has 1$\sigma=190$ pc. Thus our measurement of extended local clustering CO emission on $\sim 200$~pc scales around bright peaks appears robust to resolution and changes in data sensitivity. This structure appears to be a general feature across the PHANGS--ALMA data set.


\section{Summary}
\label{sec:summary}

We apply techniques common in studies of large-scale structure to analyse the spatial distribution of \numGMCs\ giant molecular clouds (CO peaks) drawn from \numGalaxies\ nearby star-forming galaxies observed by PHANGS--ALMA. We measure cloud spacing via the nearest and fifth-nearest neighbour distance and the number of neighbours within fixed size apertures (\S~\ref{sec:spacing}). Next we measure the two-point correlation function (2PCF) of GMCs relative to several controls (\S~\ref{sec:tpcf}). Finally, we construct stacked radial profiles of integrated CO(2-1) intensity about GMC peaks (\S~\ref{sec:stacking}). These methods provide complementary information about the structure of the molecular gas. We observe:

\begin{enumerate}
\setcounter{enumi}{0}
\item When considering heterogeneous data, the nearest neighbour distance between GMCs depends on the resolution and sensitivity of the data used to construct the GMC catalogue (\S~\ref{subsec:spacingres}, Fig.~\ref{fig:spacing_and_res}). For the native resolution PHANGS--ALMA data, the median nearest neighbour distance is $3.0$ times the beam size. The fact that this number is a multiple of the beam size appears to reflect a limit to how closely common GMC identification algorithms will ``pack'' clouds in rich regions.

\item Sensitivity correlates with cloud spacing because more sensitive measurements recover GMCs in lower-density outer disks and other faint regions, which typically have larger cloud-cloud spacing.
\end{enumerate}

Analyzing GMC catalogues derived from cubes homogenized to share a common $150$~pc physical resolution and fixed 46~mK noise level, we find:

\begin{enumerate}
\setcounter{enumi}{2}
\item The nearest neighbour and fifth-nearest neighbour distance between GMCs correlate with galactocentric radius and anti-correlate with the kpc-scale CO intensity of the disk, \Icokpc. GMCs are more strongly clustered in the gas-rich, inner regions of galaxies (\S~\ref{subsec:homogspacing}, Fig.~\ref{fig:spacing_and_nn}).

\item The number of GMCs within a fixed aperture centred on each cloud anti-correlates with galactocentric radius and correlates with \Icokpc . The number of neighbours is a more sensitive metric than cloud spacing in dense regions, while the cloud spacing is a more sensitive metric in diffuse regions.
\end{enumerate}

This suggests that the GMC distribution largely reflects the large-scale organisation of gas in the galaxy. This is reinforced by our analysis of the two-point correlation function:

\begin{enumerate}
\setcounter{enumi}{4}
\item The two-point correlation function (2PCF) of GMC locations shows strong clustering when calculated relative to a uniform control. It rises from large to small scales. The observed correlation function peaks near $\approx 500$~pc and stays flat or declines at smaller scales (\S~\ref{sec:tpcf}, Fig.~\ref{fig:twopt_multipanel}). This behaviour on small scales reflects the limited ability of the cloud-finding algorithms to pack clouds more closely, so that the main utility of the 2PCF is on scales greater than several times the beam size.

\item The amplitude of the 2PCF decreases when calculated relative to any control that accounts for the large-scale CO distribution (an exponential disk, the observed CO radial profile, or a kpc-scale CO map, \S~\ref{subsec:2pcf_results}, Fig.~\ref{fig:twopt_multipanel}). This indicates that most of the clustering of GMCs identified at $150$~pc resolution reflects the large-scale gas organization. 


\item The remaining clustering signal appears to correlate with the presence of gas-rich centres, spiral arms and bar lanes. Galaxies with these features show higher 2PCF amplitude at a fiducial \textbf{$500$}~pc scale compared to those with more flocculent structure (\S~\ref{subsec:galtogal2pcf}, Fig.~\ref{fig:mom0_sort_by_tpcf}).

\item The ``Flat'' 2PCF has a power-law slope of -0.25 (scatter range of -0.1 -- 0.5) across scales of 500 -- 1000 pc. This is consistent with many of the slopes of the 2PCF for stellar clusters measured on these larger scales (their outer power law) in recent studies. In contrast, the ``$I_{\rm CO}$'' 2PCF is almost constant across this scale range with a slope close to 0, which suggests that the second power-law tail in these and likely other extra-galactic 2PCF measurements arises from the impact of large-scale structures. 

\end{enumerate}

Since the cloud spacing and 2PCF approximate GMCs as point sources, they miss information on scales between the resolution and nearest neighbour distance. To address this:

\begin{enumerate}
\setcounter{enumi}{8}
\item We construct stacked profiles of CO(2-1) intensity centred on the locations of GMCs. The resulting profile is well-described by a combination of two Gaussians and a constant offset (\S~\ref{sec:stacking}, Tab.~\ref{tab:gaus_fit}, Fig.~\ref{fig:stacking_fitting}). The large constant offset suggests a typical filling factor of $\sim$ 50\% at 150 pc resolution out to $\sim$500 pc for regions centred on detected GMCs. Above this constant offset, the broad Gaussian component accounts for $\sim$70\% of the total over-density power. This suggests that clustered distributions of neighbouring gas around bright CO peaks represent a general feature of CO emission from galaxies. 

\item We compare our measured stacks to stacks derived from model images constructed using the GMC catalogue locations and cloud sizes. The results agree well out to $\approx 200$~pc scales, validating that the catalogued cloud sizes capture the spatial distribution of the CO emission on these scales.

\item We stack subsets of the data after splitting the peak by \Icokpc , $R_{\rm gal}$, and spiral arm vs. interarm environment. We observe more extended stacked intensity profiles in gas-rich (high \Icokpc ) regions, low $R_{\rm gal}$ and spiral arm regions. However, the spiral arm and interarm stacks show very similar profiles, which potentially indicates a similar physical process regulating GMC formation in these two regions. We caution that our measurements might be biased towards bright clouds and could be affected by our environmental masking strategy, but recognise this as a result that merits future, more careful investigation with better quality data.

\item We create stacked CO profiles about peaks after splitting by GMC mass and virial parameter. We find that higher-mass GMCs ($M \geq 4\times 10^6$ M$_{\odot}$) and more gravitationally bound GMCs ($\alpha_{\rm vir}\leq 2$) are associated with a stronger broad Gaussian component ($\sim$80\% of the over-density power), implying stronger clustering of neighbouring gas for these structures. This might imply a link between such clustering and the formation of massive, gravitationally bound GMCs. We propose that higher resolution data to enable a more direct comparison between this clustering signal, gas kinematics and simulations of GMC formation is a fruitful avenue for future investigation. 

\item We compare the stacked profiles at 60~pc and 150~pc resolutions for seven galaxies where sensitive, high resolution observations are available. The narrow Gaussian component is more compact and pronounced at 60~pc ($\sigma_1$), consistent with bright, central unresolved emission in both profiles that has a measured extent proportional to the beam size ($l$). The size of the broad Gaussian component ($\sigma_2$) agrees within $\approx 30\%$ for both profiles.

\end{enumerate}
We propose several future directions for this work. The utility of these metrics can be gauged by comparison to simulations that aim to produce realistic molecular ISM phases. If these metrics are difficult for current simulations to reproduce, they could be adopted as an informative benchmark that future simulation should aim to satisfy. Our analyses remain limited by the 150~pc resolution of our main dataset. Subsets of the PHANGS--ALMA CO maps achieve up to $\sim3\times$ higher physical resolution (albeit for only a handful of galaxies and with less diversity of galactic environments). 
A wide-field CO mapping survey of galaxies at $\sim10$pc resolution would be needed to clearly resolve the spatial scales of interest, and disentangle cloud-cloud clustering from decomposition biases towards beam-scale structures. Very nearby galaxies would be a preferred target for this work. Another possibility would be to apply our analysis to JWST PAH emission maps, which achieve higher spatial resolution and sensitivity than ALMA and appear to trace the neutral gas structure well \citep[e.g.,][]{chown_polycyclic_2025}.

\begin{acknowledgements}

This work was carried out as part of the PHANGS collaboration.

AKL thanks Paul Martini for valuable discussions during the development of this project.

A.K.L. gratefully acknowledges support from NSF AST AWD 2205628, JWST-GO-02107.009-A, and JWST-GO-03707.001-A and a Humboldt Research Award.

ER acknowledges the support of the Natural Sciences and Engineering Research Council of Canada (NSERC), funding reference number RGPIN-2022-03499.

AH acknowledges support by the Programme National Cosmology et Galaxies (PNCG) of CNRS/INSU with INP and IN2P3, co-funded by CEA and CNES, and by the Programme National Physique et Chimie du Milieu Interstellaire (PCMI) of CNRS/INSU with INC/INP co-funded by CEA and CNES.

J.S. acknowledges support by the National Aeronautics and Space Administration (NASA) through the NASA Hubble Fellowship grant HST-HF2-51544 awarded by the Space Telescope Science Institute (STScI), which is operated by the Association of Universities for Research in Astronomy, Inc., under contract NAS~5-26555.

M.C.\ gratefully acknowledges funding from the Deutsche Forschungsgemeinschaft (DFG, German Research Foundation) through an Emmy Noether Research Group (grant number CH2137/1-1).
COOL Research DAO \citep{cool_whitepaper} is a Decentralised Autonomous Organisation supporting research in astrophysics aimed at uncovering our cosmic origins.

This paper makes use of the following ALMA data, which have been processed as part of the PHANGS--ALMA CO(2-1) survey: \\
\noindent ADS/JAO.ALMA\#2012.1.00650.S, \linebreak 
ADS/JAO.ALMA\#2013.1.00803.S, \linebreak 
ADS/JAO.ALMA\#2013.1.01161.S, \linebreak 
ADS/JAO.ALMA\#2015.1.00121.S, \linebreak 
ADS/JAO.ALMA\#2015.1.00782.S, \linebreak 
ADS/JAO.ALMA\#2015.1.00925.S, \linebreak 
ADS/JAO.ALMA\#2015.1.00956.S, \linebreak 
ADS/JAO.ALMA\#2016.1.00386.S, \linebreak 
ADS/JAO.ALMA\#2017.1.00392.S, \linebreak 
ADS/JAO.ALMA\#2017.1.00766.S, \linebreak 
ADS/JAO.ALMA\#2017.1.00886.L, \linebreak 
ADS/JAO.ALMA\#2018.1.01321.S, \linebreak 
ADS/JAO.ALMA\#2018.1.01651.S, \linebreak 
ADS/JAO.ALMA\#2018.A.00062.S, \linebreak 
ADS/JAO.ALMA\#2019.1.01235.S, \linebreak 
ADS/JAO.ALMA\#2019.2.00129.S, \linebreak 
ALMA is a partnership of ESO (representing its member states), NSF (USA), and NINS (Japan), together with NRC (Canada), NSC and ASIAA (Taiwan), and KASI (Republic of Korea), in cooperation with the Republic of Chile. The Joint ALMA Observatory is operated by ESO, AUI/NRAO, and NAOJ. The National Radio Astronomy Observatory is a facility of the National Science Foundation operated under cooperative agreement by Associated Universities, Inc.

\end{acknowledgements}

%

\bibliographystyle{aa/aa} 
\bibliography{ref_v1,refs2} 

\begin{thebibliography}{59}
\expandafter\ifx\csname natexlab\endcsname\relax\def\natexlab#1{#1}\fi

\bibitem[{Anand {et~al.}(2021)Anand, Lee, Van~Dyk, Leroy, Rosolowsky, Schinnerer, Larson, Kourkchi, Kreckel, Scheuermann, Rizzi, Thilker, Tully, Bigiel, Blanc, Boquien, Chandar, Dale, Emsellem, Deger, Glover, Grasha, Groves, S.~Klessen, Kruijssen, Querejeta, Sánchez-Blázquez, Schruba, Turner, Ubeda, Williams, \& Whitmore}]{anand_distances_2021}
Anand, G.~S., Lee, J.~C., Van~Dyk, S.~D., {et~al.} 2021, MNRAS, 501, 3621, publisher: OUP ADS Bibcode: 2021MNRAS.501.3621A

\bibitem[{{Balogh} {et~al.}(2004){Balogh}, {Eke}, {Miller}, {Lewis}, {Bower}, {Couch}, {Nichol}, {Bland-Hawthorn}, {Baldry}, {Baugh}, {Bridges}, {Cannon}, {Cole}, {Colless}, {Collins}, {Cross}, {Dalton}, {de Propris}, {Driver}, {Efstathiou}, {Ellis}, {Frenk}, {Glazebrook}, {Gomez}, {Gray}, {Hawkins}, {Jackson}, {Lahav}, {Lumsden}, {Maddox}, {Madgwick}, {Norberg}, {Peacock}, {Percival}, {Peterson}, {Sutherland}, \& {Taylor}}]{nearest_nbrs_galaxy}
{Balogh}, M., {Eke}, V., {Miller}, C., {et~al.} 2004, \mnras, 348, 1355

\bibitem[{Bigiel {et~al.}(2008)Bigiel, Leroy, Walter, Brinks, de~Blok, Madore, \& Thornley}]{bigiel_star_2008}
Bigiel, F., Leroy, A., Walter, F., {et~al.} 2008, AJ, 136, 2846, aDS Bibcode: 2008AJ....136.2846B

\bibitem[{Brown {et~al.}(2021)Brown, Wilson, Zabel, Davis, Boselli, Chung, Ellison, Lagos, Stevens, Cortese, Bahé, Bisaria, Bolatto, Cashmore, Catinella, Chown, Diemer, Elahi, Hani, Jiménez-Donaire, Lee, Leidig, Mok, Olsen, Parker, Roberts, Smith, Spekkens, Thorp, Tonnesen, Vienneau, Villanueva, Vogel, Wadsley, Welker, \& Yoon}]{brown_vertico_2021}
Brown, T., Wilson, C.~D., Zabel, N., {et~al.} 2021, ApJS, 257, 21, arXiv: 2111.00937

\bibitem[{{Chevance} {et~al.}(2025){Chevance}, {Kruijssen}, \& {Longmore}}]{cool_whitepaper}
{Chevance}, M., {Kruijssen}, J.~M.~D., \& {Longmore}, S.~N. 2025, arXiv e-prints, arXiv:2501.13160

\bibitem[{{Chevance} {et~al.}(2023){Chevance}, {Krumholz}, {McLeod}, {Ostriker}, {Rosolowsky}, \& {Sternberg}}]{chevance_life_2023}
{Chevance}, M., {Krumholz}, M.~R., {McLeod}, A.~F., {et~al.} 2023, in Astronomical Society of the Pacific Conference Series, Vol. 534, Protostars and Planets VII, ed. S.~{Inutsuka}, Y.~{Aikawa}, T.~{Muto}, K.~{Tomida}, \& M.~{Tamura}, 1

\bibitem[{Chown {et~al.}(2025)Chown, Leroy, Sandstrom, Chastenet, Sutter, Koch, Koziol, Neumann, Sun, Williams, Baron, Anand, Barnes, Bazzi, Belfiore, Bigiel, Bolatto, Boquien, Cao, Chevance, Colombo, Dale, den Brok, Egorov, Eibensteiner, Emsellem, Hassani, Henshaw, He, Kim, Klessen, Kreckel, Larson, Lee, Meidt, Murphy, Oakes, Ostriker, Pan, Pathak, Rosolowsky, Sarbadhicary, Schinnerer, Teng, Thilker, Weinbeck, \& Watkins}]{chown_polycyclic_2025}
Chown, R., Leroy, A.~K., Sandstrom, K., {et~al.} 2025, ApJ, 983, 64, publisher: IOP ADS Bibcode: 2025ApJ...983...64C

\bibitem[{{Clark} \& {Evans}(1954)}]{nearest_nbrs}
{Clark}, P.~J. \& {Evans}, F.~C. 1954, Ecology, 35, 445

\bibitem[{Colombo {et~al.}(2014)Colombo, Hughes, Schinnerer, Meidt, Leroy, Pety, Dobbs, García-Burillo, Dumas, Thompson, Schuster, \& Kramer}]{colombo_pdbi_2014}
Colombo, D., Hughes, A., Schinnerer, E., {et~al.} 2014, ApJ, 784, 3, aDS Bibcode: 2014ApJ...784....3C

\bibitem[{{Dobbs}(2008)}]{dobbs2008}
{Dobbs}, C.~L. 2008, \mnras, 391, 844

\bibitem[{{Dobbs} {et~al.}(2014){Dobbs}, {Krumholz}, {Ballesteros-Paredes}, {Bolatto}, {Fukui}, {Heyer}, {Mac Low}, {Ostriker}, \& {V{\'a}zquez-Semadeni}}]{dobbs_formation_2014}
{Dobbs}, C.~L., {Krumholz}, M.~R., {Ballesteros-Paredes}, J., {et~al.} 2014, in Protostars and Planets VI, ed. H.~{Beuther}, R.~S. {Klessen}, C.~P. {Dullemond}, \& T.~{Henning}, 3--26

\bibitem[{{Dobbs} \& {Pringle}(2013)}]{dobbs2013}
{Dobbs}, C.~L. \& {Pringle}, J.~E. 2013, \mnras, 432, 653

\bibitem[{{Elmegreen} \& {Elmegreen}(1983)}]{elmegreen_1983}
{Elmegreen}, B.~G. \& {Elmegreen}, D.~M. 1983, \mnras, 203, 31

\bibitem[{Espada {et~al.}(2018)Espada, Martin, Verley, Pettitt, Matsushita, Argudo-Fernández, Randriamanakoto, Hsieh, Saito, Miura, Kawana, Sabater, Verdes-Montenegro, Ho, Kawabe, \& Iono}]{espada_molecular_2018}
Espada, D., Martin, S., Verley, S., {et~al.} 2018, ApJ, 866, 77, \_eprint: 1809.04788

\bibitem[{Grasha {et~al.}(2019)Grasha, Calzetti, Adamo, Kennicutt, Elmegreen, Messa, Dale, Fedorenko, Mahadevan, Grebel, Fumagalli, Kim, Dobbs, Gouliermis, Ashworth, Gallagher, Smith, Tosi, Whitmore, Schinnerer, Colombo, Hughes, Leroy, \& Meidt}]{grasha_spatial_2019}
Grasha, K., Calzetti, D., Adamo, A., {et~al.} 2019, MNRAS, 483, 4707, aDS Bibcode: 2019MNRAS.483.4707G

\bibitem[{Grasha {et~al.}(2017)Grasha, Calzetti, Adamo, Kim, Elmegreen, Gouliermis, Dale, Fumagalli, Grebel, Johnson, Kahre, Kennicutt, Messa, Pellerin, Ryon, Smith, Shabani, Thilker, \& Ubeda}]{grasha_hierarchical_2017}
Grasha, K., Calzetti, D., Adamo, A., {et~al.} 2017, ApJ, 840, 113, publisher: IOP ADS Bibcode: 2017ApJ...840..113G

\bibitem[{Grasha {et~al.}(2018)Grasha, Calzetti, Bittle, Johnson, Donovan~Meyer, Kennicutt, Elmegreen, Adamo, Krumholz, Fumagalli, Grebel, Gouliermis, Cook, Gallagher, Aloisi, Dale, Linden, Sacchi, Thilker, Walterbos, Messa, Wofford, \& Smith}]{grasha_connecting_2018}
Grasha, K., Calzetti, D., Bittle, L., {et~al.} 2018, MNRAS, 481, 1016, aDS Bibcode: 2018MNRAS.481.1016G

\bibitem[{{Grishunin} {et~al.}(2024){Grishunin}, {Weiss}, {Colombo}, {Chevance}, {Chen}, {G{\"u}sten}, {Rubio}, {Hunt}, {Wyrowski}, {Harrington}, {Menten}, \& {Herrera-Camus}}]{grishunin_lmc_2024}
{Grishunin}, K., {Weiss}, A., {Colombo}, D., {et~al.} 2024, \aap, 682, A137

\bibitem[{Gusev(2014)}]{gusev_hierarchy_2014}
Gusev, A.~S. 2014, MNRAS, 442, 3711, publisher: OUP ADS Bibcode: 2014MNRAS.442.3711G

\bibitem[{Guszejnov {et~al.}(2018)Guszejnov, Hopkins, \& Grudić}]{guszejnov_universal_2018}
Guszejnov, D., Hopkins, P.~F., \& Grudić, M.~Y. 2018, MNRAS, 477, 5139, aDS Bibcode: 2018MNRAS.477.5139G

\bibitem[{Heyer \& Dame(2015)}]{heyer_molecular_2015}
Heyer, M. \& Dame, T.~M. 2015, ARA\&A, 53, 583

\bibitem[{Hirota {et~al.}(2024)Hirota, Koda, Egusa, Sawada, Sakamoto, Heyer, Lee, Maeda, Boissier, Calzetti, Elmegreen, Harada, Ho, Kobayashi, Kuno, Madore, Martín, Donovan~Meyer, Muraoka, \& Watanabe}]{hirota_whole-disk_2024}
Hirota, A., Koda, J., Egusa, F., {et~al.} 2024, ApJ, 976, 198, publisher: IOP ADS Bibcode: 2024ApJ...976..198H

\bibitem[{{Hopkins}(2012)}]{hopkins2012}
{Hopkins}, P.~F. 2012, \mnras, 423, 2016

\bibitem[{Hughes {et~al.}(2013)Hughes, Meidt, Colombo, Schinnerer, Pety, Leroy, Dobbs, García-Burillo, Thompson, Dumas, Schuster, \& Kramer}]{hughes_comparative_2013}
Hughes, A., Meidt, S.~E., Colombo, D., {et~al.} 2013, ApJ, 779, 46, aDS Bibcode: 2013ApJ...779...46H

\bibitem[{{Inutsuka} {et~al.}(2015){Inutsuka}, {Inoue}, {Iwasaki}, \& {Hosokawa}}]{inutsuka2015}
{Inutsuka}, S.-i., {Inoue}, T., {Iwasaki}, K., \& {Hosokawa}, T. 2015, \aap, 580, A49

\bibitem[{{Kainulainen} {et~al.}(2017){Kainulainen}, {Stutz}, {Stanke}, {Abreu-Vicente}, {Beuther}, {Henning}, {Johnston}, \& {Megeath}}]{isf_spacing}
{Kainulainen}, J., {Stutz}, A.~M., {Stanke}, T., {et~al.} 2017, \aap, 600, A141

\bibitem[{Kennicutt \& Evans(2012)}]{kennicutt_star_2012}
Kennicutt, R.~C. \& Evans, N.~J. 2012, ARA\&A, 50, 531, \_eprint: 1204.3552

\bibitem[{Kerscher {et~al.}(2000)Kerscher, Szapudi, \& Szalay}]{kerscher_comparison_2000}
Kerscher, M., Szapudi, I., \& Szalay, A.~S. 2000, ApJ, 535, L13, publisher: IOP ADS Bibcode: 2000ApJ...535L..13K

\bibitem[{{Keto}(2024)}]{radial_profiles_mw}
{Keto}, E. 2024, Astronomische Nachrichten, 345, e20240044

\bibitem[{Kim {et~al.}(2022)Kim, Chevance, Kruijssen, Leroy, Schruba, Barnes, Bigiel, Blanc, Cao, Congiu, Dale, Faesi, Glover, Grasha, Groves, Hughes, Klessen, Kreckel, McElroy, Pan, Pety, Querejeta, Razza, Rosolowsky, Saito, Schinnerer, Sun, Tomičić, Usero, \& Williams}]{kim_environmental_2022}
Kim, J., Chevance, M., Kruijssen, J. M.~D., {et~al.} 2022, MNRAS, 516, 3006, publisher: OUP ADS Bibcode: 2022MNRAS.516.3006K

\bibitem[{{Kobayashi} {et~al.}(2017){Kobayashi}, {Inutsuka}, {Kobayashi}, \& {Hasegawa}}]{kobayashi2017}
{Kobayashi}, M. I.~N., {Inutsuka}, S.-i., {Kobayashi}, H., \& {Hasegawa}, K. 2017, \apj, 836, 175

\bibitem[{Landy \& Szalay(1993)}]{landy_bias_1993}
Landy, S.~D. \& Szalay, A.~S. 1993, ApJ, 412, 64, publisher: IOP ADS Bibcode: 1993ApJ...412...64L

\bibitem[{Lang {et~al.}(2020)Lang, Meidt, Rosolowsky, Nofech, Schinnerer, Leroy, Emsellem, Pessa, Glover, Groves, Hughes, Kruijssen, Querejeta, Schruba, Bigiel, Blanc, Chevance, Colombo, Faesi, Henshaw, Herrera, Liu, Pety, Puschnig, Saito, Sun, \& Usero}]{lang_phangs_2020}
Lang, P., Meidt, S.~E., Rosolowsky, E., {et~al.} 2020, ApJ, 897, 122, publisher: IOP ADS Bibcode: 2020ApJ...897..122L

\bibitem[{Leroy {et~al.}(2021{\natexlab{a}})Leroy, Hughes, Liu, Pety, Rosolowsky, Saito, Schinnerer, Schruba, Usero, Faesi, Herrera, Chevance, Hygate, Kepley, Koch, Querejeta, Sliwa, Will, Wilson, Anand, Barnes, Belfiore, Bešlić, Bigiel, Blanc, Bolatto, Boquien, Cao, Chandar, Chastenet, Chiang, Congiu, Dale, Deger, den Brok, Eibensteiner, Emsellem, García-Rodríguez, Glover, Grasha, Groves, Henshaw, Jiménez~Donaire, Kim, Klessen, Kreckel, Kruijssen, Larson, Lee, Mayker, McElroy, Meidt, Mok, Pan, Puschnig, Razza, Sánchez-Bl'azquez, Sandstrom, Santoro, Sardone, Scheuermann, Sun, Thilker, Turner, Ubeda, Utomo, Watkins, \& Williams}]{leroy_phangs-alma_2021-1}
Leroy, A.~K., Hughes, A., Liu, D., {et~al.} 2021{\natexlab{a}}, ApJS, 255, 19, publisher: IOP ADS Bibcode: 2021ApJS..255...19L

\bibitem[{Leroy {et~al.}(2016)Leroy, Hughes, Schruba, Rosolowsky, Blanc, Bolatto, Colombo, Escala, Kramer, Kruijssen, Meidt, Pety, Querejeta, Sandstrom, Schinnerer, Sliwa, \& Usero}]{leroy_portrait_2016}
Leroy, A.~K., Hughes, A., Schruba, A., {et~al.} 2016, ApJ, 831, 16, aDS Bibcode: 2016ApJ...831...16L

\bibitem[{Leroy {et~al.}(2021{\natexlab{b}})Leroy, Schinnerer, Hughes, Rosolowsky, Pety, Schruba, Usero, Blanc, Chevance, Emsellem, Faesi, Herrera, Liu, Meidt, Querejeta, Saito, Sandstrom, Sun, Williams, Anand, Barnes, Behrens, Belfiore, Benincasa, Bešlić, Bigiel, Bolatto, Brok, Cao, Chandar, Chastenet, Chiang, Congiu, Dale, Deger, Eibensteiner, Egorov, García-Rodríguez, Glover, Grasha, Henshaw, Ho, Kepley, Kim, Klessen, Kreckel, Koch, Kruijssen, Larson, Lee, Lopez, Machado, Mayker, McElroy, Murphy, Ostriker, Pan, Pessa, Puschnig, Razza, Sánchez-Blázquez, Santoro, Sardone, Scheuermann, Sliwa, Sormani, Stuber, Thilker, Turner, Utomo, Watkins, \& Whitmore}]{leroy_phangs-alma_2021}
Leroy, A.~K., Schinnerer, E., Hughes, A., {et~al.} 2021{\natexlab{b}}, ApJS, 257, 43, arXiv: 2104.07739

\bibitem[{Leroy {et~al.}(2008)Leroy, Walter, Brinks, Bigiel, De~Blok, Madore, \& Thornley}]{leroy_star_2008}
Leroy, A.~K., Walter, F., Brinks, E., {et~al.} 2008, AJ, 136, 2782, \_eprint: 0810.2556

\bibitem[{{Meidt}(2022)}]{meidt_molecular_2022}
{Meidt}, S.~E. 2022, \apj, 937, 88

\bibitem[{{Meidt} \& {van der Wel}(2024)}]{meidt_bottom_2024}
{Meidt}, S.~E. \& {van der Wel}, A. 2024, \apj, 966, 62

\bibitem[{Menon {et~al.}(2021)Menon, Grasha, Elmegreen, Federrath, Krumholz, Calzetti, Sánchez, Linden, Adamo, Messa, Cook, Dale, Grebel, Fumagalli, Sabbi, Johnson, Smith, \& Kennicutt}]{menon_dependence_2021}
Menon, S.~H., Grasha, K., Elmegreen, B.~G., {et~al.} 2021, MNRAS, 507, 5542, aDS Bibcode: 2021MNRAS.507.5542M

\bibitem[{Peebles(1980)}]{peebles_large-scale_1980}
Peebles, P. J.~E. 1980, The large-scale structure of the universe, publication Title: Large-Scale Structure of the Universe by Phillip James Edwin Peebles. Princeton University Press ADS Bibcode: 1980lssu.book.....P

\bibitem[{Peltonen {et~al.}(2023)Peltonen, Rosolowsky, Johnson, Seth, Dalcanton, Bell, Braine, Koch, Lazzarini, Leroy, Skillman, Smercina, Wainer, \& Williams}]{peltonen_clusters_2023}
Peltonen, J., Rosolowsky, E., Johnson, L.~C., {et~al.} 2023, MNRAS, 522, 6137, publisher: OUP ADS Bibcode: 2023MNRAS.522.6137P

\bibitem[{Pineda {et~al.}(2009)Pineda, Rosolowsky, \& Goodman}]{pineda_perils_2009}
Pineda, J.~E., Rosolowsky, E.~W., \& Goodman, A.~A. 2009, ApJ, 699, L134, publisher: IOP ADS Bibcode: 2009ApJ...699L.134P

\bibitem[{Querejeta {et~al.}(2024)Querejeta, Leroy, Meidt, Schinnerer, Belfiore, Emsellem, Klessen, Sun, Sormani, Bešlić, Cao, Chevance, Colombo, Dale, García-Burillo, Glover, Grasha, Groves, Koch, Neumann, Pan, Pessa, Pety, Pinna, Ramambason, Razza, Romanelli, Rosolowsky, Ruiz-García, Sánchez-Blázquez, Smith, Stuber, Ubeda, Usero, \& Williams}]{querejeta_spiral_2024}
Querejeta, M., Leroy, A.~K., Meidt, S.~E., {et~al.} 2024, A\&A, 687, A293, publisher: EDP ADS Bibcode: 2024A\&A...687A.293Q

\bibitem[{Querejeta {et~al.}(2021)Querejeta, Schinnerer, Meidt, Sun, Leroy, Emsellem, Klessen, Muñoz-Mateos, Salo, Laurikainen, Bešlić, Blanc, Chevance, Dale, Eibensteiner, Faesi, García-Rodríguez, Glover, Grasha, Henshaw, Herrera, Hughes, Kreckel, Kruijssen, Liu, Murphy, Pan, Pety, Razza, Rosolowsky, Saito, Schruba, Usero, Watkins, \& Williams}]{querejeta_stellar_2021}
Querejeta, M., Schinnerer, E., Meidt, S., {et~al.} 2021, Astronomy \&amp; Astrophysics, Volume 656, id.A133, {\textless}NUMPAGES{\textgreater}27{\textless}/NUMPAGES{\textgreater} pp., 656, A133

\bibitem[{Rosolowsky {et~al.}(2021)Rosolowsky, Hughes, Leroy, Sun, Querejeta, Schruba, Usero, Herrera, Liu, Pety, Saito, Bešlić, Bigiel, Blanc, Chevance, Dale, Deger, Faesi, Glover, Henshaw, Klessen, Kruijssen, Larson, Lee, Meidt, Mok, Schinnerer, Thilker, \& Williams}]{rosolowsky_giant_2021}
Rosolowsky, E., Hughes, A., Leroy, A.~K., {et~al.} 2021, MNRAS, 502, 1218, aDS Bibcode: 2021MNRAS.502.1218R

\bibitem[{Rosolowsky \& Leroy(2006)}]{rosolowsky_bias-free_2006}
Rosolowsky, E. \& Leroy, A. 2006, PASP, 118, 590, aDS Bibcode: 2006PASP..118..590R

\bibitem[{Saintonge \& Catinella(2022)}]{saintonge_cold_2022}
Saintonge, A. \& Catinella, B. 2022, ARA\&A, 60, 319, aDS Bibcode: 2022ARA\&A..60..319S

\bibitem[{Schinnerer \& Leroy(2024)}]{schinnerer_molecular_2024}
Schinnerer, E. \& Leroy, A.~K. 2024, ARA\&A, 62, 369, aDS Bibcode: 2024ARA\&A..62..369S

\bibitem[{{Shu} {et~al.}(1987){Shu}, {Adams}, \& {Lizano}}]{shu_87}
{Shu}, F.~H., {Adams}, F.~C., \& {Lizano}, S. 1987, \araa, 25, 23

\bibitem[{Stark \& Lee(2006)}]{stark_giant_2006}
Stark, A.~A. \& Lee, Y. 2006, ApJ, 641, L113, publisher: IOP ADS Bibcode: 2006ApJ...641L.113S

\bibitem[{Sun {et~al.}(2022)Sun, Leroy, Rosolowsky, Hughes, Schinnerer, Schruba, Koch, Blanc, Chiang, Groves, Liu, Meidt, Pan, Pety, Querejeta, Saito, Sandstrom, Sardone, Usero, Utomo, Williams, Barnes, Benincasa, Bigiel, Bolatto, Boquien, Chevance, Dale, Deger, Emsellem, Glover, Grasha, Henshaw, Klessen, Kreckel, Kruijssen, Ostriker, \& Thilker}]{sun_molecular_2022}
Sun, J., Leroy, A.~K., Rosolowsky, E., {et~al.} 2022, AJ, 164, 43, aDS Bibcode: 2022AJ....164...43S

\bibitem[{Sun {et~al.}(2020)Sun, Leroy, Schinnerer, Hughes, Rosolowsky, Querejeta, Schruba, Liu, Saito, Herrera, Faesi, Usero, Pety, Kruijssen, Ostriker, Bigiel, Blanc, Bolatto, Boquien, Chevance, Dale, Deger, Emsellem, Glover, Grasha, Groves, Henshaw, Jimenez-Donaire, Kim, Klessen, Kreckel, Lee, Meidt, Sandstrom, Sardone, Utomo, \& Williams}]{sun_molecular_2020}
Sun, J., Leroy, A.~K., Schinnerer, E., {et~al.} 2020, ApJL, 901, L8

\bibitem[{Turner {et~al.}(2022)Turner, Dale, Lilly, Boquien, Deger, Lee, Whitmore, Anand, Benincasa, Bigiel, Blanc, Chevance, Emsellem, Faesi, Glover, Grasha, Hughes, Klessen, Kreckel, Kruijssen, Leroy, Pan, Rosolowsky, Schruba, \& Williams}]{turner_phangs_2022}
Turner, J.~A., Dale, D.~A., Lilly, J., {et~al.} 2022, MNRAS, 516, 4612, aDS Bibcode: 2022MNRAS.516.4612T

\bibitem[{{V{\'a}zquez-Semadeni} {et~al.}(2019){V{\'a}zquez-Semadeni}, {Palau}, {Ballesteros-Paredes}, {G{\'o}mez}, \& {Zamora-Avil{\'e}s}}]{vazquez-semadeni_global_2019}
{V{\'a}zquez-Semadeni}, E., {Palau}, A., {Ballesteros-Paredes}, J., {G{\'o}mez}, G.~C., \& {Zamora-Avil{\'e}s}, M. 2019, \mnras, 490, 3061

\bibitem[{{White} {et~al.}(2007){White}, {Helfand}, {Becker}, {Glikman}, \& {de Vries}}]{qso_stacking}
{White}, R.~L., {Helfand}, D.~J., {Becker}, R.~H., {Glikman}, E., \& {de Vries}, W. 2007, \apj, 654, 99

\bibitem[{Williams {et~al.}(1994)Williams, de~Geus, \& Blitz}]{williams_determining_1994}
Williams, J.~P., de~Geus, E.~J., \& Blitz, L. 1994, ApJ, 428, 693, publisher: IOP ADS Bibcode: 1994ApJ...428..693W

\bibitem[{Young \& Scoville(1991)}]{young_molecular_1991}
Young, J.~S. \& Scoville, N.~Z. 1991, ARA\&A, 29, 581, aDS Bibcode: 1991ARA\&A..29..581Y

\bibitem[{Zhang {et~al.}(2001)Zhang, Fall, \& Whitmore}]{zhang_antennae_2001}
Zhang, Q., Fall, S.~M., \& Whitmore, B.~C. 2001, Observatory, 10

\end{thebibliography}







   
  



\begin{appendix}

\onecolumn
\section{Measured GMC clustering and number density}

We record our median and mean values of \Icokpc, cloud spacing $d_{\mathrm{sep, 1st}}$ and number of neighbours at different \rgal in Table~\ref{tab:spacing_nn}. This table can be used to reproduce the trends seen in Fig. \ref{fig:spacing_and_nn} and \ref{fig:Ico_rgal}. 

\begin{table*}[h!]
\caption{\label{tab:spacing_nn} \Icokpc, spacing and number of neighbours at different \rgal}
\centering
\begin{tabularx}{0.7\textwidth}{cccccccc}
\hline\hline 
\rgal [$R_{\mathrm{eff}}$] & \multicolumn{3}{c}{$\log_{10}$\Icokpc [K km s$^{-1}$]} & \multicolumn{3}{c}{$\log_{10} d_{\mathrm{sep, 1st}}$ [pc]} & Number of Neighbours \\
& 50th & 16th & 84th & 50th & 16th & 84th & mean \\
(1) & (2) & (3) & (4) & (5) & (6) & (7) & (8) \\
\hline
0.0 & 1.7 & 0.8 & 2.1 & 2.0 & 1.6 & 2.4 & 5.7 \\
0.1 & 1.6 & 0.8 & 2.0 & 2.3 & 1.9 & 2.6 & 4.3 \\
0.2 & 1.3 & 0.8 & 1.7 & 2.4 & 2.0 & 2.7 & 3.1 \\
0.3 & 1.2 & 0.6 & 1.6 & 2.5 & 2.1 & 2.8 & 2.2 \\
0.4 & 1.0 & 0.4 & 1.5 & 2.6 & 2.2 & 2.8 & 1.6 \\
0.5 & 0.9 & 0.3 & 1.4 & 2.6 & 2.3 & 2.8 & 1.3 \\
0.6 & 0.9 & 0.4 & 1.3 & 2.6 & 2.3 & 2.8 & 1.3 \\
0.7 & 0.8 & 0.2 & 1.3 & 2.6 & 2.3 & 2.9 & 1.2 \\
0.8 & 0.7 & 0.2 & 1.3 & 2.6 & 2.4 & 2.9 & 1.1 \\
0.9 & 0.6 & 0.1 & 1.2 & 2.6 & 2.4 & 2.9 & 1.1 \\
1.0 & 0.6 & 0.1 & 1.2 & 2.6 & 2.4 & 2.9 & 1.0 \\
1.1 & 0.6 & 0.1 & 1.1 & 2.6 & 2.4 & 2.9 & 1.0 \\
1.2 & 0.6 & 0.2 & 1.0 & 2.6 & 2.4 & 2.9 & 1.0 \\
1.3 & 0.5 & 0.1 & 0.9 & 2.6 & 2.4 & 2.9 & 1.0 \\
1.4 & 0.5 & 0.1 & 0.9 & 2.6 & 2.4 & 2.9 & 0.9 \\
1.5 & 0.5 & 0.1 & 0.9 & 2.7 & 2.4 & 2.9 & 0.9 \\
1.6 & 0.4 & -0.1 & 0.9 & 2.7 & 2.4 & 3.0 & 0.9 \\
1.7 & 0.4 & -0.1 & 0.8 & 2.7 & 2.4 & 2.9 & 0.9 \\
1.8 & 0.4 & -0.1 & 0.8 & 2.7 & 2.4 & 3.0 & 0.7 \\
1.9 & 0.4 & -0.1 & 0.8 & 2.7 & 2.5 & 3.0 & 0.8 \\
2.0 & 0.5 & 0.1 & 0.7 & 2.7 & 2.4 & 3.0 & 0.7 \\
2.1 & 0.3 & -0.0 & 0.6 & 2.7 & 2.4 & 3.1 & 0.6 \\
2.2 & 0.3 & -0.1 & 0.6 & 2.7 & 2.5 & 3.0 & 0.6 \\
2.3 & 0.3 & 0.0 & 0.7 & 2.7 & 2.4 & 3.0 & 0.8 \\
2.4 & 0.2 & -0.0 & 0.7 & 2.8 & 2.5 & 3.2 & 0.6 \\
\hline
\end{tabularx}
\tablefoot{
(1) Galactocentric radius, normalised by the stellar mass effective radius (2) Median of kpc-scale integrated CO intensity (\Icokpc) (3) 16th percentile value of \Icokpc (4) 84th percentile value of \Icokpc  (5) Median of the nearest neighbour distance ($d_{\mathrm{sep, 1st}}$) (6) 16th percentile value of $d_{\mathrm{sep, 1st}}$ (7) 84th percentile value of $d_{\mathrm{sep, 1st}}$ (8) Average number of neighbours inside a XXpc aperture 
}
\end{table*}

\section{Ancillary data from two-point correlation analyses}
\label{app:unc}

\begin{table*}[h!]
\renewcommand{\arraystretch}{1.1}
\caption{\label{tab:2pcf_indvd} 2PCF of individual galaxies}
\centering
\begin{tabularx}{\textwidth}{ccrrrrrrrrrrrr}
\hline \hline
Galaxy & $\log_{10}d_{\mathrm{sep}}$ [pc] & \multicolumn{3}{c}{$\log_{10} (1+\omega)_{\mathrm{flat}}$} & \multicolumn{3}{c}{$\log_{10} (1+\omega)_{\mathrm{exp}}$} & \multicolumn{3}{c}{$\log_{10} (1+\omega)_{\mathrm{rad}}$} & \multicolumn{3}{c}{$\log_{10} (1+\omega)_{I_\mathrm{co}}$} \\
 &  & med & lower & upper & med & lower & upper & med & lower & upper & med & lower & upper \\
 (1) & (2) & (3) & (4) & (5) & (6) & (7) & (8) & (9) & (10) & (11) & (12) & (13) & (14) \\
\hline
IC1954 & 2.15 & 0.46 & -0.07 & 0.7 & 0.02 & -0.51 & 0.25 & -0.09 & -0.62 & 0.14 & -0.11 & -0.64 & 0.13 \\
IC1954 & 2.25 & 0.3 & -0.23 & 0.53 & -0.2 & -0.73 & 0.03 & -0.03 & -0.56 & 0.21 & -0.11 & -0.64 & 0.12 \\
IC1954 & 2.35 & 0.6 & 0.38 & 0.75 & 0.15 & -0.08 & 0.3 & 0.13 & -0.1 & 0.27 & 0.08 & -0.14 & 0.23 \\
$\vdots$ & $\vdots$ & $\vdots$ & $\vdots$ & $\vdots$ & $\vdots$ & $\vdots$ & $\vdots$ & $\vdots$ & $\vdots$ & $\vdots$ & $\vdots$ & $\vdots$& $\vdots$  \\
NGC7793* & 3.25 & 0.33 & 0.27 & 0.39 & 0.06 & -0.0 & 0.11 & 0.05 & -0.01 & 0.1 & 0.05 & -0.02 & 0.1 \\
NGC7793* & 3.35 & 0.09 & 0.03 & 0.15 & 0.0 & -0.06 & 0.05 & -0.02 & -0.08 & 0.03 & -0.03 & -0.09 & 0.02 \\
NGC7793* & 3.45 & -0.22 & -0.29 & -0.17 & -0.02 & -0.08 & 0.03 & -0.03 & -0.09 & 0.02 & -0.04 & -0.1 & 0.01 \\
\hline
\end{tabularx}
\tablefoot{This is a subset of the full table. The full table is online in a machine-readable format. \newline
(1) Galaxy name (2) The separation distance \newline
(3) (4) (5)  2PCF amplitudes, lower and upper limit with ``Flat" control. \newline 
(6) (7) (8) 2PCF amplitudes, lower and upper limit with ``Exp" control. \newline 
(9) (10) (11) 2PCF amplitudes, lower and upper limit with ``Rad" control. \newline
(12) (13) (14) 2PCF amplitudes, lower and upper limit with ``Ico" control.}
\end{table*}

\begin{table*}[h!]
\renewcommand{\arraystretch}{1.3}
\caption{\label{tab:tpcf_meds} Median, 16th and 84th values of two-point correlation functions in our sample}
\centering
\begin{tabularx}{\textwidth}{crrrrrrrrrrrr}
\hline \hline
$\log_{10}d_{\mathrm{sep}}$ [pc] & \multicolumn{3}{c}{$\log_{10} (1+\omega)_{\mathrm{flat}}$} & \multicolumn{3}{c}{$\log_{10} (1+\omega)_{\mathrm{exp}}$} & \multicolumn{3}{c}{$\log_{10} (1+\omega)_{\mathrm{rad}}$} & \multicolumn{3}{c}{$\log_{10} (1+\omega)_{I_\mathrm{co}}$}\\
 & median & 16th & 84th & median & 16th & 84th & median & 16th & 84th & median & 16th & 84th \\
 (1) & (2) & (3) & (4) & (5) & (6) & (7) & (8) & (9) & (10) & (11) & (12) & (13) \\
 \hline
2.15 & 0.23 & -0.12 & 0.57 & 0.01 & -0.15 & 0.25 & 0.07 & -0.11 & 0.22 & 0.02 & -0.18 & 0.17 \\
2.25 & 0.27 & -0.08 & 0.59 & 0.06 & -0.06 & 0.17 & 0.12 & -0.04 & 0.2 & 0.04 & -0.15 & 0.15 \\
2.35 & 0.28 & 0.05 & 0.62 & 0.07 & -0.01 & 0.17 & 0.12 & -0.04 & 0.21 & 0.03 & -0.13 & 0.17 \\
2.45 & 0.33 & 0.17 & 0.53 & 0.12 & 0.0 & 0.22 & 0.14 & 0.01 & 0.22 & 0.09 & -0.06 & 0.17 \\
2.55 & 0.33 & 0.14 & 0.56 & 0.11 & 0.01 & 0.22 & 0.13 & -0.03 & 0.22 & 0.05 & -0.06 & 0.18 \\
2.65 & 0.33 & 0.15 & 0.54 & 0.11 & -0.01 & 0.25 & 0.14 & 0.0 & 0.21 & 0.08 & -0.04 & 0.18 \\
2.75 & 0.34 & 0.15 & 0.56 & 0.1 & 0.04 & 0.19 & 0.13 & 0.03 & 0.22 & 0.09 & 0.01 & 0.18 \\
2.85 & 0.3 & 0.14 & 0.54 & 0.09 & 0.04 & 0.17 & 0.11 & 0.04 & 0.2 & 0.07 & 0.01 & 0.17 \\
2.95 & 0.27 & 0.14 & 0.52 & 0.07 & 0.02 & 0.16 & 0.1 & 0.02 & 0.2 & 0.07 & -0.0 & 0.18 \\
3.05 & 0.26 & 0.08 & 0.51 & 0.06 & 0.02 & 0.13 & 0.08 & 0.02 & 0.18 & 0.08 & 0.01 & 0.18 \\
3.15 & 0.22 & 0.07 & 0.47 & 0.04 & 0.02 & 0.12 & 0.04 & 0.02 & 0.15 & 0.05 & 0.01 & 0.15 \\
3.25 & 0.22 & 0.03 & 0.42 & 0.04 & 0.0 & 0.09 & 0.03 & -0.01 & 0.09 & 0.03 & -0.0 & 0.1 \\
3.35 & 0.17 & 0.05 & 0.42 & 0.02 & -0.0 & 0.08 & 0.01 & -0.03 & 0.05 & 0.01 & -0.01 & 0.05 \\
3.45 & 0.13 & 0.02 & 0.39 & 0.01 & -0.01 & 0.06 & 0.0 & -0.07 & 0.03 & 0.0 & -0.04 & 0.03 \\
\hline
\end{tabularx}\
\tablefoot{(1) The separation distance. (2) The median of 2PCF amplitudes of all galaxies with $N_{\mathrm{GMC}}>50$ with its uncertainty (3) The 16th values of 2PCF of all the selected galaxies (4) The 84th values of 2PCF of all the selected galaxies. \newline
(5) (6) (7) Same as (3), (4) and (5) but for 2PCF with ``Exp" control \newline
(8) (9) (10) Same as (3), (4) and (5) but for 2PCF with ``Rad" control \newline
(11) (12) (13) Same as (3), (4) and (5) but for 2PCF with ``$I_{\mathrm{CO}}$" control 
}
\end{table*}

{\bfseries
\begin{table*}[h!]
\renewcommand{\arraystretch}{1.1}
\caption{\label{tab:clustering_summary} Summary of GMC clustering statistics for individual galaxies}
\centering
\begin{tabularx}{\textwidth}{rccccrrrrrrrr}
\hline \hline
Galaxy & N$_{\mathrm{GMC}}$ & $l_{\star}$ [kpc]  & $\log_{10} d_{\mathrm{sep, 1st}}$ [pc] & $N_{\mathrm{neighb, 500pc}}$ & \multicolumn{4}{c}{$\log_{10} d_{\omega, \mathrm{max}}$ [pc]} & \multicolumn{4}{c}{$\log_{10}(1+\omega)_{500 \mathrm{pc}}$} \\
 & & & Median & Mean & flat & exp & rad  & Ico  & flat & exp & rad & Ico  \\
 (1) & (2) & (3) & (4) & (5) & (6) & (7) & (8) & (9) & (10) & (11) & (12) & (13) \\
\hline
IC1954 & 94 & 1.5 & 2.68 & 0.9 & 2.35 & 2.35 & 2.35 & 2.35 & 0.44 & -0.02 & -0.01 & -0.03 \\
IC5273 & 79 & 1.3 & 2.74 & 0.7 & 2.75 & 2.75 & 2.75 & 2.75 & 0.5 & 0.15 & 0.19 & 0.1 \\
NGC0253 & 481 & 2.8 & 2.65 & 1.4 & 2.25 & 2.25 & 2.85 & 2.25 & 0.42 & 0.08 & 0.22 & 0.2 \\
NGC0300* & 9 & 1.3 & 3.05 & 0.1 & 2.55 & 2.55 & 2.55 & 2.55 & 0.48 & -0.15 & -0.47 & 0.29 \\
NGC0628 & 271 & 2.9 & 2.67 & 0.8 & 2.45 & 2.45 & 2.45 & 2.45 & 0.07 & 0.06 & 0.04 & 0.0 \\
NGC0685 & 151 & 3.1 & 2.71 & 0.7 & 2.55 & 2.55 & 2.55 & 2.55 & 0.22 & -0.04 & -0.01 & -0.14 \\
NGC1097 & 344 & 4.3 & 2.62 & 1.2 & 2.55 & 2.55 & 2.55 & 2.55 & 0.43 & 0.24 & 0.22 & 0.2 \\
NGC1511 & 110 & 1.7 & 2.71 & 0.7 & 2.25 & 2.25 & 2.25 & 2.65 & 0.71 & 0.23 & 0.18 & 0.09 \\
NGC1546 & 117 & 2.1 & 2.63 & 1.0 & 2.45 & 2.45 & 2.45 & 2.45 & 0.69 & 0.02 & -0.01 & -0.01 \\
NGC1637 & 168 & 1.8 & 2.59 & 1.2 & 2.55 & 2.55 & 2.55 & 2.65 & 0.22 & 0.07 & 0.16 & 0.11 \\
NGC1792 & 539 & 2.4 & 2.64 & 1.1 & 2.25 & 2.25 & 2.25 & 2.25 & 0.43 & 0.12 & 0.11 & 0.1 \\
NGC2903 & 419 & 3.5 & 2.68 & 0.8 & 2.85 & 2.85 & 2.45 & 2.85 & 0.23 & 0.04 & 0.17 & 0.14 \\
NGC2997 & 917 & 4.0 & 2.59 & 1.3 & 2.55 & 2.55 & 2.55 & 2.55 & 0.15 & 0.12 & 0.21 & 0.15 \\
NGC3137 & 116 & 3.0 & 2.73 & 0.6 & 2.75 & 2.75 & 2.25 & 2.25 & 0.5 & 0.1 & -0.01 & -0.04 \\
NGC3351 & 170 & 2.1 & 2.65 & 0.9 & 2.65 & 2.65 & 2.85 & 2.95 & 0.17 & 0.22 & 0.25 & 0.23 \\
NGC3489* & 14 & 1.4 & 2.48 & 1.4 & 2.35 & 2.35 & 2.25 & 2.35 & 0.54 & 0.21 & -0.05 & -0.16 \\
NGC3511 & 208 & 2.4 & 2.75 & 0.6 & 2.95 & 2.95 & 2.25 & 2.95 & 0.52 & -0.01 & 0.07 & 0.03 \\
NGC3521 & 688 & 4.9 & 2.69 & 0.8 & 2.75 & 2.75 & 2.75 & 3.05 & 0.24 & 0.03 & 0.05 & 0.04 \\
NGC3621 & 208 & 2.0 & 2.69 & 0.8 & 2.95 & 2.95 & 2.95 & 2.95 & 0.2 & -0.02 & -0.0 & -0.01 \\
NGC3627 & 497 & 3.7 & 2.57 & 1.4 & 2.25 & 2.35 & 2.85 & 2.85 & 0.2 & 0.06 & 0.13 & 0.11 \\
NGC4254 & 610 & 1.8 & 2.56 & 1.4 & 2.55 & 2.55 & 2.55 & 2.95 & 0.14 & 0.13 & 0.11 & 0.09 \\
NGC4298 & 236 & 1.6 & 2.63 & 1.0 & 2.65 & 2.65 & 2.65 & 3.05 & 0.29 & 0.09 & 0.08 & 0.03 \\
NGC4459* & 30 & 3.3 & 2.06 & 5.3 & 2.35 & 2.35 & 2.95 & 2.35 & 0.91 & 0.82 & -0.09 & 0.0 \\
NGC4476* & 11 & 1.2 & 2.48 & 1.6 & 2.45 & 2.45 & 2.45 & 2.45 & 1.17 & 0.72 & -0.14 & -0.02 \\
NGC4477* & 8 & 2.1 & 1.76 & 3.5 & 2.25 & 2.25 & 2.25 & 2.25 & 1.2 & 1.08 & 0.01 & 0.13 \\
NGC4536 & 321 & 2.7 & 2.7 & 1.1 & 2.55 & 2.45 & 2.45 & 2.45 & 0.66 & 0.14 & 0.23 & 0.2 \\
NGC4548 & 199 & 3.0 & 2.67 & 0.9 & 2.45 & 2.45 & 2.75 & 2.75 & 0.33 & 0.29 & 0.19 & 0.09 \\
NGC4569 & 251 & 4.3 & 2.67 & 1.0 & 2.35 & 2.35 & 2.35 & 2.35 & 0.54 & 0.24 & 0.18 & 0.16 \\
NGC4596* & 12 & 3.8 & 1.84 & 5.5 & 2.35 & 2.25 & 2.25 & 2.35 & 0.64 & 0.55 & -0.06 & -0.78 \\
NGC4731* & 40 & 3.0 & 2.83 & 0.5 & 2.25 & 2.25 & 2.25 & 2.25 & 0.34 & 0.22 & 0.06 & -0.76 \\
NGC4781 & 108 & 1.1 & 2.59 & 1.2 & 2.55 & 2.55 & 2.55 & 2.55 & 0.5 & 0.04 & 0.02 & -0.04 \\
NGC4826* & 26 & 1.1 & 2.47 & 1.7 & 2.25 & 2.25 & 2.25 & 2.25 & 0.52 & 0.14 & -0.05 & -0.09 \\
NGC4941 & 153 & 2.2 & 2.65 & 1.5 & 2.45 & 2.45 & 2.45 & 2.45 & 0.41 & 0.2 & 0.01 & 0.02 \\
NGC5068 & 69 & 1.3 & 2.71 & 0.7 & 2.35 & 2.45 & 2.45 & 2.35 & 0.09 & 0.11 & 0.08 & -0.01 \\
NGC5128 & 251 & 4.1 & 2.31 & 5.4 & 2.55 & 2.35 & 2.35 & 2.65 & 0.7 & 0.47 & 0.12 & 0.03 \\
NGC5236 & 456 & 2.4 & 2.54 & 1.5 & 2.45 & 2.45 & 2.45 & 2.45 & 0.06 & 0.07 & 0.2 & 0.16 \\
NGC5643 & 360 & 1.6 & 2.6 & 1.2 & 2.75 & 2.75 & 2.65 & 2.75 & 0.21 & 0.16 & 0.2 & 0.17 \\
NGC7456* & 27 & 2.9 & 2.94 & 0.3 & 2.35 & 2.25 & 2.25 & 2.95 & 0.53 & 0.32 & 0.24 & -0.01 \\
NGC7496 & 172 & 1.5 & 2.67 & 0.9 & 2.55 & 2.55 & 2.75 & 2.75 & 0.34 & 0.16 & 0.23 & 0.21 \\
NGC7793* & 44 & 1.1 & 2.81 & 0.6 & 2.75 & 2.55 & 2.75 & 2.75 & 0.48 & -0.04 & 0.02 & -0.01 \\
\hline
\end{tabularx}
\tablefoot{(1) Galaxy name, * indicates galaxies with less than 50~GMCs (2) Number of GMCs in the galaxy. (3) Stellar disk exponential scale (4) Median of the nearest neighbour distance (5) Mean of the number of neighbours within $r=500$~pc aperture (6) Max 2PCF amplitude scale within 150 -- 1000~pc ($d_{\omega, \mathrm{max}}$) with flat control (7) $d_{\omega, \mathrm{max}}$ with exp control (8) $d_{\omega, \mathrm{max}}$ with rad control (9) $d_{\omega, \mathrm{max}}$ with Ico control (10) Fiducial 2PCF amplitude at 300~pc ($(1+\omega)_{300 \mathrm{pc}}$) with ``Flat" control (11) $(1+\omega)_{300 \mathrm{pc}}$ with ``Exp" control (12) $(1+\omega)_{300 \mathrm{pc}}$ with ``Rad" control (13) $(1+\omega)_{300 \mathrm{pc}}$ with ``$I_{\mathrm{CO}}$'' control.
}
\end{table*}
}

We record the values of the 2PCFs of each galaxy with different null hypotheses in Table~\ref{tab:2pcf_indvd}. The uncertainty of 2PCF amplitude is calculated using error propagation equation from Poisson statistics:
\begin{equation}
    \sigma_{\mathrm{poisson}}(\theta) = (1+\omega(\theta)) / \sqrt{N_{\mathrm{dd}}(\theta)},
\end{equation}
where $N_{\mathrm{dd}}(\theta)$ is the number of data-data pairs at the given separation $\theta$. This table can be used to reproduce 2PCFs of individual galaxies displayed in Fig. \ref{fig:twopt_multipanel}.  We also calculate the median, 16th and 84th percentile values of 2PCFs of a sub-sample of galaxies with $N_{\mathrm{GMC}}>50$, which is recorded in Table \ref{tab:tpcf_meds}. The median uncertainty is calculated based on the asymptotic variance formula in normal distribution case: 
\begin{equation}
\sigma_{\mathrm{med}}(\theta) = \sqrt{\frac{\pi}{2}} \frac{\sigma_{\mathrm{16-84\%}}}{\sqrt{N_{\mathrm{gal}}(\theta)}},
\end{equation}
where $N_{\mathrm{gal}}(\theta)$ is the number of galaxies that comes into median calculation at given separation $\theta$. Table~\ref{tab:tpcf_meds} can be used to reproduce the overall trend of 2PCF of our sample displayed in Fig.~\ref{fig:twopt_multipanel} and~\ref{fig:tpcf_zoomin}. 

In Section \ref{subsec:galtogal2pcf}, we measure the fiducial amplitudes of 2PCFs of each galaxy at 300 pc. The 300~pc amplitude is obtained from linear interpolation in log-log space. We record these values along with other clustering metrics for individual galaxies in Table~\ref{tab:clustering_summary}. This table can be used to reproduce the 2PCF amplitude comparisons in Fig.~\ref{fig:tpcf_zoomin}.

\section{Stacked intensity profiles of individual galaxies}

\begin{figure*}
\gridline{
\fig{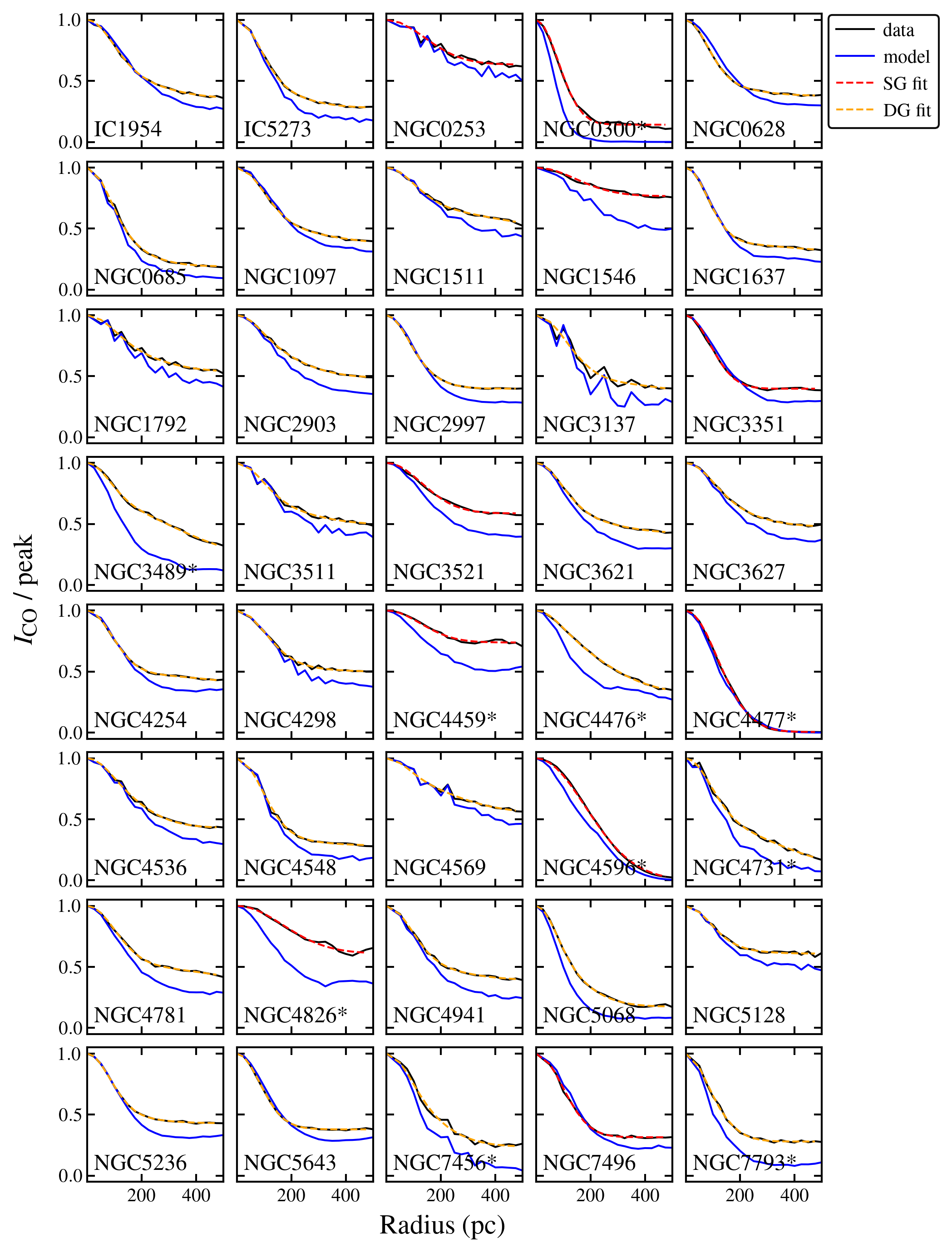}{0.9\textwidth}{}
}
\vspace{-2\baselineskip}
\caption{Stacked GMC radial profiles for individual galaxies. Black solid lines indicate the stacked intensity profiles measured from data and the blue solid lines indicate profiles measured from ``Elliptical GMC" model maps. The red and orange dashed lines indicate fit using either single or double Gaussian functions. Galaxies with a * next to their names have less than 50~GMCs.   }
\label{fig:stacking_indvd}
\end{figure*}

We record our measurements from stacked intensity profiles in Table~\ref{tab:stacking_summary}. These metrics can be used to reproduce the comparison between the model and the data displayed in Fig.~\ref{fig:stacking_radial_galaxies}. We also present the fitting parameters of the stacked profile of each galaxy in Table~\ref{tab:stacking_summary} as a reference for future studies to reproduce these profiles. For the fitting, we tried both single and double Gaussian functions (Eq.~\ref{eq:doublegauss}).
In general, a double Gaussian gives a better fit to the data. However, it sometime gives unrealistic fitting parameters. In our double Gaussian fit, we set $0.1 < A_1, A_2 < 1$, $0 < C < 1$ and $25 < \sigma_1, \sigma_2 < 500$. We further excluded the cases where $(A_1 \sigma_1^2)/(A_2\sigma_2^2) > 10$, which arises when the second Gaussian component becomes insignificant. We then compare RMS from both fits and pick the model that results in smaller RMS values. We show the stacked data and model profiles along with their fits in Fig.~\ref{fig:stacking_indvd}. 

\begin{table*}[h!]
\renewcommand{\arraystretch}{1.1}
\caption{\label{tab:stacking_summary} Summary of stacked CO intensity profiles for individual galaxies}
\centering
\begin{tabularx}{\textwidth}{cccccccccc}
\hline \hline
Galaxy & Peak  & bkg$_{\mathrm{data}}$ & bkg$_{\mathrm{model}}$ & $R_{\mathrm{bkg}}$ & $F_{\mathrm{model}}/F_{\mathrm{data}}$ & $f_{\mathrm{GMC}}$ & $d_\mathrm{10\%\   diff}$ & function & ($A$, $\sigma$, $C$) or \\
 & [K km s$^{-1}$] & & & & &  & [pc] & & ($A_1$, $\sigma_1$, $A_2$, $\sigma_2$, $C$) \\
(1) & (2) & (3) & (4) & (5) & (6) & (7) & (8) & (9) & (10)  \\
\hline
IC1954 & 6.1 & 0.39 & 0.3 & 0.79 & 0.72 & 0.73 & 275 & double & [0.39, 90, 0.25, 198, 0.36] \\
IC5273 & 3.8 & 0.27 & 0.18 & 0.66 & 0.69 & 0.48 & 150 & double & [0.49, 82, 0.23, 168, 0.28] \\
NGC0253 & 12.4 & 0.58 & 0.48 & 0.84 & 1.08 & 0.92 & 300 & single & [0.37, 139, 0.63] \\
NGC0300* & 2.1 & 0.12 & 0.0 & 0.0 & 0.14 & 0.01 & 50 & single & [0.86, 81, 0.14] \\
NGC0628 & 5.7 & 0.38 & 0.31 & 0.82 & 0.86 & 0.83 & 250 & double & [0.49, 77, 0.14, 171, 0.37] \\
NGC0685 & 2.9 & 0.18 & 0.1 & 0.56 & 0.63 & 0.4 & 150 & double & [0.51, 80, 0.3, 151, 0.19] \\
NGC1097 & 7.1 & 0.36 & 0.27 & 0.76 & 0.86 & 0.88 & 250 & double & [0.44, 95, 0.19, 233, 0.37] \\
NGC1511 & 12.3 & 0.49 & 0.41 & 0.82 & 0.77 & 0.93 & 250 & double & [0.29, 96, 0.22, 299, 0.49] \\
NGC1546 & 24.8 & 0.77 & 0.48 & 0.62 & 0.56 & 0.95 & 125 & single & [0.23, 144, 0.77] \\
NGC1637 & 4.0 & 0.33 & 0.24 & 0.72 & 0.85 & 0.8 & 200 & double & [0.6, 80, 0.11, 334, 0.29] \\
NGC1792 & 13.6 & 0.53 & 0.42 & 0.79 & 0.73 & 0.96 & 225 & double & [0.21, 92, 0.25, 190, 0.54] \\
NGC2903 & 9.4 & 0.5 & 0.35 & 0.69 & 0.71 & 0.95 & 200 & double & [0.2, 82, 0.31, 177, 0.49] \\
NGC2997 & 7.0 & 0.39 & 0.29 & 0.75 & 0.75 & 0.86 & 200 & double & [0.31, 73, 0.29, 116, 0.4] \\
NGC3137 & 3.7 & 0.39 & 0.29 & 0.74 & 0.73 & 0.74 & 150 & double & [0.51, 108, 0.23, 500, 0.26] \\
NGC3351 & 3.6 & 0.38 & 0.29 & 0.75 & 0.7 & 0.82 & 250 & single & [0.6, 89, 0.4] \\
NGC3489* & 8.7 & 0.25 & 0.11 & 0.44 & 0.44 & 0.13 & 75 & double & [0.3, 77, 0.5, 291, 0.2] \\
NGC3511 & 7.6 & 0.47 & 0.4 & 0.84 & 0.77 & 0.83 & 225 & double & [0.29, 89, 0.21, 199, 0.5] \\
NGC3521 & 14.9 & 0.54 & 0.39 & 0.72 & 0.7 & 0.91 & 150 & single & [0.41, 128, 0.59] \\
NGC3621 & 7.6 & 0.43 & 0.3 & 0.7 & 0.68 & 0.79 & 175 & double & [0.37, 89, 0.2, 199, 0.43] \\
NGC3627 & 14.3 & 0.49 & 0.37 & 0.76 & 0.79 & 0.93 & 200 & double & [0.28, 95, 0.24, 179, 0.48] \\
NGC4254 & 11.2 & 0.44 & 0.36 & 0.83 & 0.79 & 0.9 & 225 & double & [0.5, 89, 0.18, 500, 0.32] \\
NGC4298 & 7.2 & 0.49 & 0.38 & 0.76 & 0.76 & 0.9 & 225 & double & [0.46, 105, 0.11, 500, 0.43] \\
NGC4459* & 20.9 & 0.38 & 0.31 & 0.83 & 0.81 & 0.38 & 125 & single & [0.26, 123, 0.74] \\
NGC4476* & 19.0 & 0.28 & 0.19 & 0.7 & 0.65 & 1.12 & 100 & double & [0.17, 82, 0.52, 211, 0.31] \\
NGC4477* & 52.4 & 0.0 & 0.0 & 0.0 & 0.94 & 0.19 & 125 & single & [0.99, 117, 0.01] \\
NGC4536 & 5.7 & 0.38 & 0.27 & 0.71 & 0.72 & 0.89 & 225 & double & [0.14, 76, 0.43, 154, 0.43] \\
NGC4548 & 3.8 & 0.26 & 0.15 & 0.57 & 0.55 & 0.5 & 175 & double & [0.67, 94, 0.13, 500, 0.2] \\
NGC4569 & 14.2 & 0.52 & 0.4 & 0.77 & 0.91 & 0.89 & 350 & double & [0.16, 75, 0.29, 208, 0.55] \\
NGC4596* & 29.6 & 0.0 & 0.0 & 0.01 & 0.77 & 0.98 & 75 & single & [1, 180, 0] \\
NGC4731* & 4.6 & 0.14 & 0.07 & 0.47 & 0.5 & 0.37 & 125 & double & [0.48, 88, 0.44, 289, 0.08] \\
NGC4781 & 5.7 & 0.41 & 0.27 & 0.67 & 0.59 & 0.71 & 150 & double & [0.47, 98, 0.26, 500, 0.27] \\
NGC4826* & 27.9 & 0.56 & 0.33 & 0.59 & 0.62 & 0.8 & 75 & single & [0.39, 167, 0.61] \\
NGC4941 & 3.0 & 0.37 & 0.22 & 0.59 & 0.6 & 0.56 & 175 & double & [0.44, 89, 0.17, 199, 0.39] \\
NGC5068 & 2.9 & 0.17 & 0.08 & 0.49 & 0.52 & 0.5 & 75 & double & [0.44, 71, 0.39, 144, 0.17] \\
NGC5128 & 17.0 & 0.58 & 0.42 & 0.72 & 0.9 & 0.81 & 250 & double & [0.36, 91, 0.1, 500, 0.54] \\
NGC5236 & 10.2 & 0.44 & 0.34 & 0.77 & 0.75 & 0.91 & 175 & double & [0.48, 81, 0.09, 165, 0.43] \\
NGC5643 & 6.2 & 0.38 & 0.31 & 0.82 & 0.85 & 0.85 & 250 & double & [0.48, 80, 0.14, 107, 0.38] \\
NGC7456* & 2.9 & 0.2 & 0.04 & 0.22 & 0.27 & 0.15 & 100 & double & [0.34, 83, 0.42, 158, 0.24] \\
NGC7496 & 4.5 & 0.28 & 0.21 & 0.75 & 1.07 & 0.8 & 275 & single & [0.69, 97, 0.31] \\
NGC7793* & 2.9 & 0.26 & 0.12 & 0.45 & 0.42 & 0.21 & 100 & double & [0.18, 60, 0.54, 102, 0.28] \\
\hline
\end{tabularx}
\tablefoot{(1) Galaxy name, * indicates galaxies with $N_{\mathrm{GMC}}<50$. (2) Median of the integrated intensity values ($I_{\mathrm{CO, 150pc}}$) at GMC peaks. (3) The median of the average of $I_{\mathrm{CO, 150pc}}$/peak at 500 -- 1000~pc (normalized background) measured from data. (4) The median of the normalized background measured from ``Elliptical GMC" model maps. (5) Ratio between the normalized background measured from model and data (bkg$_{\mathrm{model}}$/bkg$_{\mathrm{data}}$) (6) The total flux ratio between model and data maps (7) The fraction of GMC flux extracted from \texttt{CPROPS} in total flux (from Hughes et al. in prep.). (8) The scale length when the stacked profiles of model and data differ by 10\%. (9) The function form chosen to fit the stacked data profiles, 'single' means single Gaussian plus a constant background and 'double' means double Gaussian plus a constant background (Eq. \ref{eq:gaus_fit_apdx}). (10) The preferred fitting parameters for a single Gaussian ($A$, $\sigma$, $C$) or a double Gaussian ($A_1$, $\sigma_1$, $A_2$, $\sigma_2$, $C$).}
\end{table*}

\end{appendix}
\end{document}